\documentclass[letterpaper,twocolumn,10pt]{article}
\usepackage{ligroup}
\usepackage{xspace}
\usepackage{subcaption}  
\usepackage{pifont}
\usepackage{multirow} 
\usepackage{tcolorbox}
\usepackage{graphicx}
\usepackage{placeins}
\usepackage{xcolor}
\usepackage{colortbl}
\usepackage{amsmath}
\usepackage{amssymb} 
\usepackage[absolute]{textpos}

\AtBeginDocument{
  }

\newcommand{\mypara}[1]{\smallskip\noindent\textbf{#1}}
\newcommand{\attack}{\textit{DCMI}\xspace}
\newcommand{\member}{\textit{member-retrieved documents}\xspace}
\newcommand{\nomember}{\textit{non-member-retrieved documents}\xspace}

\begin{document}
\begin{textblock}{15}(1.5,1)
To Appear in 2025 ACM SIGSAC Conference on Computer and Communications Security, October 13-17, 2025
\end{textblock}

\date{}

\title{\bf \textit{DCMI}: A Differential Calibration Membership Inference Attack Against Retrieval-Augmented Generation}

\author{
Xinyu Gao\textsuperscript{1} \and
Xiangtao Meng\textsuperscript{1}\thanks{Corresponding authors} \and
Yingkai Dong\textsuperscript{2} \and
Zheng Li\textsuperscript{1,3,4}\footnotemark[1] \and
Shanqing Guo\textsuperscript{1,3,4}\footnotemark[1]
\\[1ex]
\textsuperscript{1}\textit{School of Cyber Science and Technology, Shandong University} \\
\textsuperscript{2}\textit{Department of Engineering Software, School of Civil Engineering, Shandong University} \\
\textsuperscript{3}\textit{State Key Laboratory of Cryptography and Digital Economy Security, Shandong University} \\
\textsuperscript{4}\textit{Shandong Key Laboratory of Artificial Intelligence Security, Shandong University}
}

\maketitle

\begin{abstract}
While Retrieval-Augmented Generation (RAG) effectively reduces hallucinations by integrating external knowledge bases, it introduces vulnerabilities to membership inference attacks (MIAs), particularly in systems handling sensitive data.  
Existing MIAs targeting RAG’s external databases often rely on model responses but ignore the interference of \nomember on RAG outputs, limiting their effectiveness.  
To address this, we propose \attack, a differential calibration MIA that mitigates the negative impact of \nomember. 
Specifically, \attack leverages the sensitivity gap between \textit{member} and \textit{non-member retrieved documents} under query perturbation. It generates perturbed queries for calibration to isolate the contribution of \member while minimizing the interference from \nomember. 
Experiments under progressively relaxed assumptions show that \attack consistently outperforms baselines—for example, achieving 97.42\% AUC and 94.35\% Accuracy against the RAG system with Flan-T5, exceeding the MBA baseline by over 40\%.  
Furthermore, on real-world RAG platforms such as Dify and MaxKB, \attack maintains a 10\%-20\% advantage over the baseline.  
These results highlight significant privacy risks in RAG systems and emphasize the need for stronger protection mechanisms.
We appeal to the community’s consideration of deeper investigations, like ours, against the data leakage risks in rapidly evolving RAG systems.\footnote{Our code is available at \url{https://github.com/Xinyu140203/RAG_MIA}.}
\end{abstract}

\section{Introduction}\label{sec:Intro}

Large language models (LLMs)~\cite{NEURIPS2020_1457c0d6,hurst2024gpt,anil2023palm,touvron2023llama} have demonstrated impressive text generation, comprehension, and reasoning abilities. 
However, LLMs' reliance on static, pre-trained knowledge inherently limits their generation of accurate, up-to-date, or domain-specific information.
To address these limitations, Retrieval-Augmented Generation (RAG)~\cite{lewis2020retrieval,shi2024replug,ram2023incontext,vanveen2024adapted,karpukhin2020dense,borgeaud2022trillions,thoppilan2022lamda} has emerged as a powerful framework that integrates the strengths of retrieval-based systems and generative models. 
By dynamically retrieving relevant information from external knowledge sources and leveraging it to enhance text generation, RAG bridges the gap between the vast knowledge stored in databases and the generative prowess of LLMs. 

While RAG has been widely adopted across various domains, its reliance on external databases—particularly those containing sensitive or high-risk data—introduces significant challenges.
Concretely, in healthcare, systems like \textit{SMART Health GPT} integrate sensitive medical data from external sources to enhance diagnostics and provide personalized recommendations, raising concerns about patient privacy and data security~\cite{wang2024potential,alghadban2023transforming}. 
Similarly, financial institutions use RAG to analyze transaction data for risk assessment, which may expose sensitive financial information~\cite{loukas2023banking}. 
In the legal domain, tools like \textit{AutoLaw} leverage case details and private data stored externally to improve document review and advisory services, posing risks of unauthorized access or misuse~\cite{mahari2021autolaw,kuppa2023chain}.
These challenges prompt the need to understand privacy risks and unauthorized data usage in RAG systems.

To this end, we focus on membership inference attacks (MIAs), which have proven to be a critical lens for studying privacy risks and unauthorized data usage in traditional machine learning (ML) and LLMs~\cite{carlini2022membership,amit2024sok,hu2022membership,liu2022membership,mattern2023membership}. 
These attacks reveal if a data sample is in a model's training set, exposing vulnerabilities. 
In this work, we investigate a distinct angle: \textit{whether a specific data sample is part of RAG's retrieval database.}

\begin{figure}[t!]
  \centering
  \includegraphics[width=\linewidth]{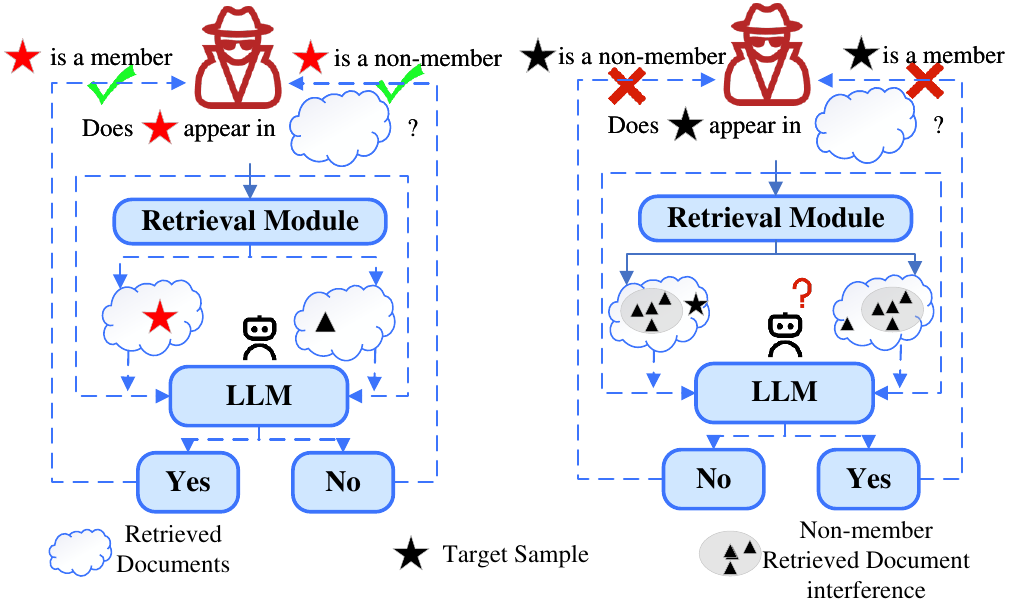}
  \caption{
The interference from \nomember compromises the reliability of membership inference signals in RAG systems.
For member queries, the presence of \nomember can degrade response quality, leading to misclassification as a non-member. For non-member queries, the retrieved \nomember may provide sufficient context to generate high-quality responses, causing misclassification as a member.}
  \label{fig:motivation}
\end{figure}

Specifically, given a query, RAG first retrieves the top-k semantically similar documents and then appends them to the query as the context for response generation.
In the MIAs scenario, these retrieved documents fall into two categories:
(1) \member, which exactly matches the query, indicating the query's presence in the retrieval database
(2) \nomember, which are semantically similar but do not exactly match the query.
If the query exists in the database, the semantic search prioritizes retrieving the matching \member along with relevant \nomember. If the query is absent, only \nomember are retrieved.
Therefore, determining whether a sample exists in the database reduces to checking whether the retrieved documents include a \member.
In theory, \member provide more accurate information because they exactly match the query, leading to higher-quality responses with greater confidence. Existing MIA methods~\cite{anderson2024membership,li2024blackbox,liu2024mask,naseh2025riddle} exploit this assumption by analyzing response characteristics to infer membership.
However, this signal is unreliable in RAG systems, as shown in \autoref{fig:motivation}. For member queries, the presence of \nomember can degrade response quality, leading to misclassification as a non-member. For non-member queries, the retrieved \nomember may provide sufficient context to generate high-quality responses, causing misclassification as a member.
These issues suggest that \nomember interfere significantly with MIAs.
\autoref{sec:analysis_long_tail} presents a quantitative analysis of this interference.

To reduce interference from \nomember, we propose a novel MIA based on Differential Calibration, called \attack.
Our method uses the RAG system's confidence in the correctness of a target sample as the inference signal, measured by the probability of the system responding ``yes''.
Our attack is based on a key observation: \member are highly sensitive to perturbations, leading to significant drops in their contribution to the confidence score. 
In contrast, \nomember remain stable, contributing consistently even after perturbation (see details in \autoref{General Paeadigm}).

Concretely, given an original query, a third-party LLM first generates its perturbed version. 
Since the perturbation significantly weakens the influence of \member, the perturbed version’s confidence score mostly reflects the contribution of \nomember. 
Subtracting this perturbed score from the original query’s score cancels out the \nomember influence, thereby isolating the unique contribution of \member and yielding a more accurate membership discrimination signal.

We evaluate three increasingly realistic adversary models on two benchmark datasets and various RAG systems.  
\attack consistently outperforms all baselines across scenarios.  
In the Adversary 1 setting, \attack achieves 97.42\% AUC and 94.35\% Accuracy on a Flan-T5-based RAG system, exceeding the MBA baseline by over 40\%.  
Analysis shows that the differential calibration module boosts performance by 7\%, while \attack remains robust to changes in perturbation magnitude and threshold.  
On real-world RAG systems (Dify, MaxKB), \attack maintains a 10\%–20\% lead, reaching up to 74\% attack success on MaxKB.

In summary, our contributions are as follows:
\begin{itemize}
    \item 
We are the first to analyze the negative impact of \nomember, a RAG-specific phenomenon, on MIA performance. This offers new insights into their influence on data-related risks in RAG frameworks.
    \item 
We introduce \attack, a differential-calibration MIA that enhances discriminative signals in RAG by mitigating the interference of \nomember. 
We progressively relax assumptions to expand the \attack's applicability, ultimately achieving a fully realistic, end-to-end attack scenario.

    \item
We conduct extensive experiments across diverse local and real-world RAG configurations, focusing on three adversary models. 
Results show that \attack significantly outperforms existing baselines and achieves 10\% better attack effectiveness on real-world RAG systems.
Additionally, we propose two defense strategies for partial risk mitigation.
\end{itemize}

\section{Preliminaries and Related Works}

\subsection{Retrieval-Augmented Generation}
\label{sec:RAG}
Retrieval-Augmented Generation (RAG) ~\cite{lewis2020retrieval,shi2023replug,ram2023context,karpukhin2020dense} has emerged as a key technique for enhancing LLMs by reducing hallucinations. 
It integrates real-time, domain-specific knowledge, offering a cost-effective way to improve the relevance, accuracy, and utility of LLMs across systems. 
A RAG system $\mathcal{S}$ consists of three core components: an external retrieval database $\mathcal{R}_{D}$, a retriever $\mathcal{R}$, and a generative module $\mathcal{G}$. 
Given a user query $q$, the system augments $q$ with the retrieved relevant documents $D_r$.
The system operates through a structured process:

\mypara{Creating External Data.} 
External data $\mathcal{R}_{D}$, sourced from APIs, databases, or document repositories, lies outside the original LLM training set. 
The retriever $\mathcal{R}$ encodes this data into numerical vectors, storing them in a vector database to create a structured knowledge base for the generative module.

\mypara{Retrieving Relevant Information.} 
During operation, the system $\mathcal{S}$ performs a similarity search when a user submits a query. 
The query $q$ is converted into a vector representation $e_q$, which is compared with the vector representations $e_{d_i}$ of samples $d_i$ in the database $\mathcal{R}_{D}$ to retrieve the most relevant records. 
Common similarity metrics, such as cosine similarity, Euclidean distance, and $L_2$-norm distance, enable the retriever to return the top-$k$ results ($D_r$) with minimal distance to the query vector. 
The retrieval step can be formulated as follows:
\[
D_r=\mathcal{R}(q, \mathcal{R}_{D}) = \{ d_i \in \mathcal{R}_{D} \mid \text{dist}(e_q, e_{d_i}) \text{ is in the top } k \},
\]

\mypara{Augmenting.} Finally, the system $\mathcal{S}$ augments the user query $q$ by integrating retrieved data $D_r$, creating an enriched prompt for the LLM. 
The LLM then processes this prompt, leveraging contextual knowledge to produce a precise and relevant response:
\[
a = \mathcal{G}(D_r \oplus q)
\]
Here, \(\oplus\) represents the integration of $D_r$ and \(q\)—typically by concatenating them together ~\cite{ram2023context}.
The subsequent sections adopt these RAG symbol definitions. Table~\ref{tab:notation} (\autoref{app:notation}) summarizes the notation introduced here along with other key symbols used throughout this paper.

\subsection{Membership Inference Attacks}
Membership inference attacks (MIAs) determine whether a specific data sample belongs to a target model’s training dataset, called a member or a non-member~\cite{li2025enhanced,hu2025membership,li2024seqmia,li2024membership,li2022auditing,he2022membership,wu2022membership}. 
This approach is crucial for investigating privacy leaks and detecting unauthorized data usage. 
From traditional ML to LLMs, researchers have proposed various attack strategies~\cite{hu2022membership,shokri2017privacy,yeom2018privacy,carlini2021extracting,mattern2023membership,shi2023detecting}.

\mypara{Membership inference attack in RAG.}  
In RAG systems, the definition of MIAs differs from that in traditional ML and LLMs. 
Unlike LLMs, which rely solely on learned knowledge, RAG systems also retrieve documents from external databases to generate responses. 
Thus, MIAs in RAG can infer not only whether a sample is part of the LLM's training dataset but also whether it exists in the retrieval database. 
In this study, we focus on the latter: \textit{determining whether a specific data sample (e.g., a text or document) is included in the RAG's retrieval database.}

Current MIAs against the RAG retrieval database~\cite{anderson2024membership,li2024blackbox,liu2024mask,naseh2025riddle} assume that retrieval results containing \member yield higher-quality or more confident responses. 
However, they overlook interference from \nomember, relying solely on the RAG system’s response as the membership signal, which reduces attack accuracy.

\mypara{Calibration-based Membership Inference Attacks.} Calibration techniques have been widely adopted to improve MIAs' accuracy, with difficulty calibration as the mainstream approach~\cite{carlini2022membership,long2018understanding,watson2021importance}. 
Watson et al.~\cite{watson2021importance} demonstrated that directly using loss values as membership signals is problematic, as low-loss samples may simply be inherently easy non-members rather than true training set members, so they proposed calibrating the raw membership scores.

Typically, such calibration techniques utilize reference or shadow models to compare membership scores under different training conditions, thereby eliminating the influence of intrinsic sample difficulty. 
Mattern et al.~\cite{mattern2023membership,fu2024membership} later enhanced the practicality of this approach by substituting neighborhood comparisons in the feature space for explicit reference models.

Unlike difficulty calibration, our differential calibration method is tailored for RAG systems.
It leverages the sensitivity gap between \textit{member} and \nomember under query perturbation to eliminate interference from \nomember and isolate the contribution of \member---thus providing a more accurate membership signal.

\section{Problem Statement}

\subsection{Threat Model}\label{sec:sec 3.1}
\mypara{Target RAG System.} The target RAG system $\mathcal{S}$ processes user queries using a private knowledge database \(\mathcal{R}_{D}\), whose contents are kept confidential to protect intellectual property.

\mypara{Adversary’s Goal.} 
The adversary's goal is to determine whether a target sample \( x \) exists in a RAG's retrieval database \( \mathcal{R}_{D} \). 
Given the adversary's external knowledge \( \Omega \), a membership inference attack \( \mathcal{A} \) against \( \mathcal{R}_{D} \) for \( x \) is defined as:
\[
\mathcal{A}: x, \mathcal{R}_{D}, \Omega \rightarrow \{0, 1\},
\]
where 0 indicates \( x \) is not in \( \mathcal{R}_{D} \), and 1 indicates it is. The specific form of \( \Omega \) is discussed in the next section.

\mypara{Adversary’s Capabilities.}
\autoref{tab:threat_models} presents three variations of the threat model, each based on different levels of adversarial capability. 
We progressively relax the assumptions used in prior work, ultimately reaching an adversary that is independent of both the model and data.

\begin{table}[t!]
\centering
\caption{Different threat models and their
assumptions.}
\label{tab:threat_models}
\small
\setlength{\tabcolsep}{2.5pt} 
\begin{tabular}{@{} l c c c c @{}}
\toprule
\textbf{Threat Model} &  
\multicolumn{1}{c}{\textbf{Partial Data}} & 
\multicolumn{1}{c}{\textbf{Data Distr.}} & 
\multicolumn{1}{c}{\textbf{Output Prob.}} &
\multicolumn{1}{c}{\textbf{Response}} \\
\midrule
Adversary 1  &  \ding{51} & \ding{51} & \ding{51} & \ding{51} \\ 
Adversary 2   & \ding{55} & \ding{51} & \ding{51} & \ding{51} \\
Adversary 3    & \ding{55} & \ding{55} & \ding{55} & \ding{51} \\
\bottomrule
\end{tabular}
\end{table}

\begin{itemize}
\item Adversary 1. Following prior work, we first consider a gray-box setting where the adversary can query the target RAG system \(\mathcal{S}\) and observe its output log probabilities. The adversary also has access to a small subset of \(\mathcal{S}\)'s retrieval database (e.g., 1,000 samples). 
Note that these assumptions are consistent with existing MIA studies on RAG systems~\cite{anderson2024membership,li2024blackbox}.

\item Adversary 2. We relax the requirement for access to a subset of \(\mathcal{S}\)'s retrieval database. Instead, the adversary is given a reference dataset that shares the same distribution but contains no overlapping content. This represents a more realistic and challenging scenario.  
We emphasize this is not a strong assumption, as many publicly available RAG systems provide introductory content and example metadata, which adversaries can use to approximate the data distribution~\cite{jiang2024rag}.

\item Adversary 3. We further assume that the adversary cannot access output log probabilities or any information about the RAG retrieval database and can only process the target query they aim to infer.  
In this scenario, the adversary relies solely on the text response returned by \(\mathcal{S}\). 
This represents the most challenging scenario and ensures that our evaluation provides meaningful real-world insights.

\end{itemize}

\subsection{The Interference of \nomember} \label{sec:analysis_long_tail}

In addition to the aforementioned qualitative analysis of the interference of \nomember in \autoref{sec:Intro}, we here conduct a quantitative analysis, both theoretical and experimental.

Concretely, unlike traditional LLMs, RAG systems integrate a query $q$ with a set of retrieved documents $D_r$, formalizing the generation process as follows ~\cite{cuconasu2024power}:
\begin{equation}
P_{\text{rag}}(y \mid q) \approx \prod_{i}
\sum_{d \in D_r} p_\eta(d \mid q)p_\theta
(y_i \mid q, d, y_{1:i-1}),
\label{eq:rag}
\end{equation}
where $p_\eta(d | q) = \frac{e^{s(d, q)}}{\sum_{d \in D_r} e^{s(d, q)}}$ is computed based on the similarity \(s(d, q)\) between a document \(d\) and the query \(q\), indicating how relevant the document is to the query~\cite{shi2024replug}. The term \(p_\theta(y_i \mid q, d, y_{1:i-1})\) denotes the conditional probability of generating the \(i\)-th token, given \(d\) as contextual support.
RAG's response is essentially a weighted sum of the outputs from all retrieved documents, where the weights reflect each document’s relevance to the query. Let \(d^*\) denote the document most semantically similar to \(q\). Thus, the MIA task reduces to determining the membership status of \(d^*\): for member samples, \(d^*\) is a \member; for non-member samples, \(d^*\) is the closest \nomember.
\autoref{eq:rag} can be simplified as:
\begin{equation}
P_{\text{rag}}(y \mid q) \approx \prod_{i}\Big[p_\eta(d^* \mid q)\cdot p_\theta(y_i \mid q, d^*, y_{1:i-1}) + \epsilon_i\Big],
\label{eq:rag_rewrite}
\end{equation}
Where $\epsilon_i$ captures the contributions from all other \nomember:
\begin{equation} \epsilon_i = \sum_{d \in D_r \setminus \{d^*\}} p_\eta(d \mid q) \cdot p_\theta(y_i \mid q,d,y_{1:i-1}) 
\label{eq:interference}
\end{equation}

\begin{figure}[t!]
    \centering
    \begin{subfigure}[b]{0.48\linewidth}
        \includegraphics[width=\linewidth]{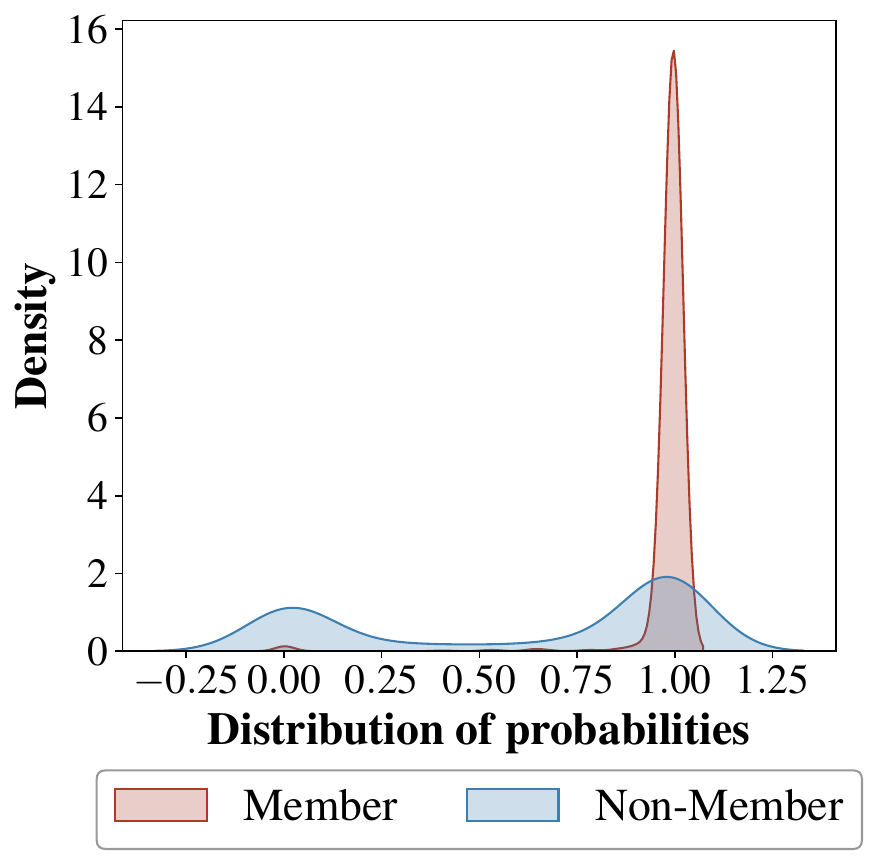}
        \caption{One document retrieved}
        \label{fig:No tail}
    \end{subfigure}
    \hfill
    \begin{subfigure}[b]{0.48\linewidth} 
        \includegraphics[width=\linewidth]{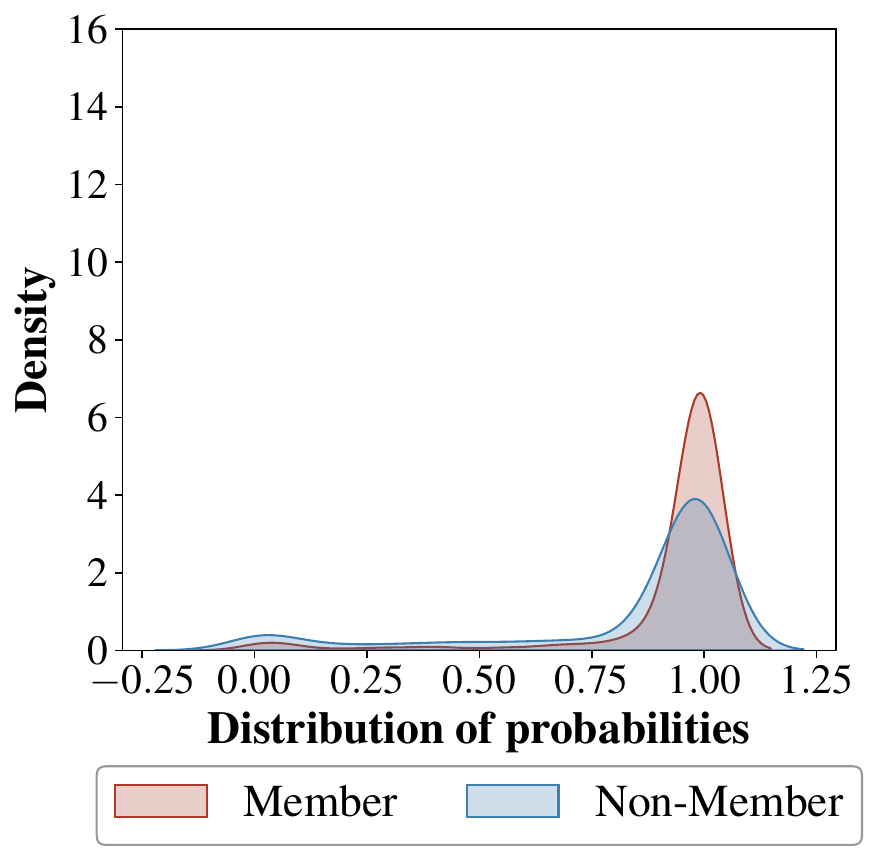} 
        \caption{Four documents retrieved}
        \label{fig:With tail}
    \end{subfigure}
    \caption{Distribution of RAG-MIA~\cite{anderson2024membership} generation probabilities $P_\text{rag}(y \mid q)$ (\autoref{eq:rag_rewrite})
    for member and non-member samples under (a) ideal conditions (one retrieved document), (b) practical conditions (four retrieved documents).
    The x-axis represents the membership score  $P_\text{rag}(y \mid q)$; the y-axis represents the probability density.
    }
    \label{fig:validate tail effect}
\end{figure}

In the ideal case, where \(\epsilon_i \approx 0\), the generation probabilities $P_{\text{rag}}(y \mid q)$ directly reflects the membership status of \(d^*\). Under this assumption, existing response-based MIAs~\cite{anderson2024membership,li2024blackbox,liu2024mask,naseh2025riddle} perform effectively.  
To simulate this scenario, we conduct experiments under the \textit{Basic RAG Setting} (see \autoref{sec:Experimental Setup}), using the RAG-MIA baseline~\cite{anderson2024membership} as an example. 
We retrieve only one document (at this point $\epsilon_i=0$), although typical settings use more than one.  
As shown in \autoref{fig:No tail}, using $P_{\text{rag}}(y \mid q)$ as the membership score clearly distinguishes between member and non-member samples.

However, in practice, the \(\epsilon_i\) term is significant, as it captures the influence of all \nomember other than \(d^*\). 
As shown in \autoref{fig:With tail}, increasing the number of retrieved documents to four substantially increases \(\epsilon_i\), causing greater interference. 
This creates significant overlap in $P_{\text{rag}}(y \mid q)$  membership scores between member and non-member samples. Such overlap blurs the decision boundary and reduces the effectiveness of traditional MIAs.

\begin{figure}[t!]
    \centering
    \begin{subfigure}[b]{0.49\linewidth} 
        \includegraphics[width=1.0\linewidth]{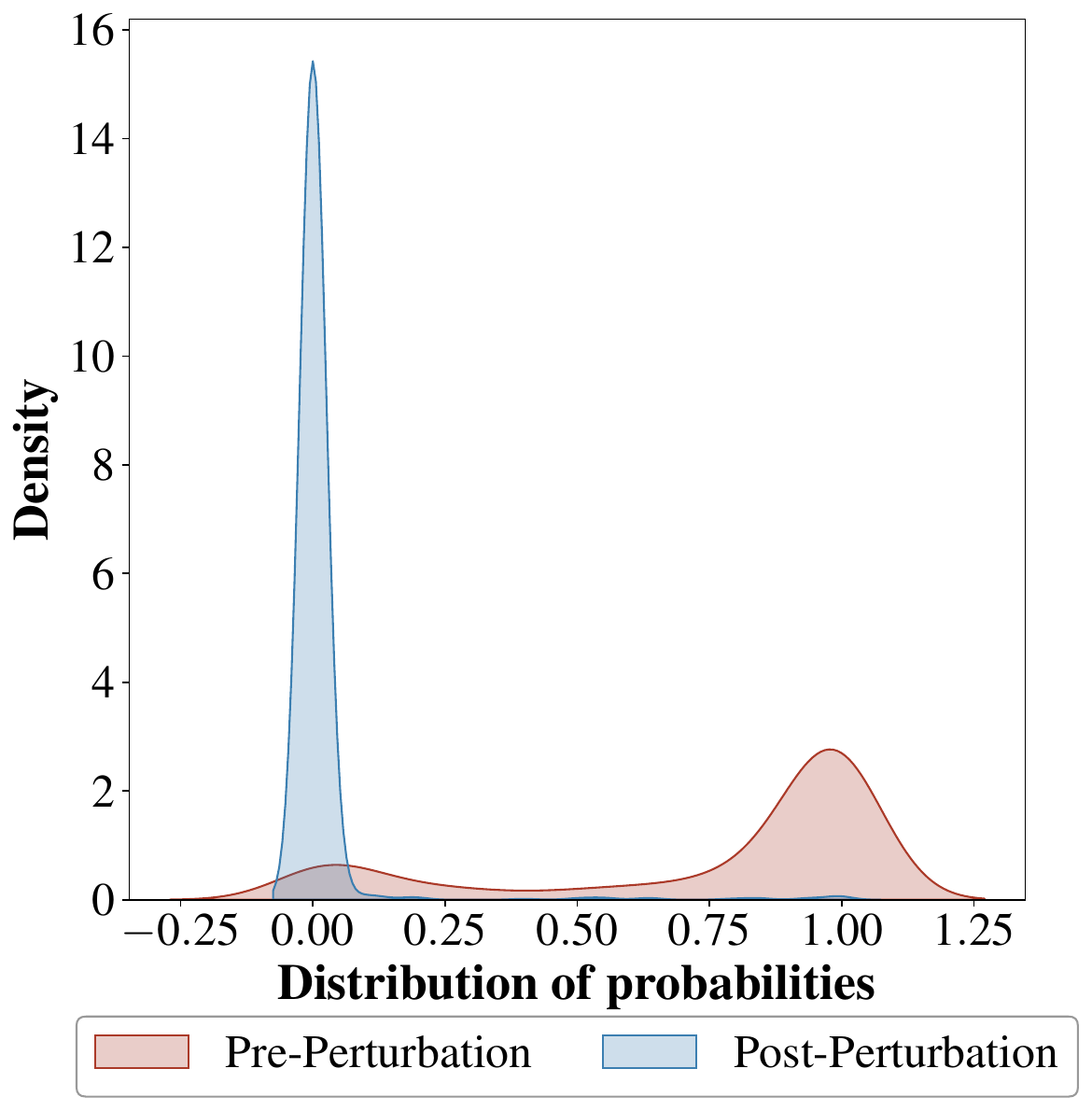} 
        \caption{Member}
        \label{fig:headdistribution}
    \end{subfigure}
    \begin{subfigure}[b]{0.49\linewidth}
        \includegraphics[width=1.0\linewidth]{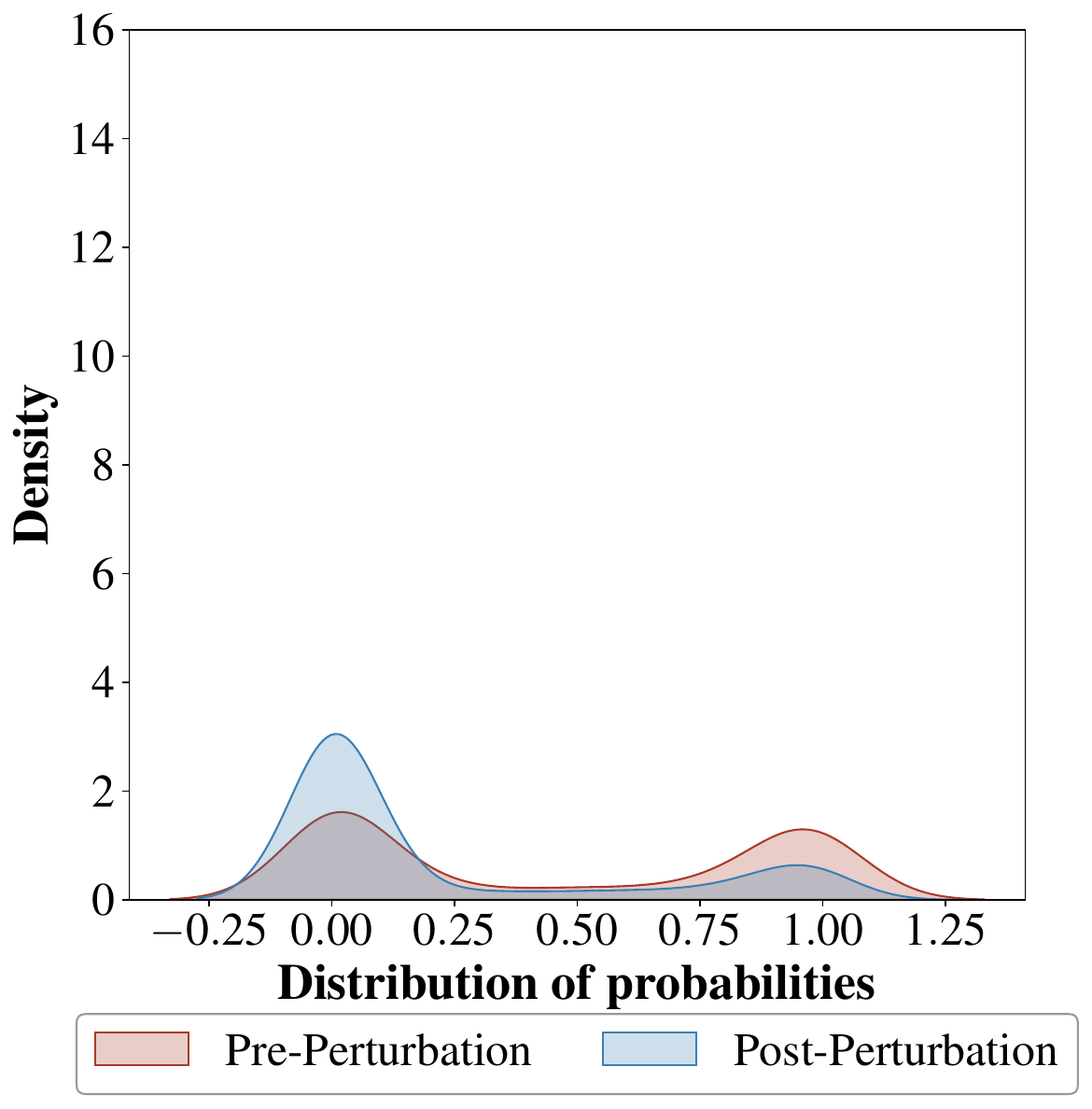} 
        \caption{Non-member}
        \label{fig:taildistribution}
    \end{subfigure}
    
    \caption{Distribution of conditional probabilities $p_\theta(\text{``Yes''} \mid q, d)$ for \member and \nomember Pre- and Post-Perturbation.
    The x-axis represents $p_\theta(\text{``Yes''} \mid q, d)$; the y-axis represents the probability density.
    }
    \label{fig:distribution}
\end{figure}

\begin{figure*}[t!]
  \centering
  \includegraphics[width=\textwidth]{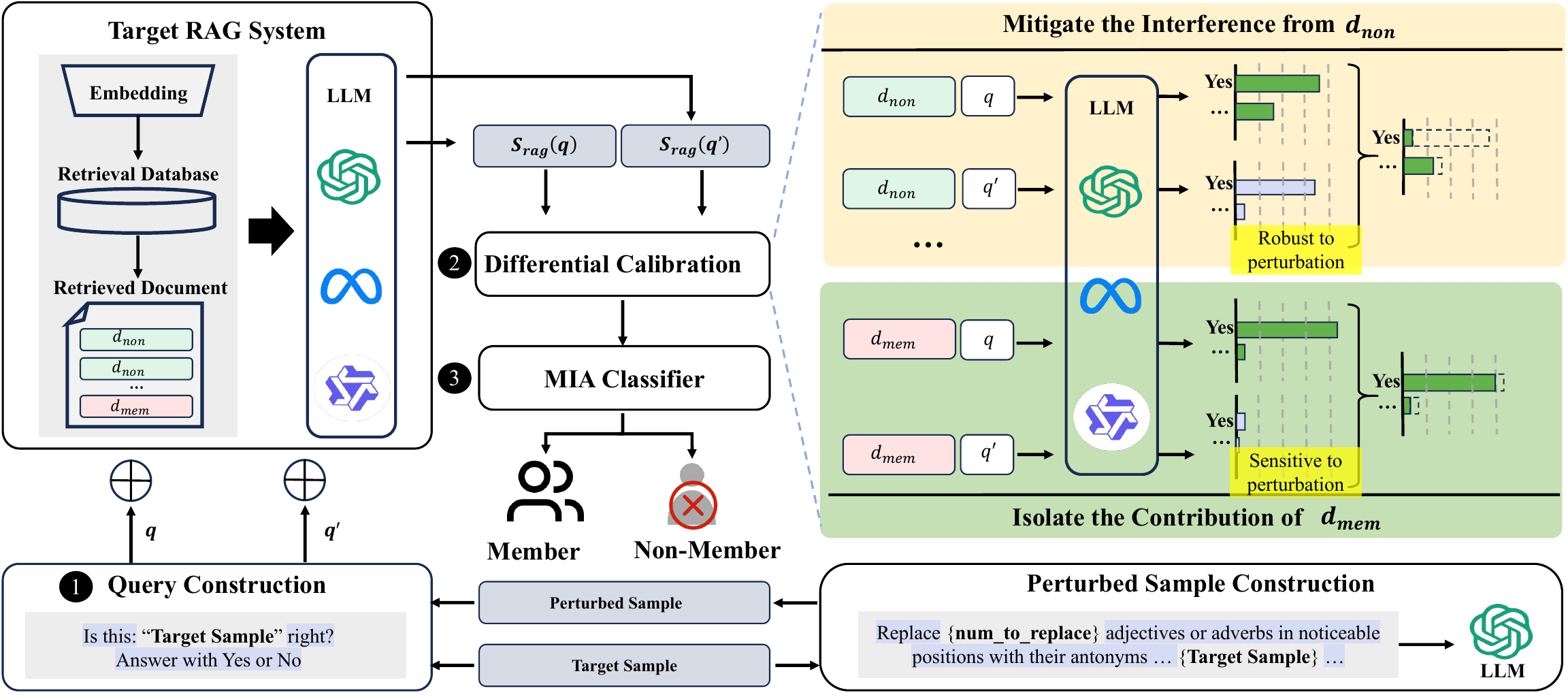}
  \caption{ The overall architecture of our proposed \attack attack. 
  $S{_\text{rag}}(\cdot)$ denotes the RAG system's output signal.
  For Adversaries 1 and 2, this signal consists of continuous probabilities $P_{\text{rag}}(\text{``Yes''} \mid \cdot)$, while Adversary 3 receives discrete binary responses (``Yes'' or ``No'').
  The differential calibration module precisely adjusts the weighted contributions of each retrieved document to the final response, mitigating the interference from \nomember ($d_{non}$) while isolating the distinctive contributions of \member ($d_{mem}$).}
  \label{fig:pipeline}
\end{figure*}

\section{Overall Attack Procedure of \attack}
This section presents our Membership Inference Attack based on Differential Calibration (\attack).

In \autoref{General Paeadigm}, we prove that differential calibration eliminates the interference from \nomember identified in \autoref{sec:analysis_long_tail}, isolating the unique contribution of \member. 
This theoretical analysis establishes a novel membership inference signal and demonstrates its effectiveness in distinguishing members from non-members. 

Building upon this foundation, \autoref{Method outline} outlines the methodology of \attack, detailing how to implement the proposed membership signal in practice.

\subsection{Theoretical Analysis of Differential Calibration}\label{General Paeadigm}

In \autoref{eq:interference}, the interference terms $\epsilon_i$ depend on previous outputs $y_{1:i-1}$. 
This chain dependency causes calibration errors to accumulate throughout the sequence, making precise calibration difficult for long outputs. 
To address this issue, we reformulate the generation task as a binary classification problem. Specifically, we design the query $q$ as a verification question that elicits a ``Yes''/``No'' response, effectively eliminating the sequential dependency of multi-token outputs. 
The generation probabilities $P_\text{rag}(y \mid q)$ from \autoref{eq:rag_rewrite} become:

\begin{equation}
P_{\text{rag}}(\text{``Yes''} \mid q) \approx p_\eta(d^* \mid q) \cdot p_\theta(\text{``Yes''} \mid q, d^*) + \epsilon,
\label{eq:rag_rewrite1}
\end{equation}
Where \(\epsilon= \sum_{d \in D_r \setminus \{d^*\}} p_\eta(d \mid q) \cdot p_\theta(\text{``Yes''} \mid q,d)\).
We create a perturbed version $q'$ by applying minimal perturbations to the original query $q$ and then perform differential calibration by subtracting $P_{\text{rag}}(\text{``Yes''} \mid q')$ from $P_{\text{rag}}(\text{``Yes''} \mid q)$.
The effectiveness of our calibration mechanism relies on two key assumptions:

First, as shown in~\cite{hsieh2019robustness}, query word perturbations exhibit sparse propagation and cause minimal semantic changes, i.e.,  
\(
\|\delta\| \ll \|q\|,
\) 
which preserves the stability of document similarity scores (see \autoref{app:robustness}),  
\(
p_{\eta}(d \mid q) \approx p_{\eta}(d \mid q').
\)

Second, inspired by~\cite{mattern2023membership},
we hypothesize the conditional probability \(p_\theta(\text{``Yes''}\mid q, d)\) is more sensitive to query perturbations for \member than for \nomember.
To verify this, we conduct experiments under the \textit{Basic RAG Setting} (details in \autoref{sec:Experimental Setup}), with a retrieval quantity of 1, at this point $\epsilon=0$, $p_{\eta}(d \mid q)=1$, so $P_{\text{rag}}(\text{``Yes''} \mid q) \approx p_\theta(\text{``Yes''} \mid q, d)$.
\autoref{fig:headdistribution} shows that for \member, the generation probabilities $P_{\text{rag}}(\text{``Yes''} \mid q)$ 
 or equivalently $p_\theta(\text{``Yes''} \mid q, d)$ differ significantly before and after perturbation, with post-perturbation probabilities concentrated near the original
values, i.e., \( p_\theta(\text{``Yes''} \mid q, d) \gg p_\theta(\text{``Yes''} \mid q', d) \), whereas for \nomember (\autoref{fig:taildistribution}), $P_{\text{rag}}(\text{``Yes''} \mid q)$ 
 or equivalently $p_\theta(\text{``Yes''} \mid q, d)$ remain essentially unchanged after perturbation, i.e., \(p_\theta(\text{``Yes''} \mid q, d) \approx p_\theta(\text{``Yes''} \mid q', d) \).

Based on the above proof, we define the calibration formula as follows:
\begin{equation}
\label{eq:calibration}
\begin{split}
P_{\text{rag, calibrated}}(\text{``Yes''} \mid q) &= P_{\text{rag}}(\text{``Yes''} \mid q) - P_{\text{rag}}(\text{``Yes''} \mid q') \\
&= p_{\eta}(d^* \mid q) \, \Delta p_\theta(\text{``Yes''} \mid q, d^*) + \Delta \epsilon
\end{split}
\end{equation}
where \(\Delta \epsilon=\sum_{d \in D_r \setminus \{d^*\}} p_\eta(d|q) [p_\theta(\text{``Yes''}|q,d)-p_\theta(\text{``Yes''}|q',d)]\). $\epsilon$ captures the contributions from all other \nomember, according to \( p_\theta(\text{``Yes''} \mid q, d) \approx p_\theta(\text{``Yes''} \mid q', d) \), \(\Delta \epsilon \approx 0\).
Besides, \(\Delta p_\theta(\text{``Yes''} \mid q, d^*)=p_\theta(\text{``Yes''} \mid q, d^*)-p_\theta(\text{``Yes''} \mid q', d^*)\).
When \( d^* \) is a \member, we have  
\(
p_\theta(\text{``Yes''} \mid q, d) \gg p_\theta(\text{``Yes''} \mid q', d),
\)  
so that  
\(
\Delta p_\theta(\text{``Yes''} \mid q, d^*) = p_\theta(\text{``Yes''} \mid q, d^*) - p_\theta(\text{``Yes''} \mid q', d^*) \approx p_\theta(\text{``Yes''} \mid q, d^*).
\)
In contrast, when \( d^* \) is a \nomember, we have  
\(
p_\theta(\text{``Yes''} \mid q, d) \approx p_\theta(\text{``Yes''} \mid q', d),
\)  
which implies  
\(
\Delta p_\theta(\text{``Yes''} \mid q, d^*) \approx 0.
\)

The final calibrated probability eliminates the influence of \nomember, isolating the unique contribution of the \member:
\begin{equation}
\label{eq:signal_Adversary1}
P_{\text{rag, calibrated}}(\text{Yes} \mid q) \approx \mathbb{I}_{\{q \in \mathcal{M}\}} \cdot p_{\eta}(d^* \mid q) \cdot p_\theta(\text{Yes} \mid q, d^*),
\end{equation}
where $\mathcal{M}$ denotes the member set, and $\mathbb{I}_{\{q \in \mathcal{M}\}}$ is the indicator function satisfying $\mathbb{I}_{\{q \in \mathcal{M}\}} = 1$ if $q \in \mathcal{M}$ else $0$.
Thus, we have derived the calibrated membership signal used by our \attack method in Adversary 1 and Adversary 2.

For practical black-box scenarios in Adversary 3, we only have access to final responses rather than generation probabilities. Under binary response settings, the RAG system outputs ``Yes'' when $P_{\text{rag}}(\text{``Yes''} \mid q) > \alpha$ (where $\alpha$ is the decision boundary, typically 0.5) and ``No'' otherwise.
Although this discretization inevitably loses fine-grained probability information, the sensitivity gap between members and non-members under query perturbation still enables calibration.

Our key finding is that for member samples, perturbation-induced probability drops---$p_{\eta}(d^* \mid q) \, \Delta p_\theta(\text{``Yes''} \mid q, d^*)$---are sufficiently large to cause $P_{\text{rag}}(\text{``Yes''} \mid q)$ and $P_{\text{rag}}(\text{``Yes''} \mid q')$ to fall on opposite sides of the decision boundary  $\alpha$, with $P_{\text{rag}}(\text{``Yes''} \mid q)>\alpha$ and $P_{\text{rag}}(\text{``Yes''} \mid q')<\alpha$, resulting in decision reversal (from ``Yes'' to ``No'').
For non-member samples, probabilities remain on the same side of the decision boundary before and after perturbation.

Based on this observation, we map outputs to binary logical values using $f_{\text{rag}}(\cdot)$ to map ``Yes'' to 1 and ``No'' to 0. We define the binary calibration score for Adversary 3:
\begin{equation}
f_{rag,calibrated}(q) = f_{\text{rag}}(q) - f_{\text{rag}}(q'),
\end{equation}
This score effectively distinguishes members from non-members. Member samples yield a score of 1 due to decision reversal, while non-member samples yield 0 due to consistent decisions or -1 occasionally.
\autoref{app:Adversary3_analysis} provides detailed theoretical analysis and empirical validation for the black-box scenario.

\subsection{Methodology Outline}\label{Method outline}
 
Based on the previous analysis, we outline the methodology of the proposed \attack.
\autoref{fig:pipeline} illustrates the overall pipeline, which consists of three key steps:

\begin{itemize}

    \item \textbf{Query Construction.} The adversary selects a target sample (e.g., a medical record), constructs a query \(q\) using a template, and inputs it into the RAG system to obtain the output signal $S_{\text{rag}}(q)$ (continuous probabilities or discrete ``Yes''/``No'' response).

    \item \textbf{Differential Calibration.} The adversary uses a large language model to perturb the target sample, constructs a perturbed query \(q'\) using the modified sample and a template, and submits it to the RAG system to obtain the perturbed output signal $S_{\text{rag}}(q')$. 
    The adversary then calibrates the original query \(q\)'s output signal by subtracting the signal from the perturbed query \(q'\).

    \item \textbf{Membership Inference} Finally, the adversary sets a classification threshold for membership inference.
\end{itemize}
The detailed \attack methodologies for Adversary 1, 2, and 3 are elaborated in \autoref{sec:Adversary1_method}, \autoref{sec:Adversary2_Mothod}, and \autoref{sec:Adversary3_method}, respectively.

\section{Gray-Box Membership Inference}
Adversary 1 operates in a gray-box setting, querying the RAG system with a target sample and observing the output log probabilities. The adversary also has access to a small subset of the retrieval database, which is used as a reference dataset to determine the perturbation magnitude and classification threshold.

\subsection{Methodology}\label{sec:Adversary1_method}

\mypara{Query Construction.}
In all three scenarios, the adversaries can only modify the query prompt to interact with the RAG system. The query must (a) ensure the retrieval module accurately retrieves the target sample, and (b) function as a verification question that produces binary ``Yes''/``No'' responses, as shown in \autoref{General Paeadigm}. 
To meet these goals, we design a simple yet effective adversarial query as follows.
\begin{tcolorbox}[
    colback=white,              
    colframe=black,             
    width=\linewidth,           
    title=Query Construction Template,        
    boxrule=0.5mm,              
    colbacktitle=black,         
    coltitle=white,             
    left=5pt,                   
    right=5pt,                  
    arc=3mm,                    
    before skip=10pt,           
    after skip=10pt            
]
\footnotesize
\textbf{Is this:}  
\texttt{``{Target Sample}''}  
\textbf{right? Answer with Yes or No.}
\end{tcolorbox}

This template represents one viable approach; \autoref{app:Prompt_Robustness} demonstrates that alternative templates meeting these criteria are equally effective.
The constructed query $q$ is then input to the RAG system to obtain output probabilities for ``Yes'' tokens.

\mypara{Differential Calibration.}
Based on the analysis in \autoref{General Paeadigm}, we propose a differential calibration module to minimize the impact of \nomember.

Specifically, we use a large language model \(L\) to generate a perturbed version of the original sample. 
The model selects a few keywords from the target sample---typically adjectives or adverbs---and replaces them with antonyms to introduce slight semantic contradictions while maintaining logical consistency. The proportion of replaced words (i.e., perturbation magnitude) is a hyperparameter selected through grid search (see next step). The perturbation template is provided in \autoref{sec:perturbation_template}.
While we focus on textual data here, \attack can be adapted to non-textual data by adjusting perturbation strategies and magnitudes according to data type characteristics (see \autoref{app:non_textual_adaptation} for details).

Next, we construct a perturbed adversarial query \( q' \) using the modified sample and the same template from the Query Construction step, and input it into the RAG system to obtain output probabilities for calibration.  
The final calibrated probability is computed by subtracting the output probability of the perturbed query \( q' \) from that of the original query \( q \), as defined in \autoref{eq:calibration}:
\begin{equation}
\label{eq:calibration_adversary1}
P_{\text{rag, calibrated}}(\text{``Yes''} \mid q) = P_{\text{rag}}(\text{``Yes''} \mid q) - P_{\text{rag}}(\text{``Yes''} \mid q')
\end{equation}

\mypara{Membership Inference.}
\label{sec:adversary1_MIA}
We use grid search to select the perturbation magnitude \(\theta\) and classification threshold \(\gamma\).  
Specifically, we sweep \(\theta\) from 0.04 to 0.12 and \(\gamma\) from 0.1 to 1.0.  
We then choose the combination that yields the highest attack accuracy on the local reference dataset.

For membership inference, given a target sample \(x\), the adversary generates both the original query \(q\) and a perturbed version \(q'\).  
Both queries are submitted to the target RAG system to compute the calibrated probability for the ``Yes'' response.  
According to the derived membership signal in \autoref{eq:signal_Adversary1} (\autoref{General Paeadigm}), if this probability exceeds the threshold \(\gamma\), the sample is classified as a member; otherwise, it is classified as a non-member.  
This inference process is formalized as:
\begin{equation}
\mathcal{I}(x) = \mathbb{I}\left\{P_{\text{rag, calibrated}}(\text{Yes} \mid q)> \gamma\right\}
\end{equation}

\subsection{Experimental Setup}
\label{sec:Experimental Setup}
\mypara{Datasets.} 
We use two widely adopted open-source datasets as retrieval databases: TREC-COVID~\cite{thakur2021beir} and HealthCareMagic-100k-en~\cite{healthcaremagic100k} (details in \autoref{app:dataset}).  
Following prior work~\cite{anderson2024membership, liu2024mask}, we randomly select 80\% of the documents as member samples (i.e., the RAG's retrieval database) and the remaining 20\% as non-member samples.  
For each dataset, we also sample 1,000 member and 1,000 non-member documents to form a reference dataset used to tune the perturbation magnitude (\(\theta\)) and classification threshold (\(\gamma\)), as commonly done in RAG-targeted MIAs~\cite{anderson2024membership,li2024blackbox,liu2024mask}.  
Finally, we select another 2,000 samples (1,000 members and 1,000 non-members) to evaluate the attack's performance.

\mypara{Target RAG System.}
To evaluate \attack, we set up a local RAG environment using the FlashRAG toolkit~\cite{jin2024flashrag}.  
The system supports various RAG frameworks, including Standard RAG~\cite{ram2023incontext}, LLMlingua~\cite{jiang2023longllmlingua}, SC-RAG~\cite{li2023compressing}, Spring~\cite{zhu2024one}, and IRCOT~\cite{trivedi2022interleaving}.  
We test four widely used generative modules: GPT-3.5-turbo, Mistral-7B-Instruct~\cite{jiang2023mistral}, Llama-2-7B-chat~\cite{touvron2023llama}, and Flan-T5-large~\cite{flan_t5_large}, along with three retrieval models---BM25, E5-base, and Contriever---covering both sparse and dense retrieval paradigms.  
\autoref{sec:rag_template} describes the prompt template used to integrate the retrieved context with the user query.

\mypara{Basic RAG Setting.} Unless specified otherwise, evaluations utilize this setting: Llama-2-7B-chat as the generative module, E5-base as the retrieval module, Standard RAG framework, and 4 retrieved documents following prior work~\cite{anderson2024membership}.

\begin{table*}[t!]
\centering
\caption{Accuracy of Adversary 1 on five RAG systems. \textbf{Bold values} indicate the best method for each system and metric.}
\label{tab:performance_system_adversary1}
\begin{tabular}{l|cccc|cccc} 
\toprule
\multirow{2}{*}{Accuracy} 
& \multicolumn{4}{c|}{HealthCareMagic-100k-en} 
& \multicolumn{4}{c}{TREC-COVID} \\
\cmidrule(lr){2-5} \cmidrule(lr){6-9} 
& \textbf{\attack} & RAG-MIA & S\textsuperscript{2}MIA & MBA 
& \textbf{\attack} & RAG-MIA & S\textsuperscript{2}MIA & MBA \\
\midrule
  Std. RAG   & \textbf{0.7945} & 0.7220 & 0.5215 & 0.6355 & \textbf{0.7255} & 0.6075 & 0.5250 & 0.5525 \\
  LLMlingua & \textbf{0.7050} & 0.6745 & 0.5425 & 0.6275 & \textbf{0.7430} & 0.6445 & 0.5570 & 0.5940 \\
  SC-RAG & \textbf{0.6795} & 0.6170 & 0.5085 & 0.6245 & \textbf{0.6700} & 0.6130 & 0.5390 & 0.6550 \\
  Spring    & \textbf{0.6885} & 0.5775 & 0.5060 & 0.6295 & \textbf{0.6965} & 0.5700 & 0.5265 & 0.6670 \\
  IRCOT     & \textbf{0.8430} & 0.6430 & 0.5335 & 0.5000 & \textbf{0.8240} & 0.8145 & 0.5984 & 0.5000 \\
   Avg.     & \textbf{0.7421} & 0.6468 & 0.5224 & 0.6030 & \textbf{0.7318} & 0.6499 & 0.5492 & 0.5937 \\
\bottomrule
\end{tabular}
\end{table*}

\mypara{Evaluation Metrics.} 
We consider two widely applied metrics:
\begin{itemize}
    \item \textbf{AUC.} 
Calculating the area under the Receiver Operating Characteristic (ROC) curve (AUC)~\cite{sankararaman2009genomic} is the most common interpretation method, reflecting the average-case success of MIAs.

    \item \textbf{Accuracy.}
This metric calculates the proportion of correctly classified instances out of the total, serving as a general standard for evaluating attack performance~\cite{liu2022membership,yeom2018privacy,choquette2021label,hayes2017logan,leino2020stolen,nasr2019comprehensive,sablayrolles2019white,song2021systematic,truex2018towards}.
\end{itemize}

\begin{table}[t!]
\centering
\caption{An overview of the different types of adversaries corresponding to baseline MIAs.}
\label{tab:baseline}
\small
\setlength{\tabcolsep}{1.5pt} 
\resizebox{\linewidth}{!}{
\begin{tabular}{l c c c c}
\toprule
\textbf{Attacks} &  
\multicolumn{1}{c}{\textbf{Partial Data}} & 
\multicolumn{1}{c}{\textbf{Data Distr.}} & 
\multicolumn{1}{c}{\textbf{Output Prob.}} &
\multicolumn{1}{c}{\textbf{Response}} \\
\midrule
RAG-MIA-gray~\cite{anderson2024membership}  &  \ding{51} & \ding{51} & \ding{51} & \ding{51} \\ 
RAG-MIA-black~\cite{anderson2024membership}  &  \ding{55} & \ding{55} & \ding{55} & \ding{51} \\ 
S\textsuperscript{2}MIA~\cite{li2024blackbox}   & \ding{51} & \ding{51} & \ding{51} & \ding{51} \\
MBA~\cite{liu2024mask}      & \ding{51} & \ding{55} & \ding{55} & \ding{51} \\
IA~\cite{naseh2025riddle}    & \ding{55} & \ding{51} & \ding{55} & \ding{51} \\
\bottomrule
\end{tabular}
}
\end{table}

\mypara{Baselines.}
We strictly adhere to the experimental settings outlined in the baseline papers and faithfully reproduce their attack strategies. An overview of the adversary capabilities corresponding to each baseline is provided in \autoref{tab:baseline}, with further details in \autoref{app:baseline}. 
For Adversary 1, we select RAG-MIA-gray~\cite{anderson2024membership}, MBA ~\cite{liu2024mask}, and S\textsuperscript{2}MIA ~\cite{li2024blackbox} as the baselines.

\subsection{Evaluation Experiment}
\label{sec:Adversary1_evaluation}
We adopt a holistic-to-specific evaluation strategy: first, benchmarking the method on diverse complete RAG system setups, then investigating the influence of altering single factors like the generative modules, retrieval modules, RAG frameworks, and retrieval quantities.

\mypara{Impact of RAG Systems.}\label{sec:Effectiveness Against Open-Source Rag Systems}
To comprehensively evaluate the effectiveness of \attack on different RAG systems, experiments are conducted on five state-of-the-art systems. These systems include the widely adopted Standard RAG as well as four complex systems that incorporate compression, fine-tuning, and iterative mechanisms (for more details, refer to \autoref{app:rag_systems}). 
All configurations---including the generative module, retrieval module, and RAG framework---are implemented strictly according to the specifications in the original papers (\autoref{tab:victim_model}, \autoref{app:rag_systems}), thereby ensuring the scientific rigor and comparability of the experimental results.

\autoref{tab:performance_system_adversary1} demonstrates the superior performance of \attack in terms of Accuracy across MIAs on two datasets. 
On average, \attack surpasses the top-performing baseline RAG-MIA-gray by approximately 10\% and S\textsuperscript{2}MIA by approximately 20\%. 
\autoref{fig:system_AUC_1} in \autoref{app:acc metric} shows AUC results that support similar conclusions.
Notably, \attack performs worse on compression RAG systems (SC-RAG and LLmlingua) than on the Standard RAG. 
Two factors may contribute to this outcome. First, compression effectively removes redundant information, reducing interference from \nomember and decreasing the need for calibration. Second, some calibration-related perturbation words may be lost during compression, which negatively impacts the calibration process.
Despite this, \attack remains optimal, delivering an approximate 10\% improvement over the best baseline on LLmlingua in the TREC-COVID dataset. 
Additionally, the RAG system with fine-tuning (Spring) offers some mitigation against MIAs, reducing the effectiveness of the RAG-MIA-gray attack by about 15\% compared to Standard RAG. 
Finally, despite the complexity of the IRCOT, its iterative process did not hinder attack performance; \attack still outperformed the best baseline by approximately 20\% on the HealthCareMagic-100k-en dataset.

\begin{figure}[t!]
  \centering
  \includegraphics[width=\linewidth]{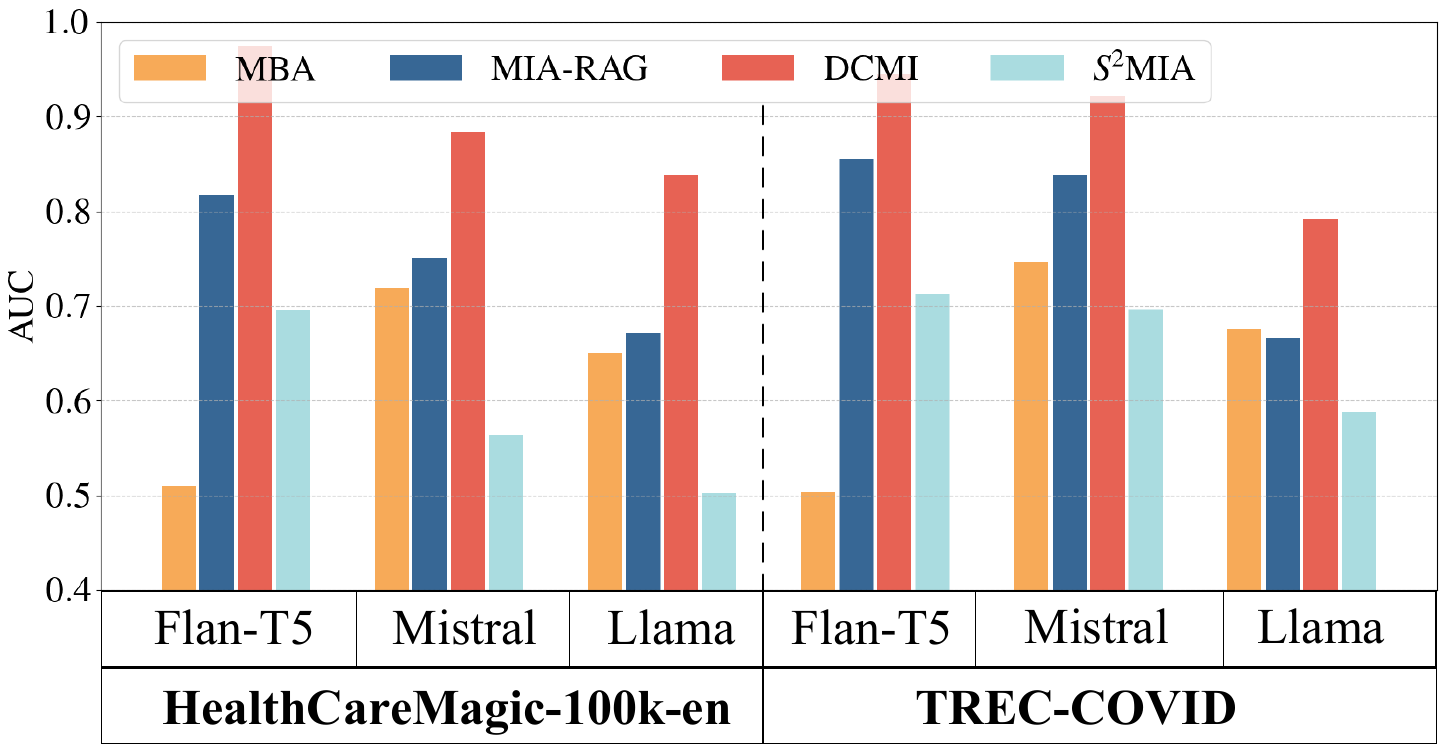}
  \caption{ AUC of Adversary 1 on three generative modules.}
  \label{fig:generative_AUC_1}
\end{figure}

\mypara{Impact of Generative
Modules.}
To evaluate our attack's effectiveness across different generative modules, we conduct experiments with a fixed retrieval module and RAG framework, following the \textit{Basic RAG Setting} in \autoref{sec:Experimental Setup}. The evaluation tests on three specific models: Llama2-7B-chat, Mistral-7B, and Flan-T5-large. 
This selection covers distinct architectures (e.g., Llama's decoder-only versus Flan-T5's encoder-decoder), facilitating a comprehensive assessment of the \attack's adaptability.

The AUC results shown in \autoref{fig:generative_AUC_1} demonstrate that \attack consistently outperforms all baselines by a wide margin on multiple generative modules.
Notably, in the RAG system incorporating Flan-T5, the Accuracy of \attack exceeds 90\%, which is over 40\% higher than MBA. 
A key observation is that our \attack performs best with Flan‑T5 integrations, compared to Llama or Mistral.
This may be attributed to Flan-T5’s optimization for question-answering tasks, rendering it more sensitive to QA-based attack patterns compared to the latter two.
The Accuracy results lead to the same conclusion (see \autoref{fig:generative_Acc_1}, \autoref{app:acc metric}).

\begin{figure}[t!]
  \centering
  \includegraphics[width=\linewidth]{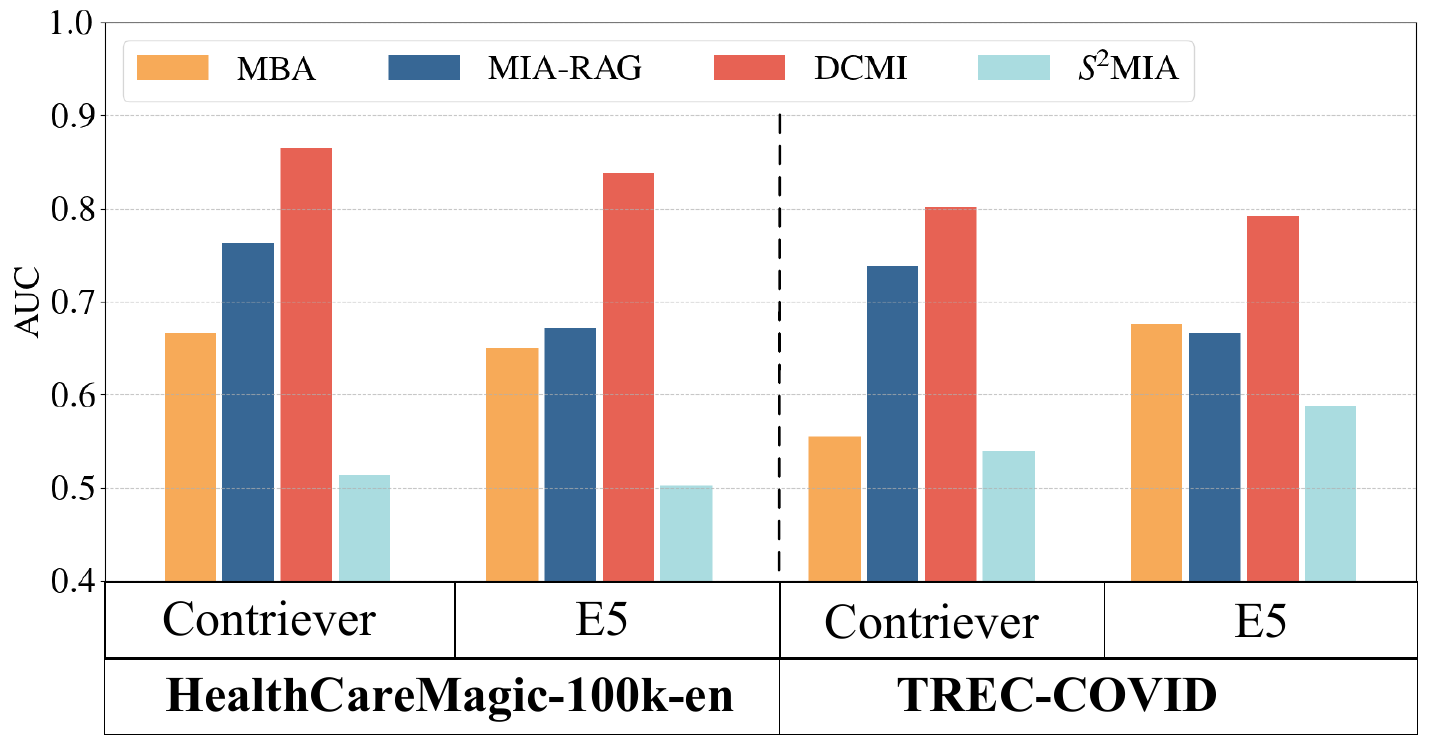}
  \caption{ AUC of Adversary 1 on two retrieval modules.}
  \label{fig:retriever_AUC_1}
\end{figure}

\mypara{Impact of Retrieval Modules.}
To evaluate our attack's effectiveness across different retrieval modules, we conduct experiments with a fixed generative module and RAG framework, following the \textit{Basic RAG Setting} in \autoref{sec:Experimental Setup}.
The evaluation
tests two specific retrieval modules: Contriever and E5-base.

\autoref{fig:retriever_AUC_1} reports the AUC performance of \attack and baseline attacks against RAG systems with different retrieval modules on two datasets. 
The results indicate
that the choice of the retrieval module does not significantly impact
\attack’s attack effectiveness.
The Accuracy results support the same conclusion (see \autoref{fig:retriever_Acc_1}, \autoref{app:acc metric}).

\begin{figure}[t!]
  \centering
  \includegraphics[width=\linewidth]{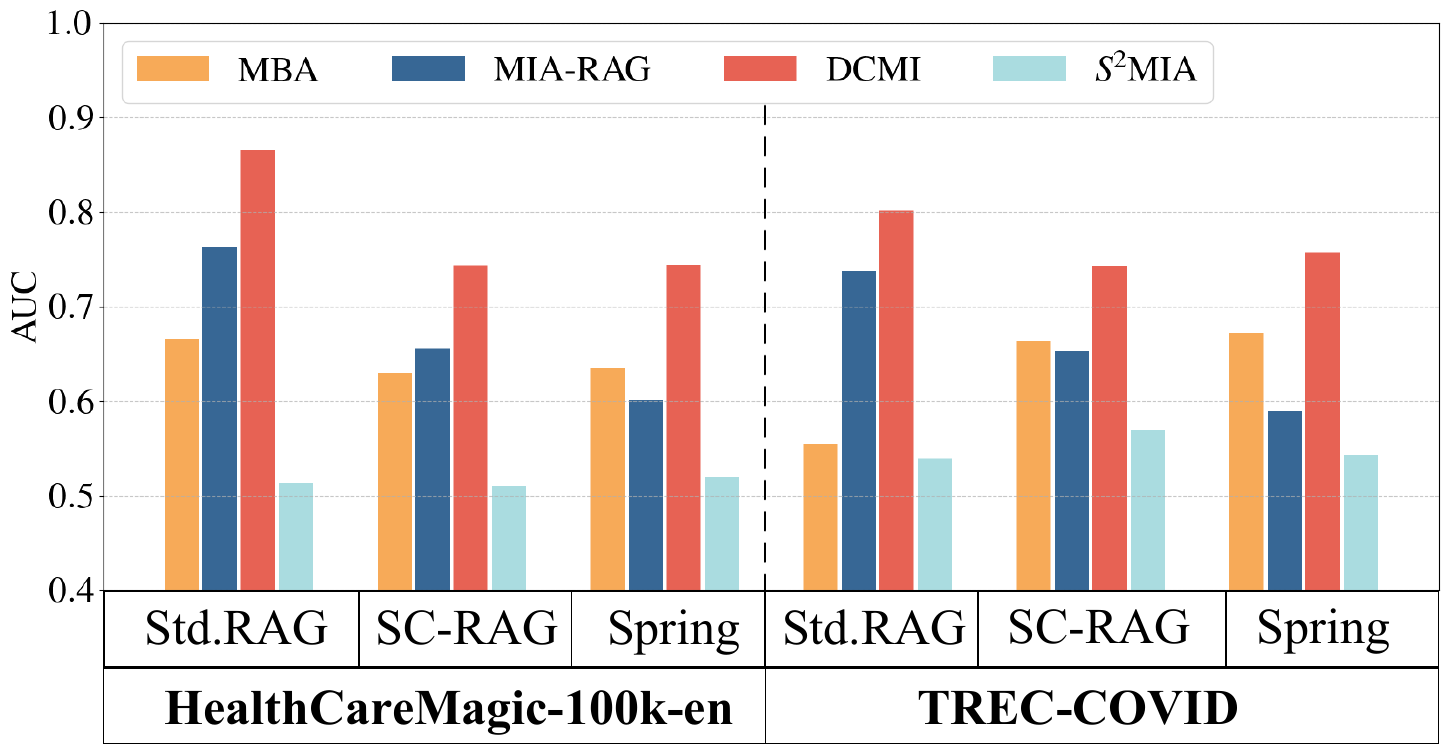}
  \caption{ AUC of Adversary 1 on three RAG frameworks.}
  \label{fig:frame_AUC_1}
\end{figure}

\mypara{Impact of RAG Frameworks.} 
To evaluate \attack's effectiveness across various RAG frameworks, we conduct experiments with a fixed generative module and retrieval module, following the \textit{Basic RAG Setting} in \autoref{sec:Experimental Setup}.
In order to further validate the negative effects of compression and fine-tuning on MIAs, we select SC-RAG and Spring for comparison, using the Standard RAG as the benchmark.

AUC results in \autoref{fig:frame_AUC_1} show that when using frameworks with compression (SC-RAG) or fine-tuning (Spring), the absolute performance is lower compared to standard RAG, which further corroborates the conclusion in Impact of RAG Systems.
The Accuracy results support the same conclusion 
(see \autoref{fig:frame_Acc_1}, \autoref{app:acc metric}).

\begin{figure}[t!]
  \centering
  \includegraphics[width=0.8\linewidth]{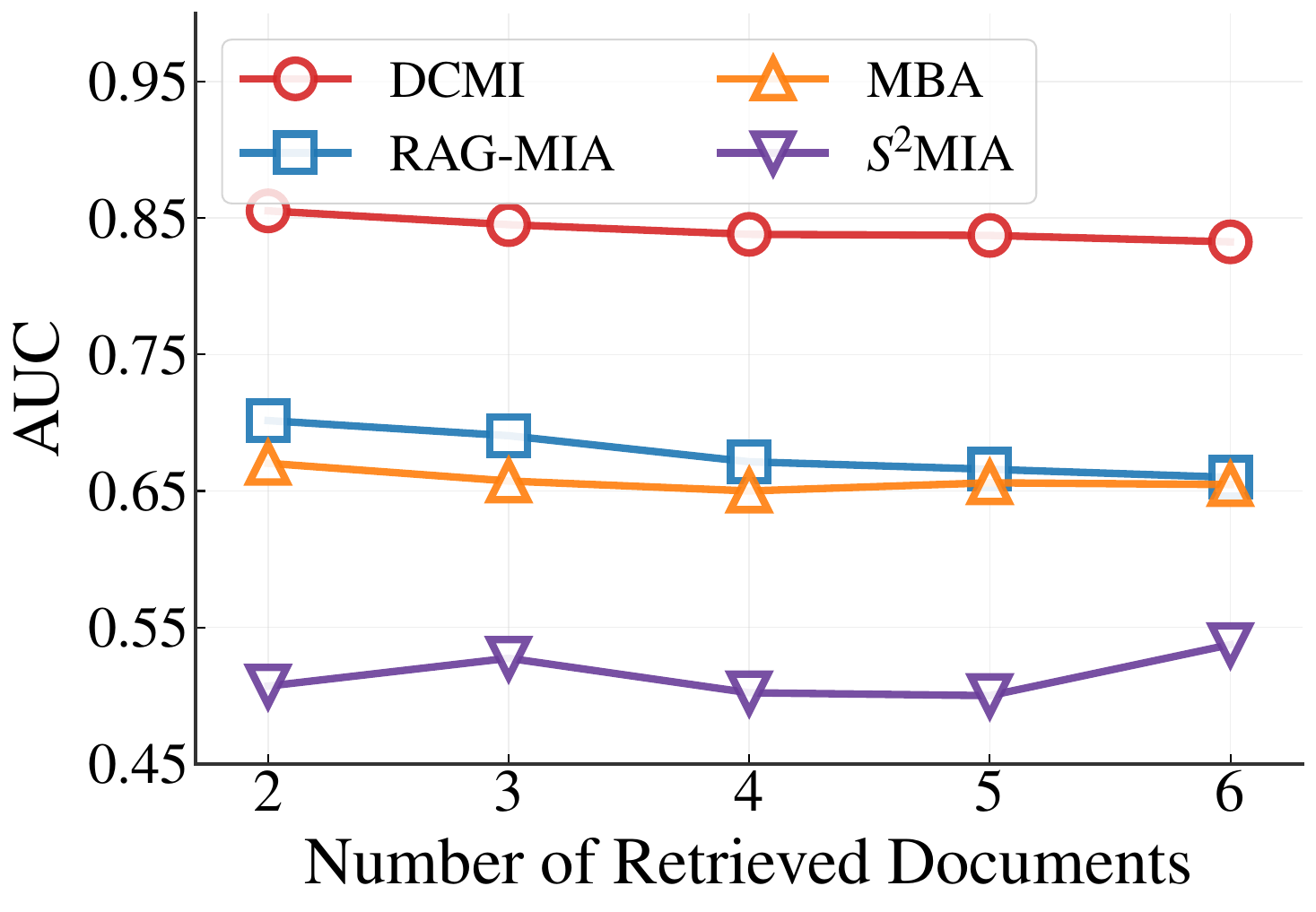}
  \caption{ AUC of Adversary 1 on different numbers of retrieved documents.}
  \label{fig:num__AUC_1}
\end{figure}

\mypara{Impact of Retrieved Document Numbers}\ 
To evaluate our attack's effectiveness across different retrieval quantities, we conduct experiments with
a fixed generative module, retrieval module, and RAG framework, following the \textit{Basic RAG Setting} in \autoref{sec:Experimental Setup}. The evaluation tests the impact of varying numbers of retrieved documents from 2 to 6, facilitating
an assessment of how retrieval quantity influences
attack performance.

The AUC results (\autoref{fig:num__AUC_1}) and the Accuracy results (\autoref{fig:num_Acc_1} in \autoref{app:acc metric}) illustrate that \attack demonstrates strong robustness, with less than a 2\% variation across different retrieval quantities. 
This minor fluctuation may result from the calibration module's coarse perturbation selection, which limits the model's ability to optimize the magnitude effectively.
In contrast, RAG-MIA-gray shows the least robustness. Its attack Accuracy drops from 66.1\% to 61.8\% as the retrieved documents increase.
Overall, \attack consistently delivers superior and stable attack performance irrespective of the retrieval quantity, which consistently maintains >76\% Accuracy and >83\% AUC outperforming the best baseline by over 10\%.

\begin{figure*}[t!]
    \centering
    \begin{subfigure}[b]{0.32\linewidth}
        \includegraphics[width=\linewidth]{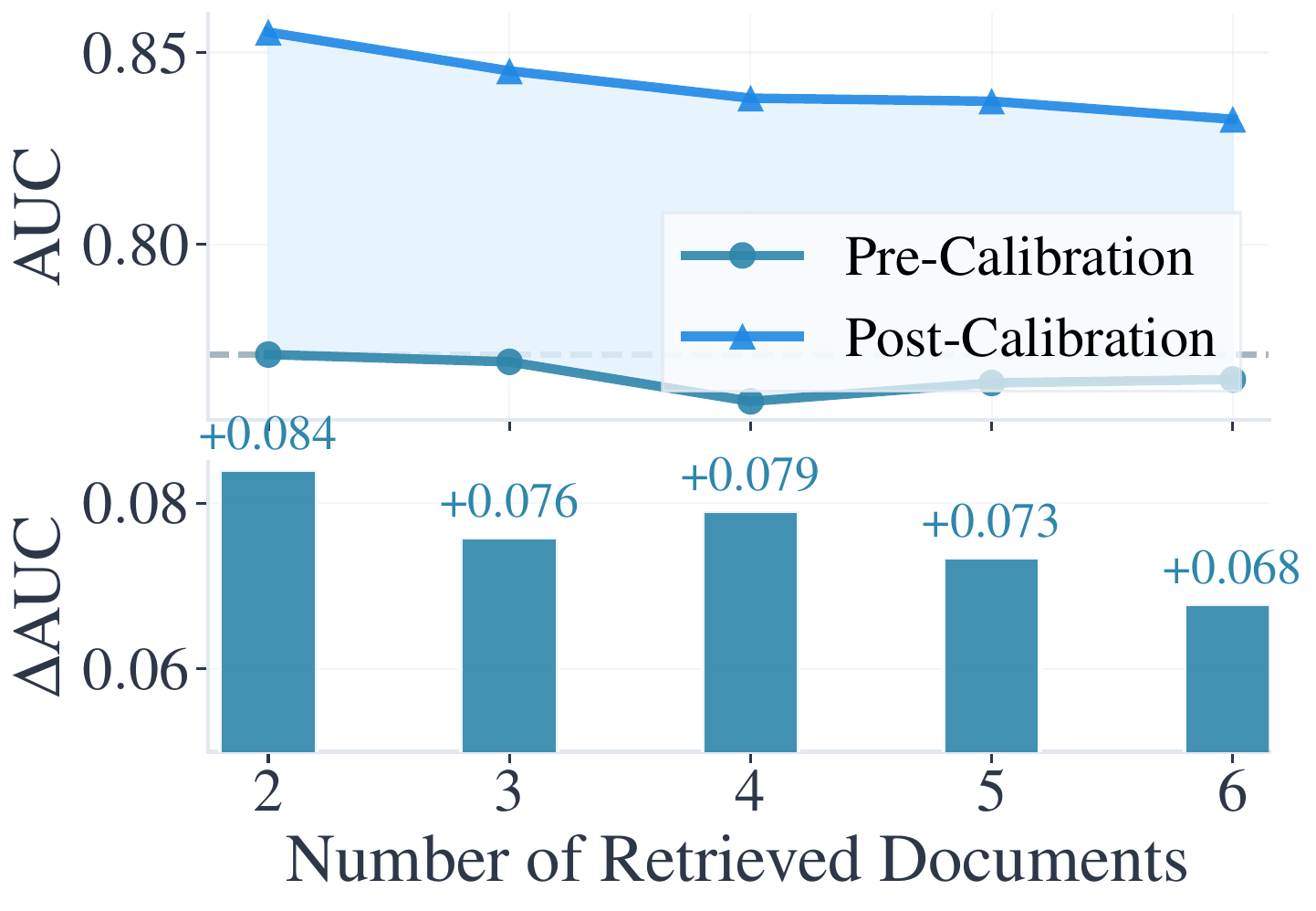}
        \caption{Impact of differential calibration module.}
        \label{fig:calibration_AUC_1}
    \end{subfigure}
    \hfill
    \begin{subfigure}[b]{0.32\linewidth}
        \includegraphics[width=\linewidth]{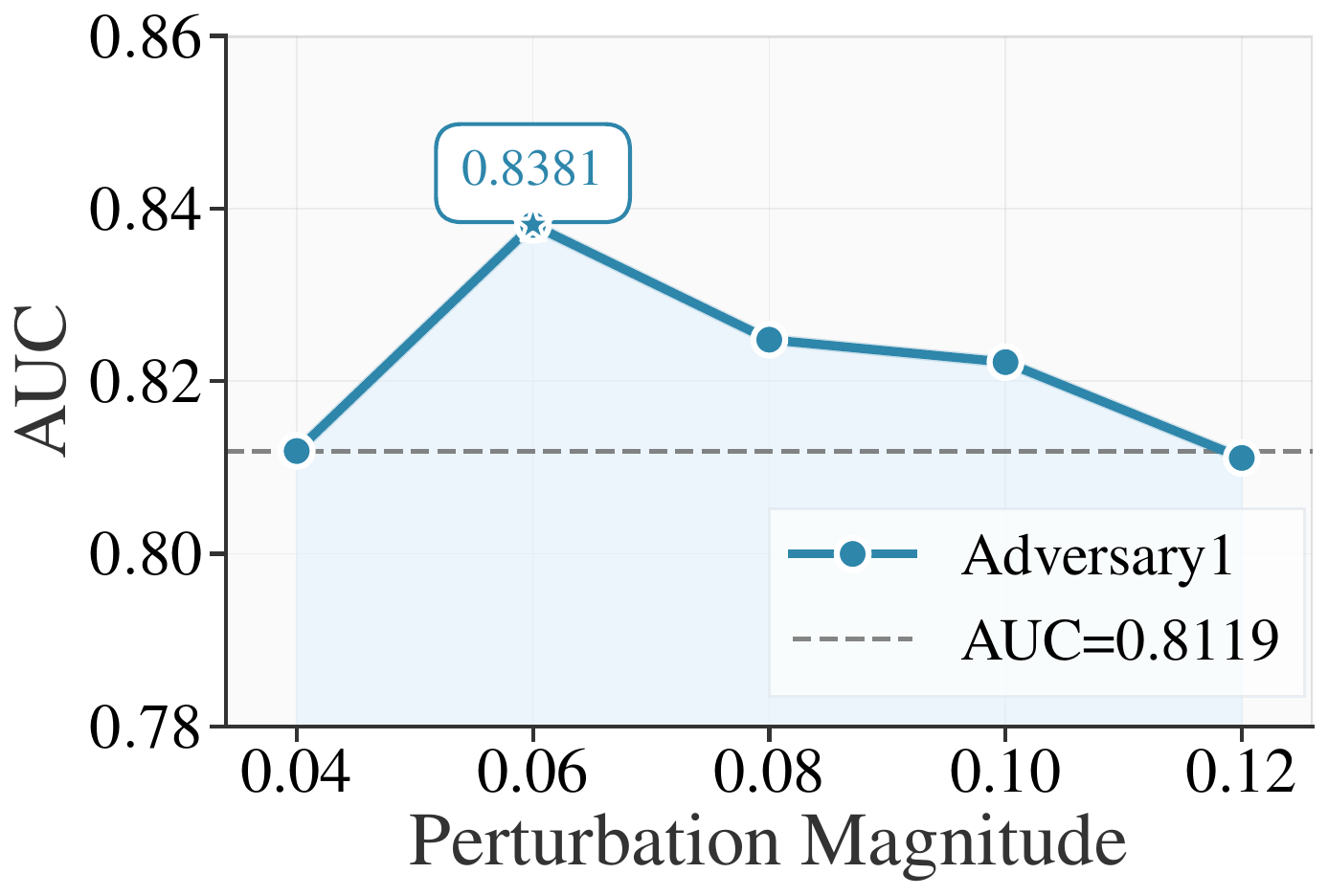}
        \caption{Impact of perturbation magnitude.}
        \label{fig:perturb_AUC_1}
    \end{subfigure}
    \hfill
    \begin{subfigure}[b]{0.32\linewidth}
        \includegraphics[width=\linewidth]{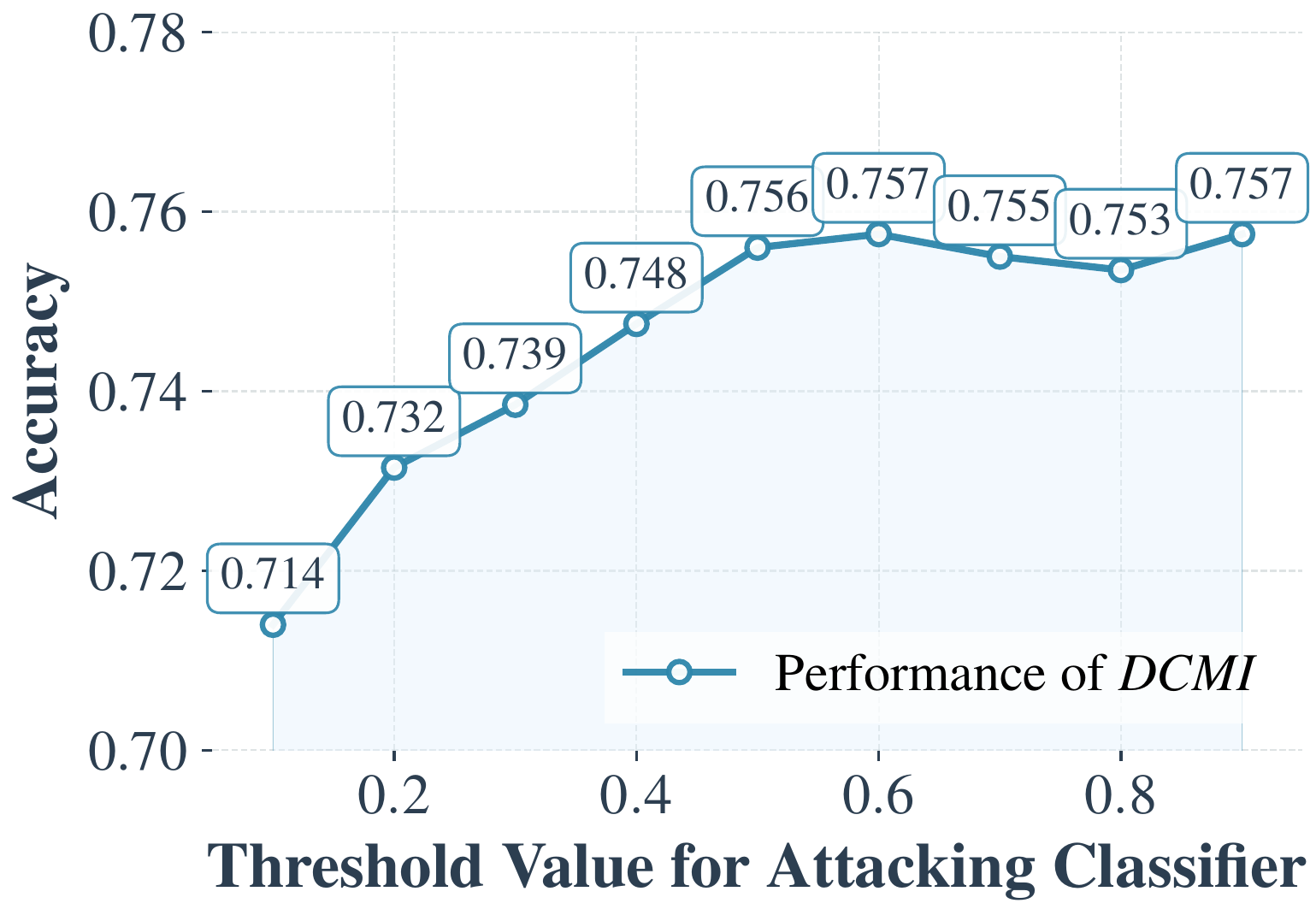}
        \caption{Impact of classification threshold.}
        \label{fig:threshold_Acc_1}
    \end{subfigure}
    
    \caption{ Performance analysis on Adversary 1.}
    \label{fig:performance_analysis_1}
\end{figure*}
\subsection{Ablation Studies}
\label{sec:Ablation_adversary1}

\mypara{Impact of Differential Calibration Module.}\ 
In this experiment, we compare two conditions: \attack with the differential calibration module and \attack without it. 
The experiment is conducted in the healthcare scenario, followed by the \textit{Basic RAG Setting} in \autoref{sec:Experimental Setup}. We perform MIAs on this RAG system and evaluate the calibration module's effectiveness across varying numbers of retrieved documents.

As shown in \autoref{fig:calibration_AUC_1}, applying the differential calibration module improves \attack's AUC by approximately 7\%. This performance gain is consistent regardless of the number of retrieved documents, highlighting the calibration module's significant role in enhancing \attack's effectiveness. The Accuracy results lead to the same conclusion (\autoref{fig:calibration_Acc_1}, \autoref{app:acc metric}).
Additionally, we conduct extensive ablation experiments under diverse settings in \autoref{app:Extended_Ablation_Study}.
The results consistently demonstrate that the differential calibration module provides significant performance improvements across all the evaluated scenarios, reinforcing our findings from the \textit{Basic RAG Setting}.

\mypara{Impact of Perturbation Magnitude.}\
In this experiment, we examine the impact of perturbation magnitude on \attack's performance. 
We vary the proportion of word replacements in the target sample from 0.04 to 0.12 and report the Accuracy and AUC results. 
The experiment is conducted in the healthcare scenario, followed by the \textit{Basic RAG Setting} in \autoref{sec:Experimental Setup}.

AUC (\autoref{fig:perturb_AUC_1}) and Accuracy (\autoref{fig:perturb_Acc_1}, \autoref{app:acc metric}) indicate that perturbation magnitudes that are too small (0.04) or too large (0.12) reduce the effectiveness of \attack, with optimal attack performance observed at a magnitude of 0.06.
Nonetheless, \attack demonstrates high robustness overall, with its attack performance fluctuating by only about 2\% across different magnitudes.
This stability demonstrates that \attack maintains effectiveness across various perturbation magnitudes, providing robust performance in scenarios where fine-tuning the perturbation magnitude is impractical.

\mypara{Impact of Classification Threshold.}
In this experiment, we examine the effect of the threshold value, \(\gamma\), on \attack's performance. 
We vary \(\gamma\) to distinguish between members and non-members and report the MIA Accuracy for each setting. 
The experiment is conducted in the healthcare scenario, followed by the \textit{Basic RAG Setting} in \autoref{sec:Experimental Setup}.

As shown in \autoref{fig:threshold_Acc_1}, \attack's accuracy remains relatively stable across a range of threshold values. 
While accuracy slightly increases as the threshold \(\gamma\) is adjusted, the overall variation is minimal, indicating low sensitivity to the choice of threshold. 
This stability ensures \attack's effectiveness across different thresholds, making it robust in scenarios where fine-tuning threshold parameters is impractical.

\section{Distribution-Aware Membership Inference}
Adversary 2 inherits the same query capabilities as Adversary 1, including access to output log probabilities. However, unlike Adversary 1, Adversary 2 has no access to any actual data from the target RAG's retrieval database. Instead, it only possesses a reference dataset that follows the same distribution but contains entirely different content with no overlap.

\subsection{Methodology}\label{sec:Adversary2_Mothod}
\mypara{Query Construction.}
The original query \(q\) and its output probabilities are constructed using the same strategy as Adversary 1.

\mypara{Differential Calibration.}
Analysis of Adversary 1 shows that \attack's performance is largely stable across perturbation magnitudes, with the best results at 0.06 (see \autoref{fig:perturb_AUC_1}).  
Therefore, Adversary 2 uses a fixed perturbation magnitude of 0.06 to generate the perturbed query \(q'\). It then follows the same procedure as Adversary 1 to obtain calibrated output probabilities.  
The differential calibration formula remains the same as in \autoref{eq:calibration_adversary1}.

\begin{figure}[t!]
  \centering
  \includegraphics[width=0.6\linewidth]{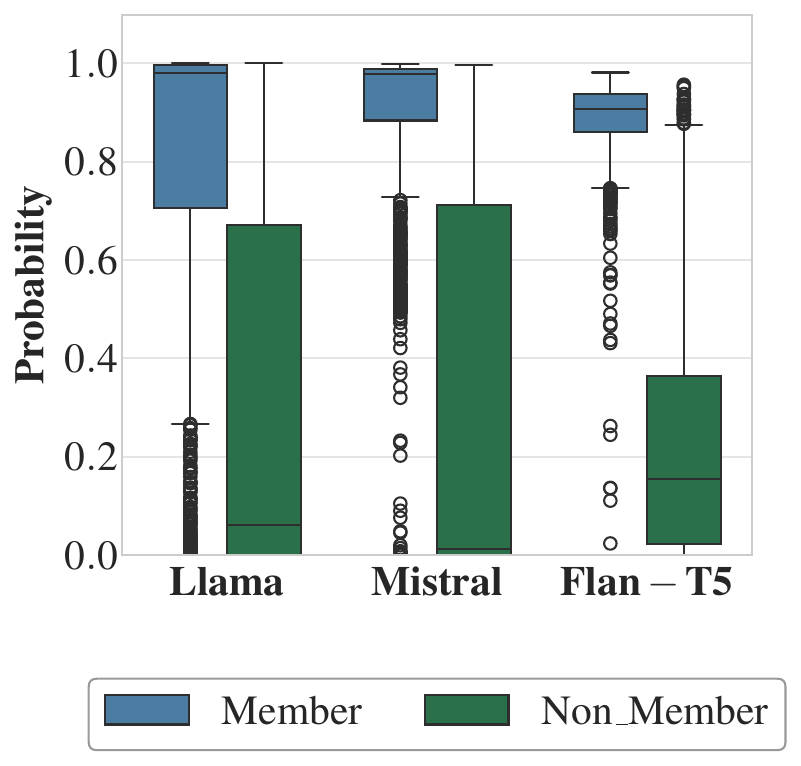}
  \caption{ Box plot for member and non-member across different generative modules.}
  \label{fig:box_generation}
\end{figure}

\mypara{Membership Inference.}
As shown in \autoref{fig:threshold_Acc_1}, \attack's performance remains stable across a wide threshold range (0.5–0.9), making threshold selection straightforward.  
\autoref{fig:box_generation} shows the output probabilities for member samples and identically distributed non-member samples from the reference dataset.  
The upper quartile (\(Q3\)) of the non-member distribution clearly separates the two groups.  
Based on this, we set the classification threshold to the \(Q3\) value of the reference dataset.

Specifically, given a reference dataset \(D_{ref}\), we generate an original query $q_{ref}$ and its perturbed version $q'_{ref}$ for each sample \(x_{ref} \in D_{ref}\).
Both $q_{ref}$ and $q'_{ref}$ are then submitted to the RAG. 
The output probability associated with the original query $q_{ref}$ is subsequently calibrated using the output probability obtained from the perturbed query $q'_{ref}$.
Finally, we use the upper quartile \(Q3\) of these calibrated probabilities as the classification threshold (\(\tau\)).
\begin{equation}
\tau = Q_3 \left( \left\{ P_{\text{rag}}(\text{Yes} \mid q_{ref}) - P_{\text{rag}}(\text{Yes} \mid q'_{ref}) \right\} \right)
\end{equation}

During the membership inference, for a target sample $x$, we apply the same method described above to obtain its calibrated probability value. If this value exceeds the threshold $\tau$, the sample is classified as a member; otherwise, it is classified as a non-member.
This MIA process can be formalized as:

\begin{equation}
\mathcal{I}(x) = \mathbb{I}\left\{P_{\text{rag, calibrated}}(\text{Yes} \mid q)> \tau\right\}
\end{equation}

\subsection{Experimental Setup}
\label{sec:Adversary2_Experimental_Setup}

\mypara{Datasets.} Unlike Adversary 1, Adversary 2 only has access to the data distribution.  
We simulate this by selecting reference data from non-member samples.  
All other dataset settings follow those described in \autoref{sec:Experimental Setup}.

\mypara{Baselines.}
In Adversary 2, where the target RAG retrieval database is inaccessible, most original baseline methods cannot be directly applied (see \autoref{tab:threat_models} and \autoref{tab:baseline}).  
Although the IA baseline~\cite{naseh2025riddle} matches the Adversary 2 setting, its original implementation does not specify how to set classification thresholds.  
The S\textsuperscript{2}MIA baseline~\cite{li2024blackbox} relies on model training and is thus incompatible with our threshold selection method; we exclude it from the comparison.  
To ensure fairness, we apply our threshold selection method to all comparable baselines (RAG-MIA-gray~\cite{anderson2024membership}, MBA~\cite{liu2024mask}, and IA~\cite{naseh2025riddle}) to determine their optimal thresholds.  
The next sections present performance comparisons between \attack and these adjusted baselines.

\mypara{Other Settings.}
All other settings of Adversary 2 are consistent with those described in \autoref{sec:Experimental Setup}.

\subsection{Evaluation Experiment}
\label{sec:evaluation_adversary2}
This section evaluates the attack effectiveness of Adversary 2. 
All evaluations herein utilize the identical settings as those corresponding to Adversary 1 in \autoref{sec:Adversary1_evaluation}.
Evaluation setting details are omitted for brevity in this section, and the results for Adversary 2 across these evaluations are presented directly below.

\mypara{Impact of RAG Systems.}\ 
\begin{figure}[t!]
  \centering
  \includegraphics[width=\linewidth]{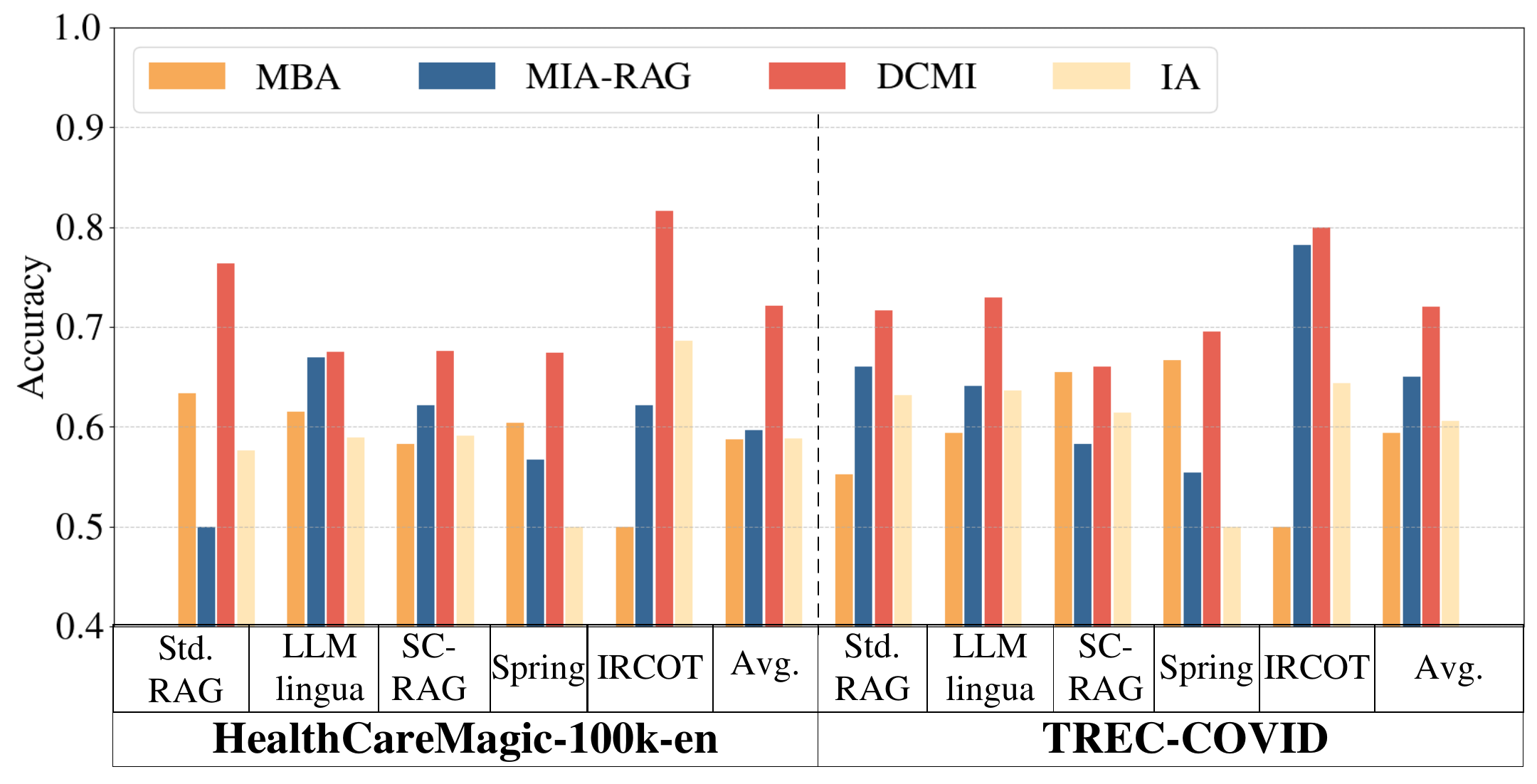}
  \caption{Accuracy of Adversary 2 on five RAG systems.}
  \label{fig:system_ACC_2}
\end{figure}
\autoref{fig:system_ACC_2} presents the evaluation results in terms of Accuracy for Adversary 2 across five RAG systems.
Compared to the corresponding Adversary 1 results (\autoref{tab:performance_system_adversary1}), our non-member sample-based threshold determination strategy incurs only a minimal decrease in attack performance, typically between 1\%-3\%.
This demonstrates the effectiveness of adapting the thresholding method to the Adversary 2 scenario. 
Furthermore, our \attack approach consistently and significantly outperforms baseline methods across most RAG system settings, despite these baselines utilizing the same threshold determination technique as \attack. 
This superiority is particularly evident on the HealthCareMagic-100k-en dataset, where baselines average below 60\% Accuracy, while our method achieves an improvement exceeding 10\%. 
These results validate both the rationale behind our proposed threshold determination for Adversary 2 and the efficacy of our differential calibration technique.
AUC metric results showing similar trends are available in \autoref{fig:system_AUC_2}, \autoref{app:acc metric}.

\begin{figure}[t!]
  \centering
  \includegraphics[width=\linewidth]{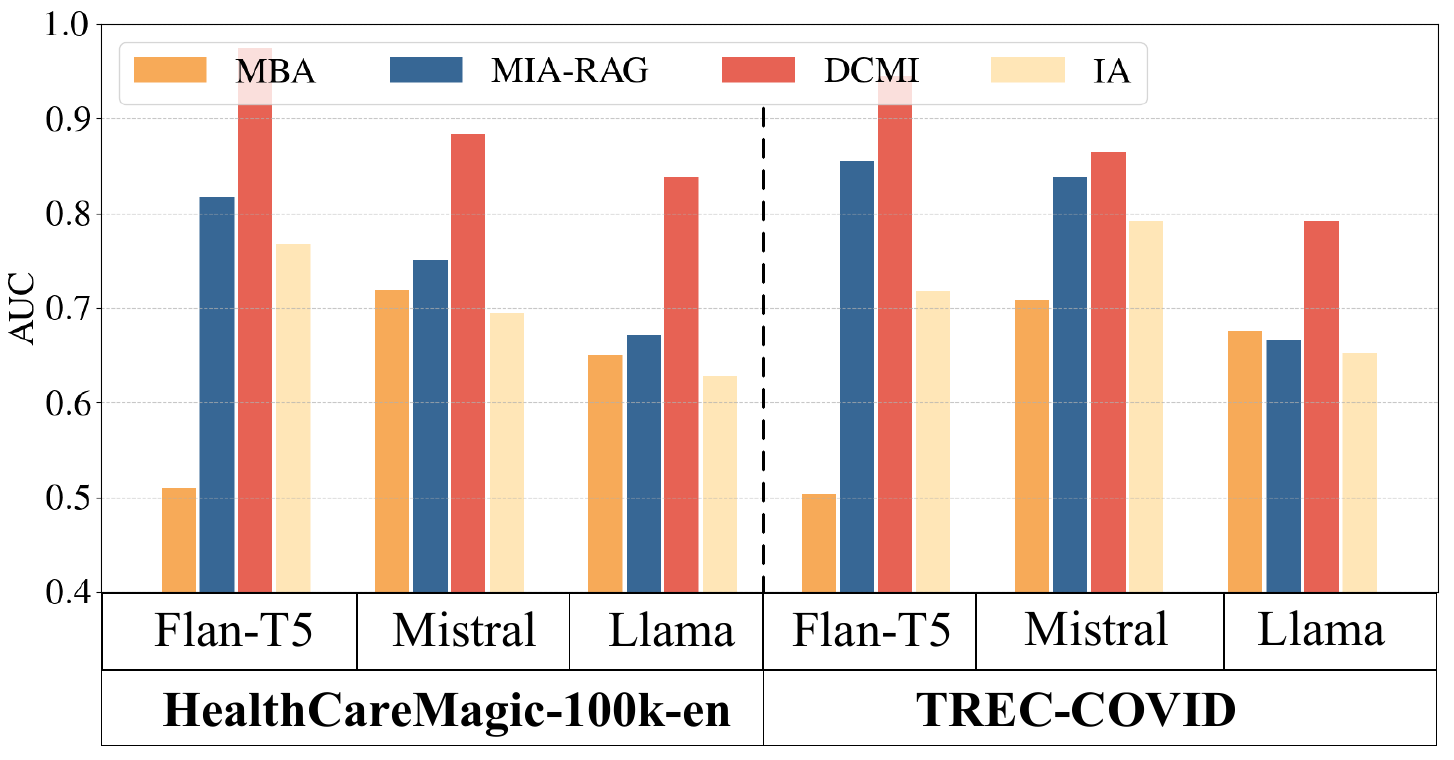}
  \caption{ AUC of Adversary 2 on three generative modules.}
  \label{fig:generative_AUC_2}
\end{figure}

\mypara{Impact of Generative
Modules.}
Consistent with Adversary 1 results, \attack performs best with Flan-T5 architecture in Adversary 2, showing 40\% improvement over MBA baseline (\autoref{fig:generative_AUC_2} and \autoref{fig:generative_Acc_2}, \autoref{app:acc metric}).

\begin{figure}[t!]
  \centering
  \includegraphics[width=\linewidth]{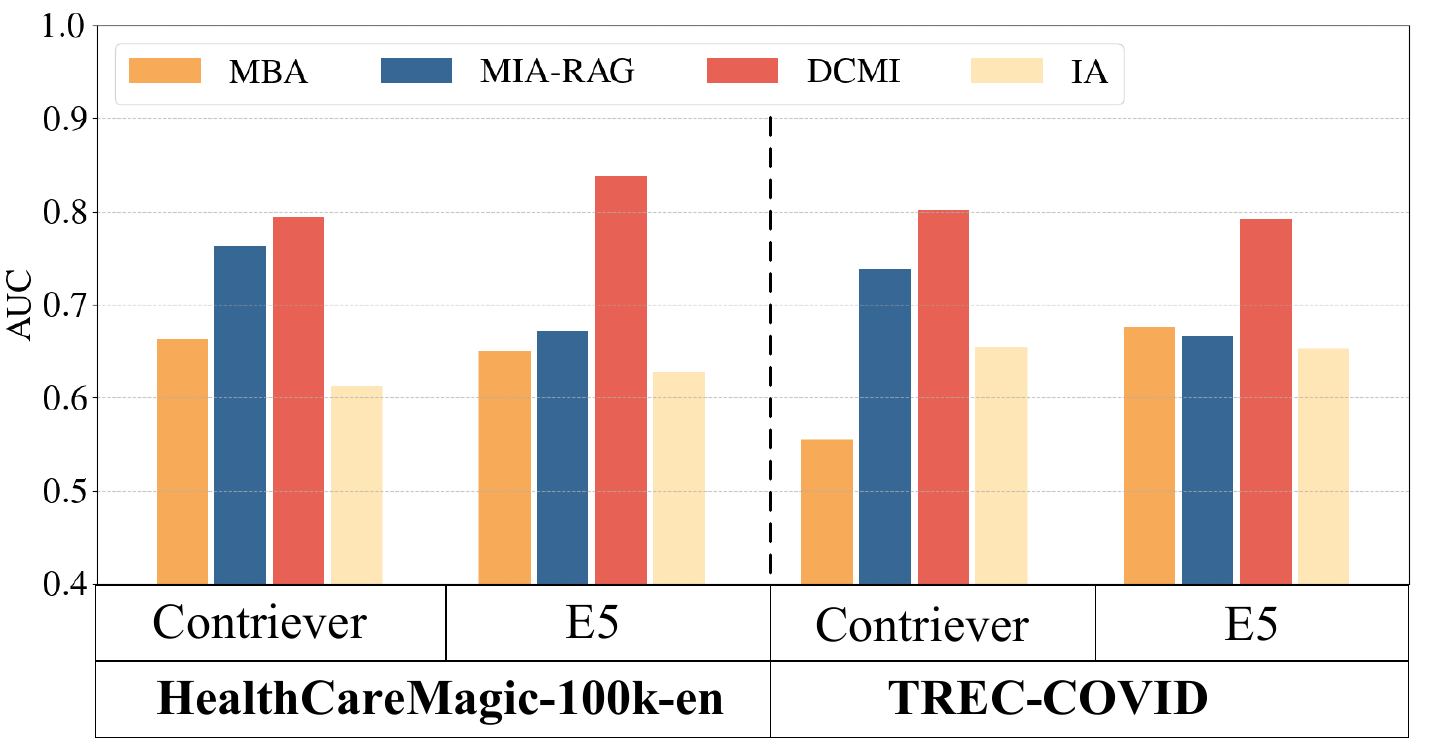}
  \caption{ AUC of Adversary 2 on two retrieval modules.}
  \label{fig:retriever_AUC_2}
\end{figure}

\mypara{Impact of Retrieval
Modules.}
As in Adversary 1, \attack's AUC remains consistent across retrieval modules (\autoref{fig:retriever_AUC_2}), with similar trends observed in Accuracy metrics (\autoref{fig:retriever_Acc_2}, \autoref{app:acc metric}).

\begin{figure}[t!]
  \centering
  \includegraphics[width=\linewidth]{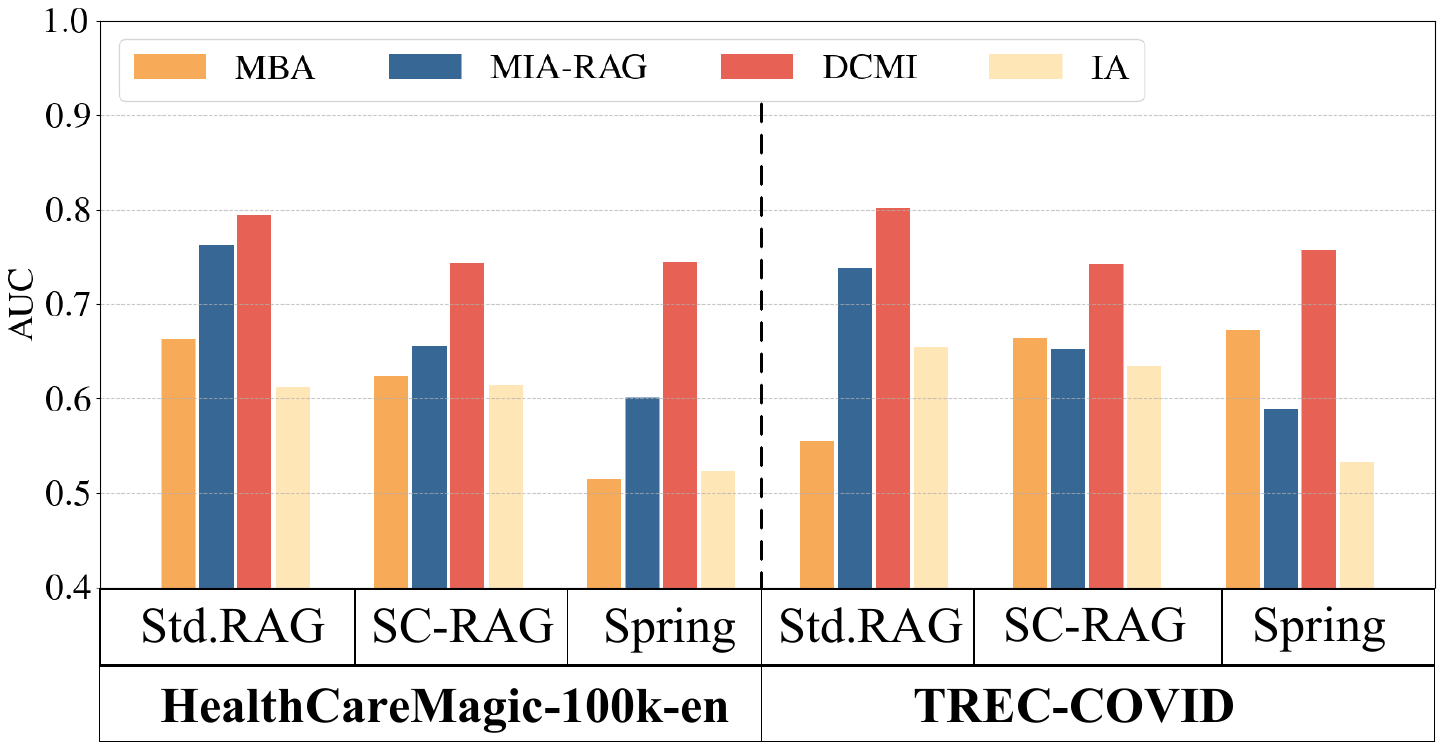}
  \caption{ AUC of Adversary 2 on three RAG frameworks.}
  \label{fig:frame_AUC_2}
\end{figure}

\mypara{Impact of RAG Frameworks.}
Mirroring the trends observed for Adversary 1, the results in \autoref{fig:frame_AUC_2} demonstrate that while advanced RAG frameworks like SC-RAG (compression) and spring (fine-tuning) reduce \attack's attack effectiveness compared to standard RAG under Adversary 2, \attack consistently maintains a significant performance advantage over baseline approaches across all evaluated framework types. 
The Accuracy results support the same conclusion (see \autoref{fig:frame_Acc_2}, \autoref{app:acc metric}).

\begin{figure*}[t!]
    \centering
    \begin{subfigure}[t]{0.32\linewidth}
        \includegraphics[width=\linewidth]{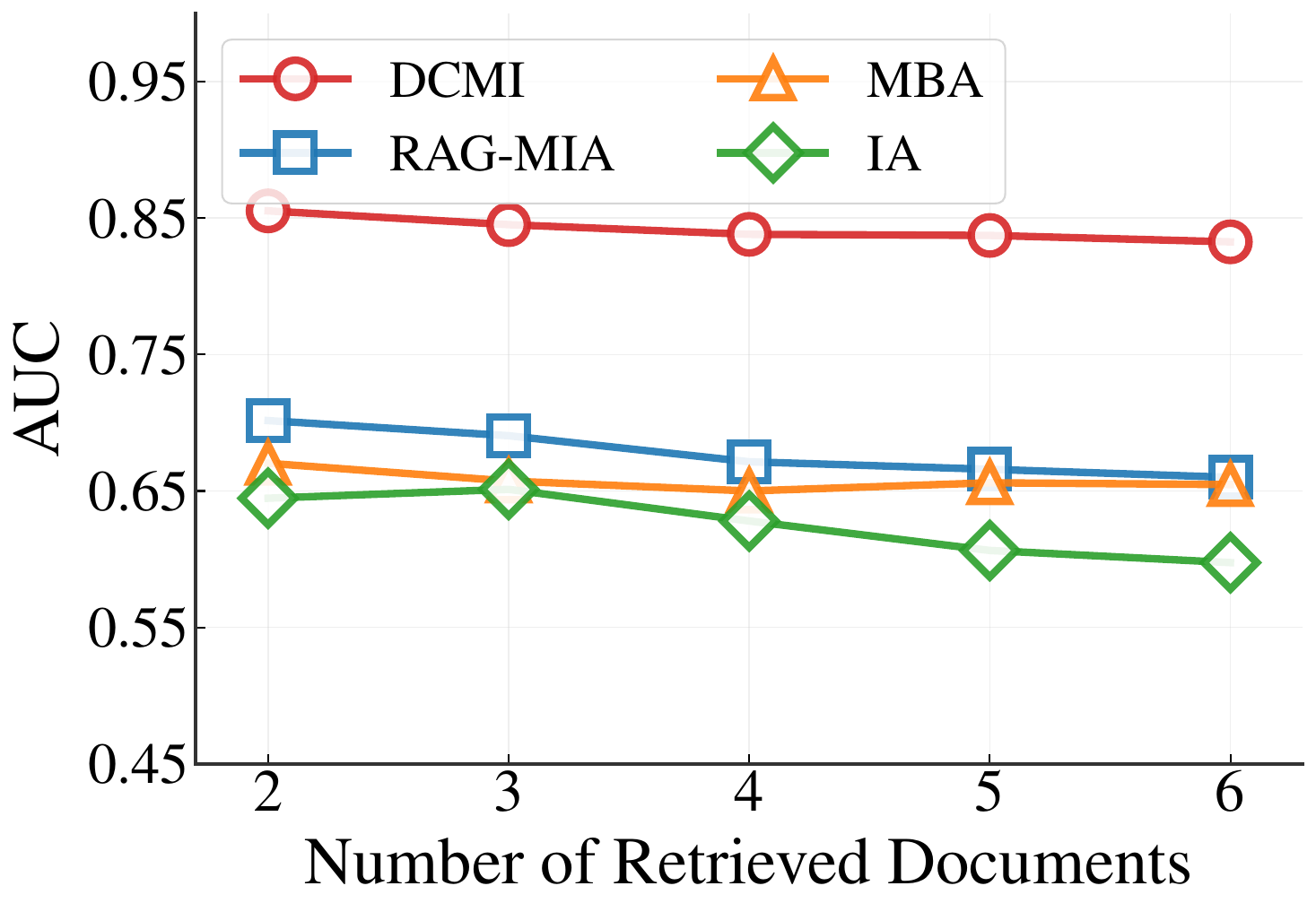}
        \caption{AUC of Adversary 2 on different numbers of retrieved documents.}
        \label{fig:num_AUC_2}
    \end{subfigure}
    \hfill
    \begin{subfigure}[t]{0.32\linewidth}
        \includegraphics[width=\linewidth]{revision/calibration.pdf}
        \caption{Impact of differential calibration module.}
        \label{fig:calibration_AUC_2}
    \end{subfigure}
    \hfill
    \begin{subfigure}[t]{0.32\linewidth}
        \includegraphics[width=\linewidth]{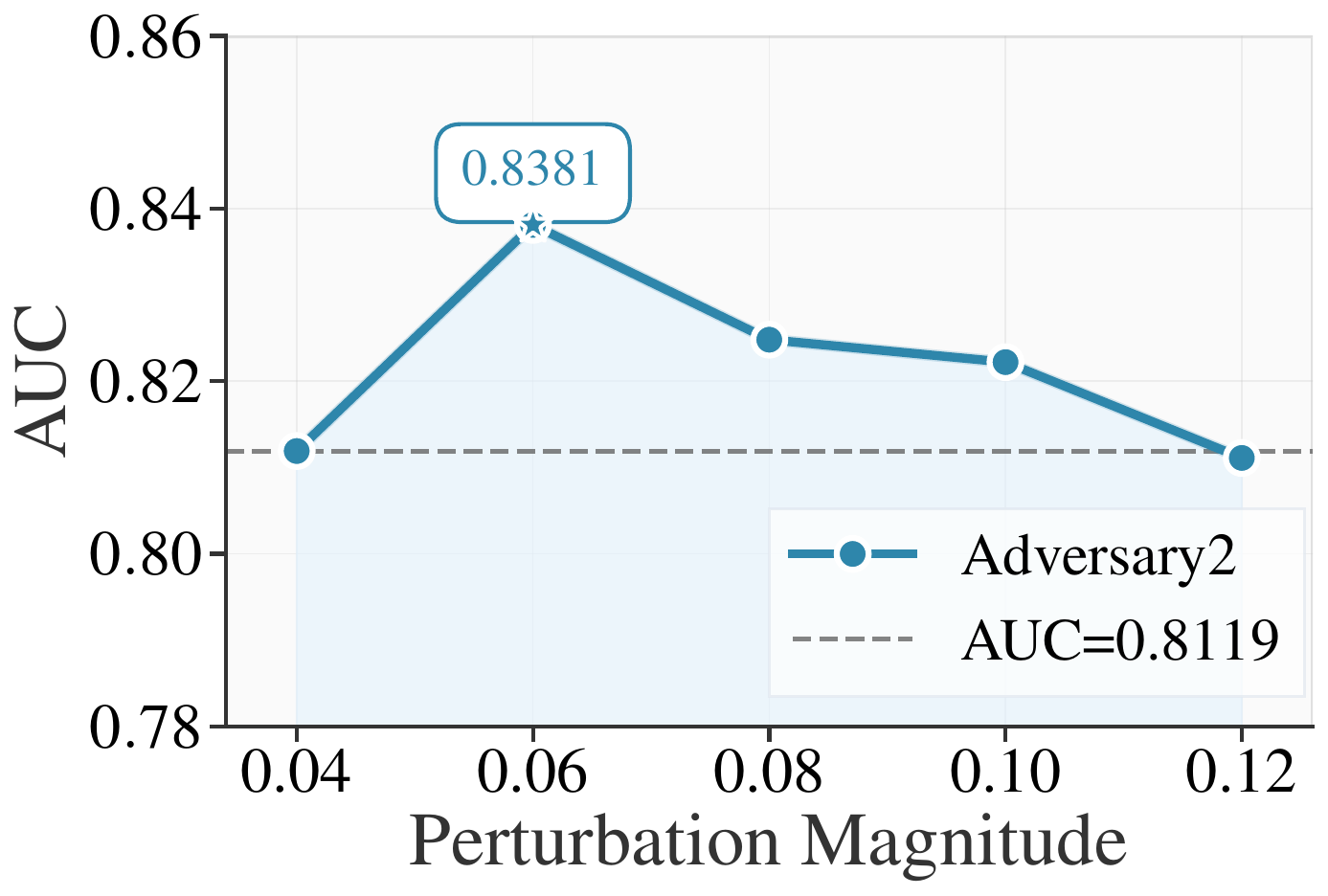}
        \caption{Impact of perturbation magnitude. }
        \label{fig:perturb_AUC_2}
    \end{subfigure}
    
    \caption{Performance Analysis on Adversary 2.}
    \label{fig:performance_analysis_2}
\end{figure*}

\mypara{Impact of Number of
Retrieved Documents.}
As shown in AUC (\autoref{fig:num_AUC_2}) and Accuracy (\autoref{fig:Num_Acc_2}, \autoref{app:acc metric}), when the number of retrieved documents increases from 2 to 6, the AUC of RAG-MIA-gray declines by 4\% while the IA baseline exhibits a 5\% performance fluctuation.
In contrast, our \attack method exhibits higher stability, with only a marginal 1\% reduction, maintaining consistent superiority over the baseline regardless of retrieval quantity.

\subsection{Ablation Studies}
\label{sec:Ablation_adversary2}
Since Adversary 1 and Adversary 2 differ only in threshold selection, their threshold ablation results are identical.  
To avoid redundancy, the Adversary 2 ablation studies focus only on the calibration module and perturbation magnitude.  
The experimental settings follow those in \autoref{sec:Ablation_adversary1}.  
Results are presented below.

\mypara{Impact of Differential Calibration Module.}\
Compared to Adversary 1, the efficacy of the calibration module in Adversary 2 is slightly reduced according to AUC (\autoref{fig:calibration_AUC_2}) and Accuracy (\autoref{fig:calibration_Acc_2}, \autoref{app:acc metric}).
Nevertheless, the calibration module still enhances the \attack attack's AUC by over 6\%, and Accuracy by over 5\%, demonstrating its crucial importance.
Extensive ablation studies in \autoref{app:Extended_Ablation_Study} further validate the differential calibration module's effectiveness across diverse settings.

\mypara{Impact of Perturbation Magnitude.}
According to AUC (\autoref{fig:perturb_AUC_2}) and Accuracy (\autoref{fig:perturb_Acc_2}, \autoref{app:acc metric}), 
\attack under Adversary 2 follows the same overall trend as under Adversary 1. 
Although Adversary 2's Accuracy fluctuates slightly, it remains within 3\%,
supporting the choice of a 0.06 perturbation magnitude in \autoref{sec:Adversary2_Mothod}.

\section{Black-Box Membership Inference}
Adversary 3 operates in a black-box setting where output probabilities and RAG retrieval database information are inaccessible.
It infers the membership solely from the RAG system's response.
\subsection{Methodology}\label{sec:Adversary3_method}
\mypara{Query Construction.}
We construct the original query \(q\) using the same strategy as Adversary 1.
In Adversary 3, we instead use the RAG system's binary (``Yes''/``No'') responses as discrimination signals.
We define the function $f_{\text{rag}}(\cdot)$ to map ``Yes'' to 1 and ``No'' to 0, and then compute \(f_{\text{rag}}(q)\) for \(q\).

\mypara{Differential Calibration.}
Based on the analysis in \autoref{General Paeadigm}, we update the differential calibration module to adapt to practical black-box scenarios. 
Specifically, Adversary 3 generates a perturbed query \(q'\) with the perturbation magnitude 0.06 (as in Adversary 2), and its corresponding \(f_{\text{rag}}(q')\).
The differential calibration formula is then defined as:
\begin{equation}
f_{rag,calibrated}(q) = f_{\text{rag}}(q) - f_{\text{rag}}(q'),
\end{equation}

\mypara{Membership Inference.}
During the membership inference, for a target sample \(x\), the adversary constructs both the original query \(q\) and its perturbed version \(q'\).
Both queries are then input into the target RAG system to obtain the calibrated numerical value $f_{rag,calibrated}(q)$. 
Based on the theoretical insight that member samples exhibit decision reversal under perturbation (\autoref{General Paeadigm}), if $f_{rag,calibrated}(q)=1$ (i.e., the original sample receives a ``Yes'' response while the perturbed version receives a ``No'' response), we classify \(x\) as a member; otherwise, when $f_{rag,calibrated}(q) \in \{0, -1\}$, we classify it as a non-member. This MIA process can be formalized as:
\begin{equation} \mathcal{I}(x) = \mathbb{I}\{f_{rag,calibrated}(q) = 1\} \end{equation}

\subsection{Experimental Setup}
\label{sec:adversary3_evaluation}

\mypara{Datasets.} 
In the black-box scenario for Adversary 3, where information regarding the RAG retrieval database is inaccessible, Adversary 3 infers the membership without using any reference dataset.
All other settings related to the datasets are consistent with those described in \autoref{sec:Experimental Setup}.

\mypara{Baselines.}
The black-box scenario for Adversary 3 assumes no access to output probabilities or retrieval database information, rendering baselines (RAG-MIA-gray~\cite{anderson2024membership}, MBA ~\cite{liu2024mask}, S\textsuperscript{2}MIA ~\cite{li2024blackbox}, and IA ~\cite{naseh2025riddle}) invalid according to \autoref{tab:threat_models} and \autoref{tab:baseline}. 
Therefore, we evaluate \attack against the RAG-MIA-black ~\cite{anderson2024membership} baseline in this section.

\mypara{Other Settings.}
All other settings of Adversary 3 are consistent with those described in \autoref{sec:Experimental Setup}.

\subsection{Evaluation Experiment}
This section evaluates the attack effectiveness of Adversary 3. Relevant evaluation settings are detailed in \autoref{sec:Experimental Setup}. 
For brevity, this section directly presents the evaluation results.

\mypara{Impact of RAG Systems.}
In Adversary 3, the Accuracy performance of all MIAs decreased compared to the gray-box scenario, according to \autoref{tab:perfomance_system_Adversary3_ACC} compared with \autoref{tab:performance_system_adversary1}.
Specifically, the Accuracy for \attack drops by approximately 5\% on average. 
The baseline method declines more sharply, with attack Accuracy falling to near random‑guessing levels ($\approx50\%$) in some cases.
Even under this more challenging scenario, \attack consistently maintains Accuracy values above 60\%, outperforming the baseline by over 10\% on average.
AUC results in
\autoref{tab:perfomance_system_Adversary3_AUC}, \autoref{app:acc metric} support the same conclusion.

\begin{table}[t!]
\centering
\renewcommand{\arraystretch}{1.1}
\caption{Accuracy of Adversary 3 on five RAG systems.}
\label{tab:perfomance_system_Adversary3_ACC}
\resizebox{\linewidth}{!}{
\begin{tabular}{lcccc}
\toprule
\multirow{2}{*}{Accuracy} & \multicolumn{2}{c}{HealthCareMagic-100k-en} & \multicolumn{2}{c}{TREC-COVID} \\
\cmidrule(lr){2-3} \cmidrule(lr){4-5}
                   & \attack      & RAG-MIA    & \attack      & RAG-MIA     \\
\midrule
Std. RAG      & \textbf{0.7500} & 0.5410  & \textbf{0.6585} & 0.5390  \\
LLMlingua         & \textbf{0.6465} & 0.6460  & \textbf{0.7065} & 0.6365 \\
SC-RAG   &  \textbf{0.6720} & 0.5375  & \textbf{0.6110} & 0.5660 \\
Spring            & \textbf{0.6360} & 0.5000 & \textbf{0.6455} & 0.5005  \\
IRCOT            & \textbf{0.7475} & 0.6350 & \textbf{0.8025} & 0.6015  \\
Avg.            & \textbf{0.6904} & 0.5719 & \textbf{0.6848} & 0.5687  \\
\bottomrule
\end{tabular}
}
\end{table}

\begin{figure}[t!]
  \centering
  \includegraphics[width=\linewidth]{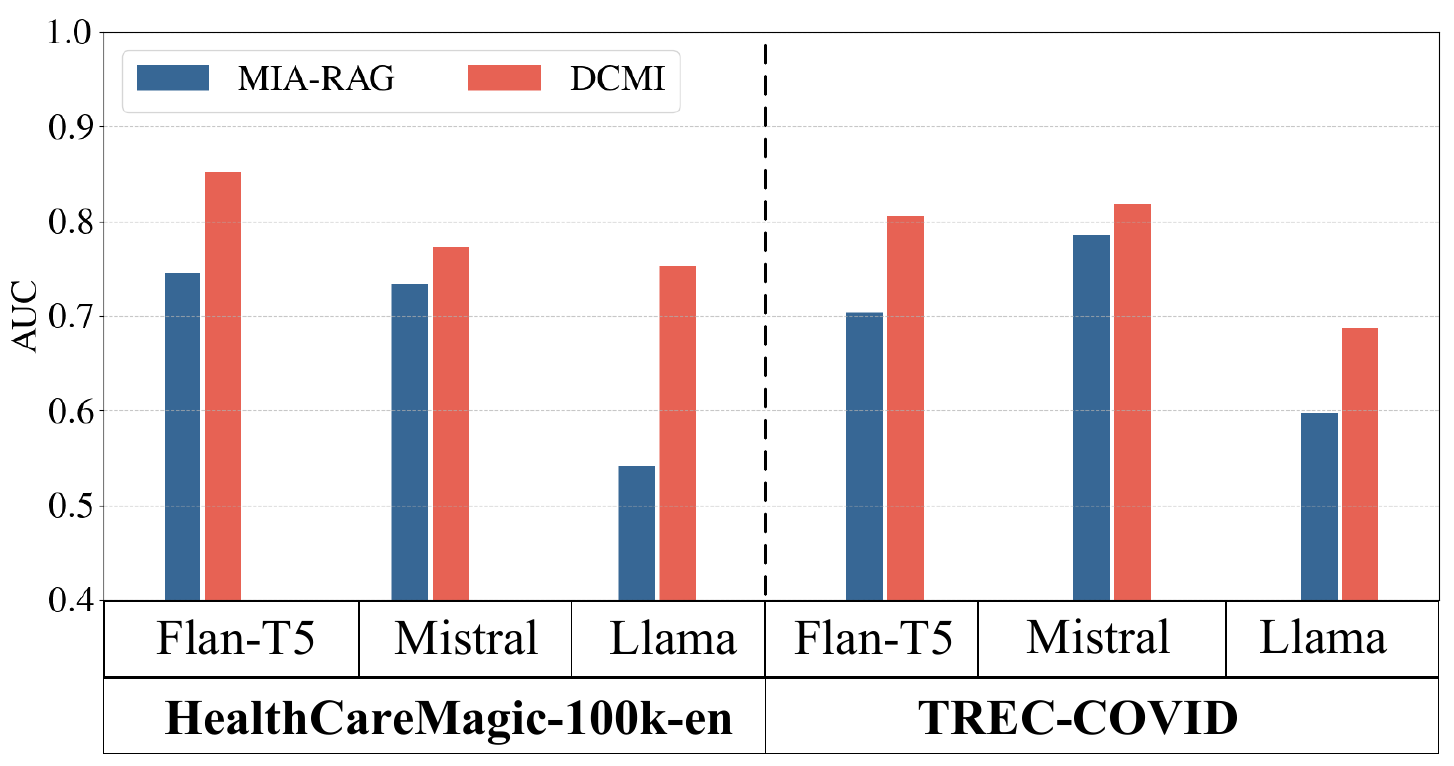}
  \caption{ AUC of Adversary 3 on three generative modules.}
  \label{fig:generative_AUC_3}
\end{figure}
\mypara{Impact of Generative
Modules.}
Consistent with Adversary 1 and Adversary 2,  \attack performs best with Flan-T5, achieving an AUC exceeding 80\% (see \autoref{fig:generative_AUC_3}).

The Accuracy results show similar trends
(see \autoref{fig:generative_Acc_3} in \autoref{app:acc metric}).

\begin{figure}[t!]
  \centering
  \includegraphics[width=\linewidth]{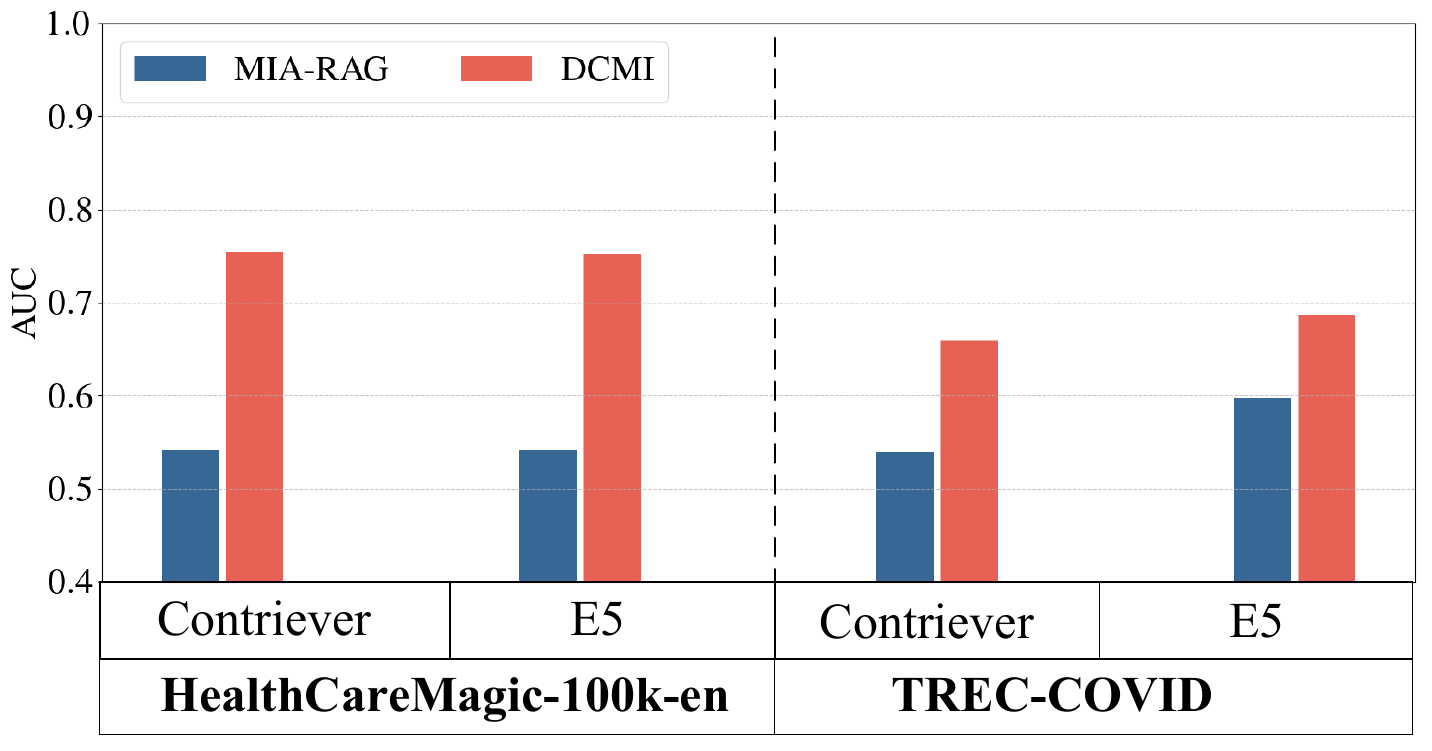}
  \caption{ AUC of Adversary 3 on two retrieval modules.}
  \label{fig:retriever_AUC_3}
\end{figure}
\mypara{Impact of Retrieval
Modules.}
\attack shows consistent AUC across retrieval modules (\autoref{fig:retriever_AUC_3}), with Accuracy exhibiting similar trends (\autoref{fig:retriever_Acc_3}, \autoref{app:acc metric}), aligning with the conclusions from Adversary 1 and Adversary 2.

\begin{figure}[t!]
  \centering
  \includegraphics[width=\linewidth]{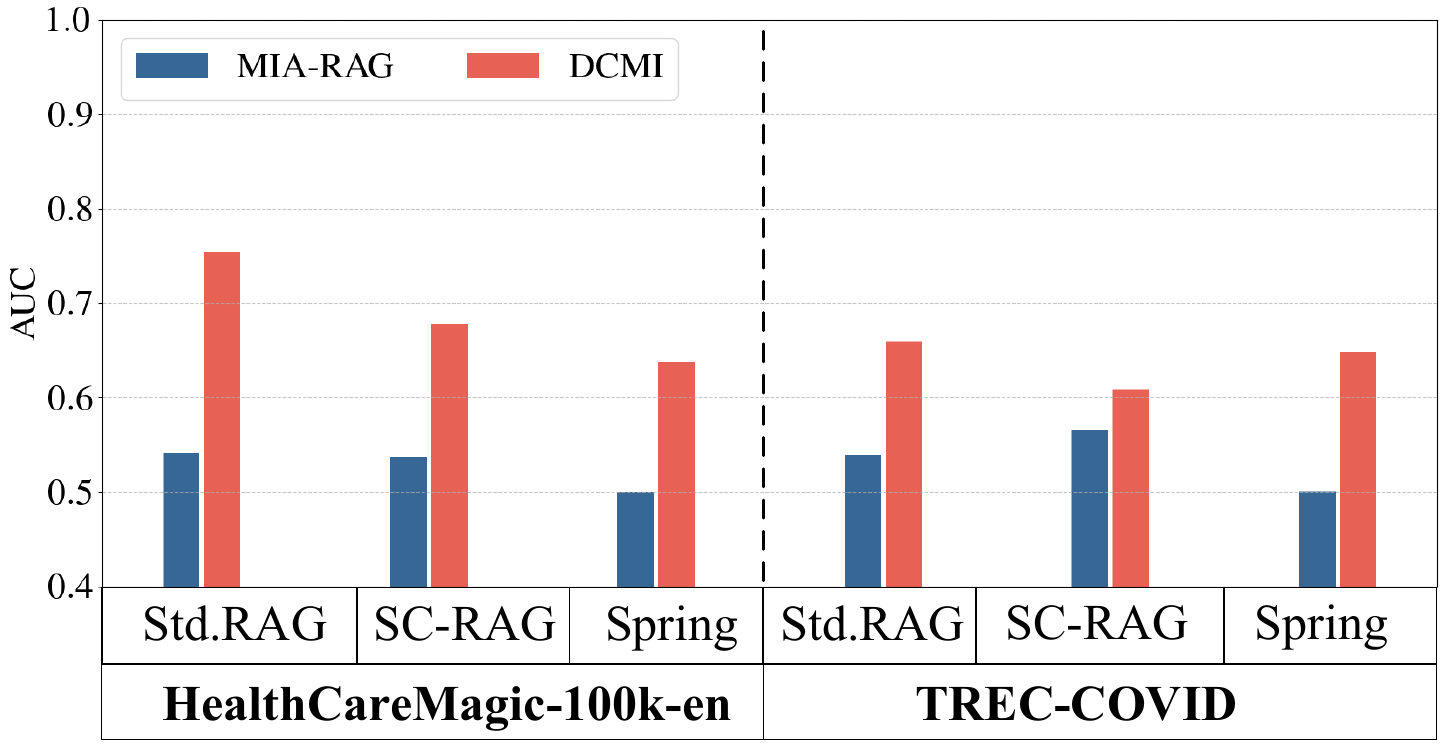}
  \caption{ AUC of Adversary 3 on three RAG frameworks.}
  \label{fig:frame_AUC_3}
\end{figure}

\mypara{Impact of RAG Frameworks.}
The AUC (\autoref{fig:frame_AUC_3}) and Accuracy (\autoref{fig:frame_Acc_3}, \autoref{app:acc metric}) results show that the Standard RAG framework is generally more vulnerable to MIAs than advanced (compressed/fine-tuned) ones.
RAG-MIA-black's attack performance drops to around 50\% in the fine-tuned Spring framework. In contrast, \attack outperforms the baseline by more than 10\% across all tested frameworks, including the challenging Spring framework.

\begin{figure*}[t!]
    \centering
    \begin{subfigure}[t]{0.32\linewidth}
        \includegraphics[width=\linewidth]{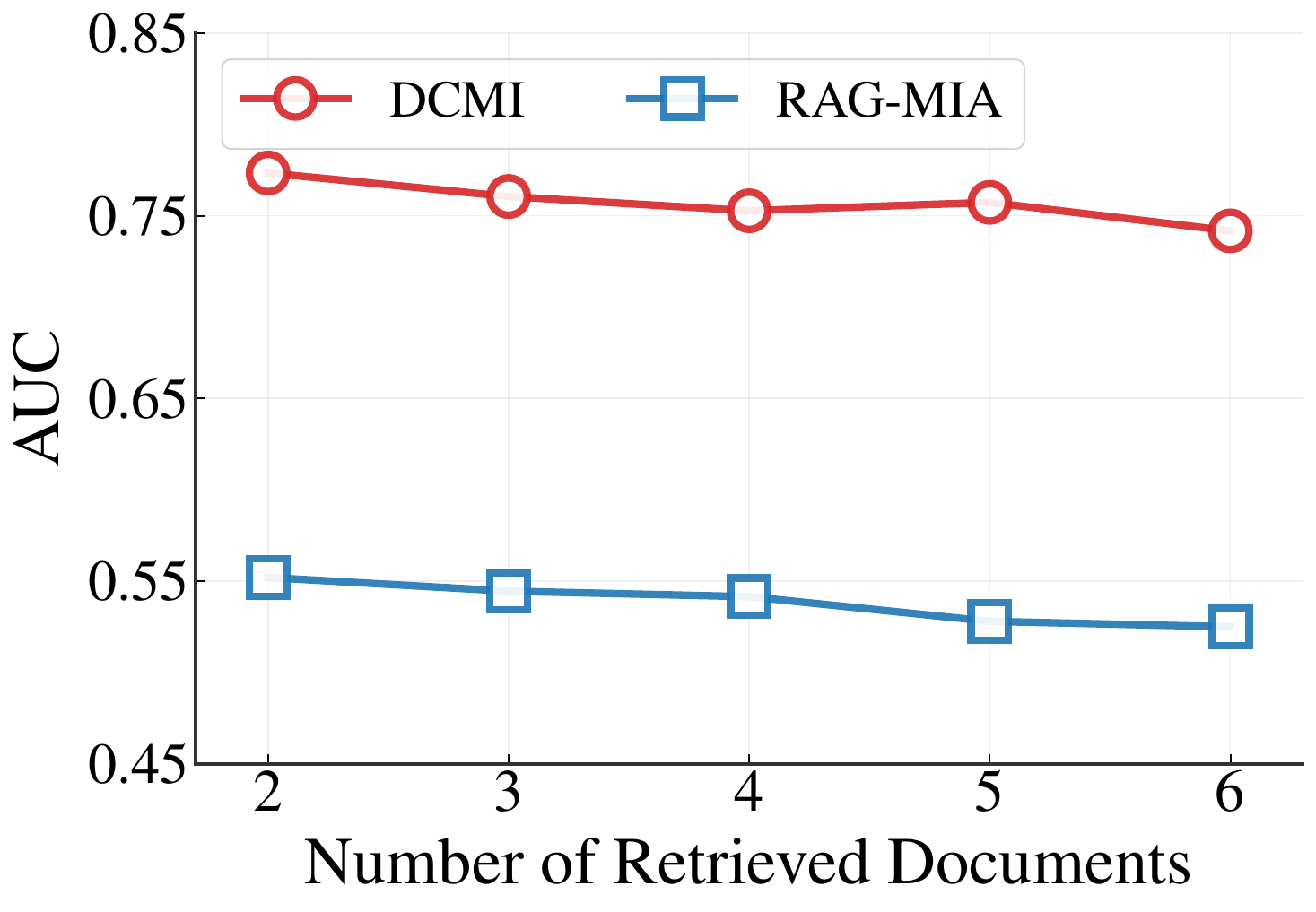}
        \caption{AUC of Adversary 3 on different numbers of retrieved documents.}
        \label{fig:num_AUC_3}
    \end{subfigure}
    \hfill
    \begin{subfigure}[t]{0.32\linewidth}
        \includegraphics[width=\linewidth]{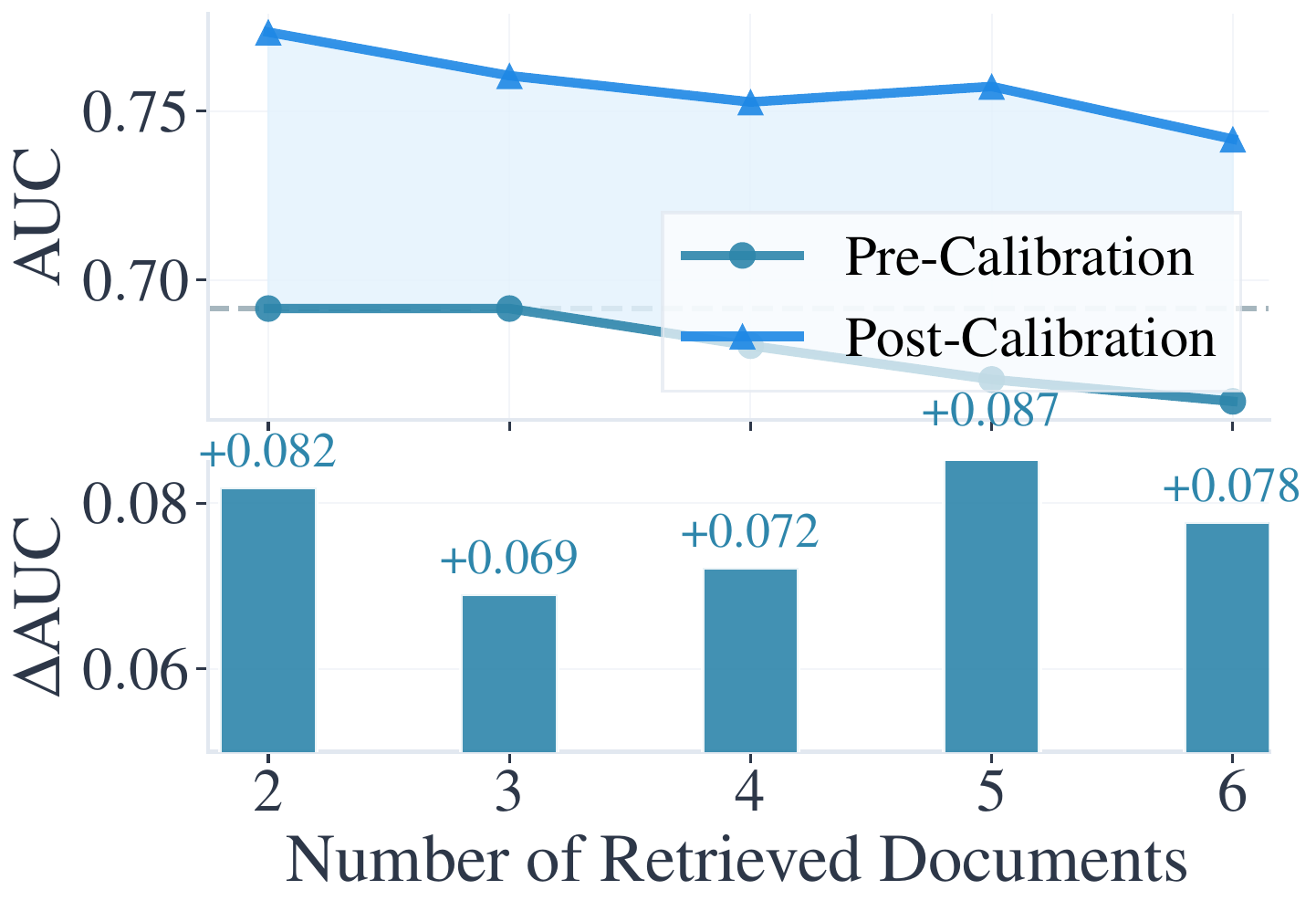}
        \caption{Impact of differential calibration module.}
        \label{fig:calibration_AUC_3}
    \end{subfigure}
    \hfill
    \begin{subfigure}[t]{0.32\linewidth}
        \includegraphics[width=\linewidth]{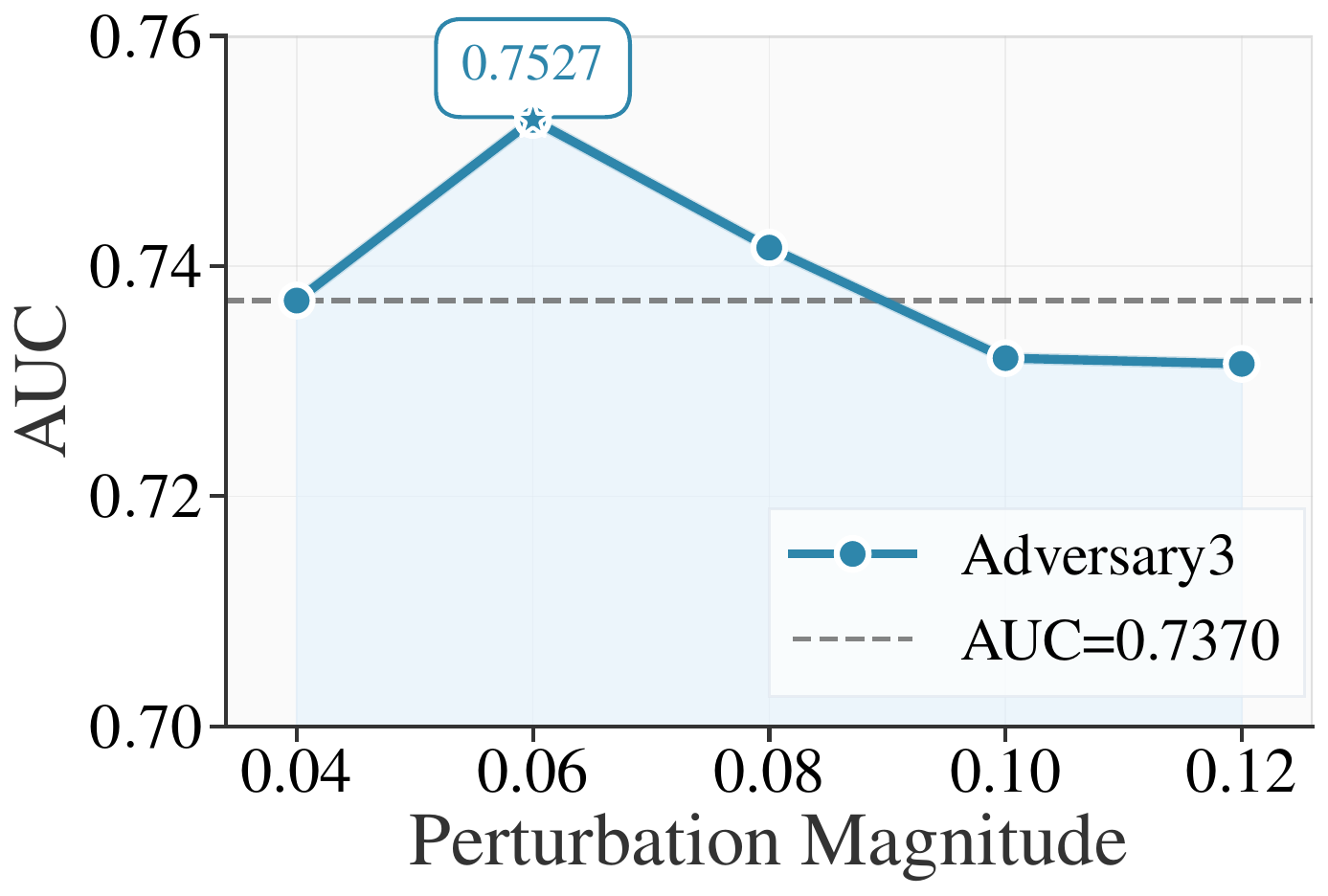}
        \caption{Impact of perturbation magnitude. }
        \label{fig:perturb_AUC_3}
    \end{subfigure}
    
    \caption{Performance Analysis on Adversary 3.}
    \label{fig:performance_analysis_3}
\end{figure*}

\mypara{Impact of Number of
Retrieved Documents.}
AUC (\autoref{fig:num_AUC_3}) and Accuracy (\autoref{fig:num_Acc_3}, \autoref{app:acc metric}) show that \attack consistently maintains a performance advantage of over 20\% relative to RAG-MIA-black as the number of retrieved documents increases.

\subsection{Ablation Studies}
\label{sec:Ablation_adversary3}
Since Adversary 3 is a black-box scenario without threshold selection, its ablation study exclusively assesses the differential calibration module and perturbation magnitude, using settings from \autoref{sec:Ablation_adversary1}. 
The results are presented below.

\mypara{Impact of Differential Calibration Module.}
AUC (\autoref{fig:calibration_AUC_3}) and Accuracy (\autoref{fig:calibration_Acc_3}, \autoref{app:acc metric}) show that the application of the differential calibration module consistently yields a stable performance improvement of 7\%-9\% for \attack across varying numbers of retrieved documents. 
This result validates the broad practicality and effectiveness of this calibration strategy, even within black-box scenarios.
\autoref{app:Extended_Ablation_Study} further validates the differential calibration module's effectiveness across diverse settings.

\mypara{Impact of Perturbation Magnitude.}
According to AUC (\autoref{fig:perturb_AUC_3}) and Accuracy (\autoref{fig:perturb_Acc_3}, \autoref{app:acc metric}), under the Adversary 3 scenario, \attack's performance demonstrates high stability as the perturbation magnitude varies within the range [0.04, 0.12], with overall fluctuations not exceeding 1\%, thereby supporting the strategy of fixing this parameter at 0.06 for the Adversary 3 scenario.

\section{Discussions}
In this section, we first perform \attack against the real-world RAG
system. We then discuss the potential defenses.

\subsection{Real-World Evaluation}
To demonstrate the practical impact of \attack, we evaluate our attack against production-grade RAG systems that operate with hidden internal components, including data preprocessing pipelines, system architectures, and privacy safeguards. 
This opacity makes them challenging to attack, but more representative of real-world scenarios.

\mypara{Setup.} We evaluate two production-grade RAG systems---MAXKB (self‑hosted) and Dify (cloud or self‑hosted)---to cover common deployment scenarios (\autoref{tab:black_rag_system}, \autoref{app:rag_systems}). 
On each system, we deploy a retrieval dataset using the HealthCareMagic‑100k‑en dataset.
Specifically, we run each RAG instance in isolation, avoiding any disruption to platforms or use of real user data.

This realistic scenario closely aligns with the adversary 3.
Therefore, the baseline is RAG-MIA-black ~\cite{anderson2024membership}, and all other experimental settings strictly adhere to the Adversary 3 specifications (\autoref{sec:adversary3_evaluation}).

\mypara{Results.} Experimental results (detailed in \autoref{tab:black_box_result}) demonstrate \attack's significant efficacy against production RAG systems. On MaxKB and Dify, \attack achieved >60\% attack success rates, improving by 20\% and 10\%, respectively, over the baseline. 
These empirical findings from operational environments conclusively establish our attack methodology as a tangible threat to existing RAG systems, revealing critical security vulnerabilities that have profound implications for future security-oriented RAG design.

\begin{table}[t!]
\centering
\caption{Performance comparison across RAG systems under real-world scenario}
\label{tab:black_box_result}
\begin{tabular}{lccccccc}
\toprule
\multirow{2}{*}{RAG System} & \multicolumn{2}{c}{Dify} & \multicolumn{2}{c}{MaxKB} \\
\cmidrule(lr){2-3} \cmidrule(lr){4-5} 
 & Accuracy & AUC & Accuracy & AUC \\
\midrule
RAG-MIA &0.5180  &0.5180  & 0.5360 & 0.5360  \\
\textbf{Ours} &\textbf{0.6190}  &\textbf{0.6330}  & \textbf{0.7415} & \textbf{0.7443} \\
\bottomrule
\end{tabular}
\end{table}

\begin{table}[t!]
\centering
\caption{Performance of AUC metric before and after defense under the Adversary 3 scenario.}
\label{tab:defense_model_AUC}

{
\begin{tabular}{l c c }
\toprule
\textbf{AUC}               & \textbf{\attack}   & \textbf{RAG-MIA}    \\ 
\midrule
No Defense                    & 0.7527              & 0.5415                           \\ 
Instruction-Based Defense                      & 0.7073              & 0.5330       \\ 
Paraphrasing-Based Defense                 & 0.5601 &0.5265   \\
Entity-relation Extraction Defense  &0.5177 &0.5150\\
\bottomrule
\end{tabular}
}
\end{table}

\subsection{Potential Defenses}
By progressively relaxing our attack assumptions, we arrive at the fully realistic black‑box Adversary 3, which demonstrates robust and efficient membership inference capabilities, revealing substantial privacy risks. 
In this section, we explore two potential defense strategies to minimize information leakage about the retrieval database.

\mypara{Setup.} 
To evaluate the effectiveness of defense strategies against various RAG MIAs, the experiment is conducted in the healthcare scenario, followed by the \textit{Basic RAG Setting} in \autoref{sec:Experimental Setup}.  
We evaluate \attack's defense performance under the Adversary 3  scenario and compare its performance to that of RAG‑MIA‑black, the baseline corresponding to Adversary 3 (\autoref{sec:adversary3_evaluation}).

\mypara{Instruction-Based Defense.} 
Inspired by techniques like prompt injection attacks, our first defense strategy integrates defensive instructions into the RAG template to prevent responses to queries that may cause information leakage. 
Using GPT-3.5, we redesigned the RAG template~\cite{wen2024membership} to ignore requests querying the retrieval database and avoid directly outputting retrieved information. 
The new defensive template is shown in \autoref{sec:defensive_template}.

The results in AUC (\autoref{tab:defense_model_AUC}) and Accuracy (\autoref{tab:defense_model_ACC}, \autoref{app:defense}) indicate that the instruction-based defense reduces the AUC of the \attack model by 4\% and the Accuracy by 7\%. In contrast, the performance drop in the RAG-MIA baseline is comparatively smaller, mainly because its attack Accuracy was already around 50\% (i.e., close to random guessing) before the defense implementation.

\mypara{Paraphrasing-Based Defense.} 
In order to further enhance the defensive effectiveness, we propose a paraphrasing-based defense strategy. This approach rewrites user queries in different styles while preserving their core meaning, thereby reducing the match between the paraphrased queries and the retrieved documents and increasing the difficulty of MIAs. 

Experimental results (AUC in \autoref{tab:defense_model_AUC} and Accuracy in \autoref{tab:defense_model_ACC}, \autoref{app:defense}) indicate that this strategy reduces the \attack effectiveness by about 20\%, achieving an attack Accuracy of 54.7\% (AUC 56.01\%), while the attack efficacy of the MIA-RAG method is only 52.65\%. 
However, this method may slightly undermine the generative ability of the RAG system, necessitating a trade-off between generative performance and privacy security.

\mypara{Post-retrieval Entity-relation Extraction.}
Based on our findings in \autoref{sec:Adversary1_evaluation},
where SC-RAG's compression operation reduces MIA performance, we propose a defense mechanism that transforms retrieved documents into entity-relation triples using GPT-3.5 before generation.

This defense addresses both traditional RAG MIA and \attack. 
Traditional RAG MIA exploits high confidence scores from exact matches between queries and member documents (\autoref{sec:Intro}). 
By converting documents to entity-relation triples, we eliminate exact text matches while preserving semantics, reducing the confidence differences that enable MIA. 
For \attack, which exploits sensitivity differences between \textit{member} and \nomember under query perturbation, our structured representation removes calibration-related words corresponding to the perturbation from \member, minimizing the sensitivity gap.

Experimental results show this defense reduces both \attack and RAG-MIA to approximately 50\% AUC (AUC in \autoref{tab:defense_model_AUC} and Accuracy in \autoref{tab:defense_model_ACC}, \autoref{app:defense}).

\section{Conclusion}
In this paper, we address the unique challenges posed by \nomember in RAG systems and their impact on membership inference attacks. We are the first to analyze how \nomember—an issue specific to RAG—interferes with MIA performance, offering new insights into its role in data privacy risks.
To address this challenge, we propose \attack, a differential calibration MIA that uses carefully crafted perturbation queries to reduce \nomember influence and enhance membership discrimination.

We evaluate \attack under three increasingly realistic adversary models on two benchmark datasets and multiple RAG configurations. Results show that \attack consistently outperforms baselines and remains robust across different perturbation magnitudes and thresholds. We also examine three defense strategies that partially reduce \attack’s effectiveness. Finally, we test \attack on real-world RAG systems (Dify and MAXKB), achieving significantly better performance than existing baselines. 
The effectiveness of \attack highlights serious privacy risks in RAG systems and underscores the need for stronger defenses. While our defenses show potential in mitigating leakage, more research is needed to develop comprehensive and robust solutions.

\bibliographystyle{plain}
\bibliography{reference}

\appendix

\clearpage

\section{Notation of symbols}
\label{app:notation}
\begin{table}[htbp]
\centering
\caption{Notation of symbols in the paper.}
\label{tab:notation}
\small
\setlength{\tabcolsep}{1.5pt} 
\begin{tabular}{ll}
\toprule
\textbf{Notation} & \textbf{Description} \\
\midrule
$\mathcal{S}$ & The target RAG system\\
$\mathcal{R}_{D}$ & The external retrieval database of RAG \\
$\mathcal{R}$ & The retriever of RAG\\
$\mathcal{G}$ & The generative module of RAG\\
$x$ & A target sample for membership inference\\
$q$ & The user query\\
$q'$ & The adversarially perturbed query derived from $q$\\
$D_r = \{d_1, d_2, \ldots, d_k\}$ & Documents retrieved for query $q$\\
$d^*$ & Most semantically similar document to $q$ in $D_r$\\
\bottomrule
\end{tabular}
\end{table}

\section{Proof of Robustness for $e^{s(d, q)}$ under Query Perturbation}
\label{app:robustness}

Assume the query \( q \in \mathbb{R}^n \) and the document \( d \in \mathbb{R}^n \). The cosine similarity is defined as:
\[
s(d,q) = \frac{d^Tq}{\|d\|\,\|q\|}.
\]
We define the function \( f(q) = \exp(s(d,q)) \) and let the perturbed query be \( q' = q + \delta \). Our goal is to show that \( |f(q') - f(q)| \) is sufficiently small when \( \|\delta\| \ll \|q\| \).

\bigskip
\textbf{Step 1: Expand the Perturbed Similarity \( s(d, q') \)}\\[1mm]
Substitute \( q' = q + \delta \) into the cosine similarity:
\[
s(d,q') = \frac{d^T (q+\delta)}{\|d\|\,\|q+\delta\|}.
\]
The norm in the denominator can be written as:
\[
\|q+\delta\| = \|q\| \sqrt{1 + \frac{2q^T\delta}{\|q\|^2} + \frac{\|\delta\|^2}{\|q\|^2}}.
\]
Since \( \|\delta\| \ll \|q\| \), we neglect the higher-order term \( \mathcal{O}\left(\frac{\|\delta\|^2}{\|q\|^2}\right) \) and approximate:
\[
\|q+\delta\| \approx \|q\|\left(1+\frac{q^T\delta}{\|q\|^2}\right).
\]
Thus,
\[
s(d,q') \approx \frac{d^Tq + d^T\delta}{\|d\|\,\|q\|\left(1+\frac{q^T\delta}{\|q\|^2}\right)}.
\]

\bigskip
\textbf{Step 2: Apply a Taylor Expansion to the Denominator}\\[1mm]
Using the first-order Taylor expansion for \( \frac{1}{1+x} \) when \( x \ll 1 \):
\[
\frac{1}{1+\frac{q^T\delta}{\|q\|^2}} \approx 1-\frac{q^T\delta}{\|q\|^2},
\]
We obtain:
\[
s(d,q') \approx \frac{d^Tq + d^T\delta}{\|d\|\,\|q\|}\left(1-\frac{q^T\delta}{\|q\|^2}\right).
\]

\bigskip
\textbf{Step 3: Separate the Original Similarity and the Perturbation Term}\\[1mm]
Expanding and keeping only first-order terms, we have:
\[
s(d,q') \approx \underbrace{\frac{d^Tq}{\|d\|\,\|q\|}}_{s(d,q)} + \underbrace{\frac{d^T\delta}{\|d\|\,\|q\|} - \frac{(d^Tq)(q^T\delta)}{\|d\|\,\|q\|^3}}_{\Delta s}.
\]
Thus, we write:
\[
s(d,q') = s(d,q) + \Delta s,
\]
with the perturbation-induced change being:
\[
\Delta s = \frac{d^T\delta}{\|d\|\,\|q\|} - s(d,q)\cdot\frac{q^T\delta}{\|q\|^2}.
\]

\bigskip
\textbf{Step 4: Bound \(\Delta s\) using the Cauchy-Schwarz Inequality}\\[1mm]
For the first term:
\[
\left|\frac{d^T\delta}{\|d\|\,\|q\|}\right| \leq \frac{\|d\|\,\|\delta\|}{\|d\|\,\|q\|} = \frac{\|\delta\|}{\|q\|}.
\]
For the second term:
\[
\left|s(d,q)\cdot\frac{q^T\delta}{\|q\|^2}\right| \leq \left|\frac{d^Tq}{\|d\|\,\|q\|}\right|\cdot\frac{\|q\|\,\|\delta\|}{\|q\|^2} = \frac{\|\delta\|}{\|q\|}.
\]
Combining these bounds gives:
\[
|\Delta s| \leq \frac{\|\delta\|}{\|q\|} + \frac{\|\delta\|}{\|q\|} = \frac{2\|\delta\|}{\|q\|}.
\]

\bigskip
\textbf{Step 5: Analyze the Difference \( f(q') - f(q) \)}\\[1mm]
Using the Taylor expansion \( \exp(a+\Delta a) \approx \exp(a)(1+\Delta a) \) for small \(\Delta a\):
\[
f(q') - f(q) = \exp(s(d,q')) - \exp(s(d,q)) \approx \exp(s(d,q)) \cdot \Delta s.
\]
Thus, we have:
\[
\left|f(q') - f(q)\right| \leq \exp(s(d,q)) \cdot |\Delta s| \leq \exp(s(d,q)) \cdot \frac{2\|\delta\|}{\|q\|}.
\]

\bigskip
\textbf{Conclusion}\\[1mm]
When \( \|\delta\| \ll \|q\| \), the change in cosine similarity satisfies
\[
|\Delta s| \le \frac{2\|\delta\|}{\|q\|},
\]
which implies
\[
\left|f(q') - f(q)\right| \le 2\,\exp(s(d,q))\cdot\frac{\|\delta\|}{\|q\|}.
\]
This shows that the function \( f(q)=\exp(s(d,q)) \) is robust to small perturbations in the query, as the difference decays linearly with the relative perturbation magnitude \( \frac{\|\delta\|}{\|q\|} \).

\section{Theoretical Analysis of black-box Scenarios}
\label{app:Adversary3_analysis}

\begin{figure*}[t!]
    \centering
    \begin{subfigure}[b]{0.32\linewidth}
        \includegraphics[width=\linewidth]{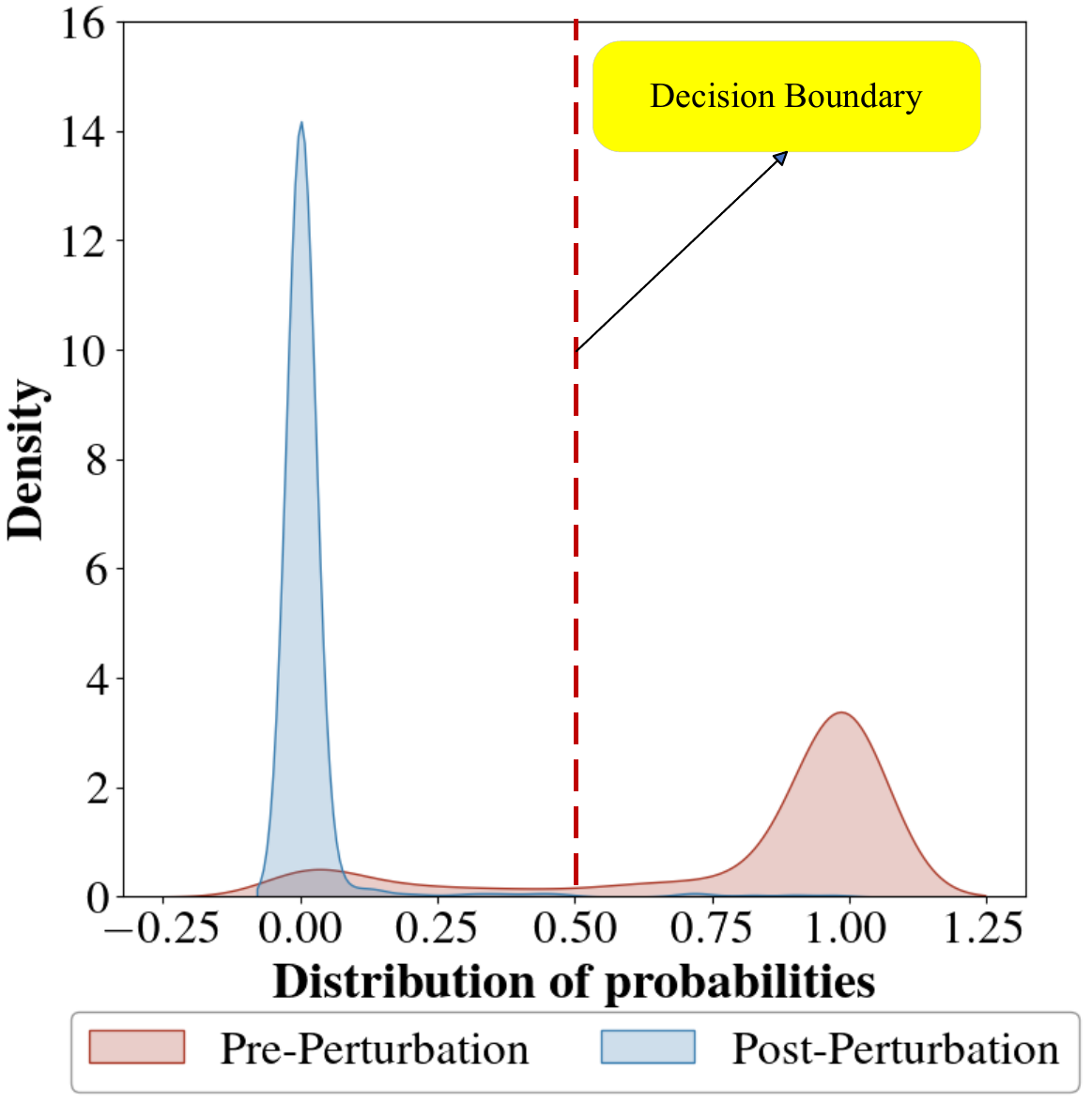}
        \caption{Llama-Member}
    \end{subfigure}
    \hfill
    \begin{subfigure}[b]{0.32\linewidth}
        \includegraphics[width=\linewidth]{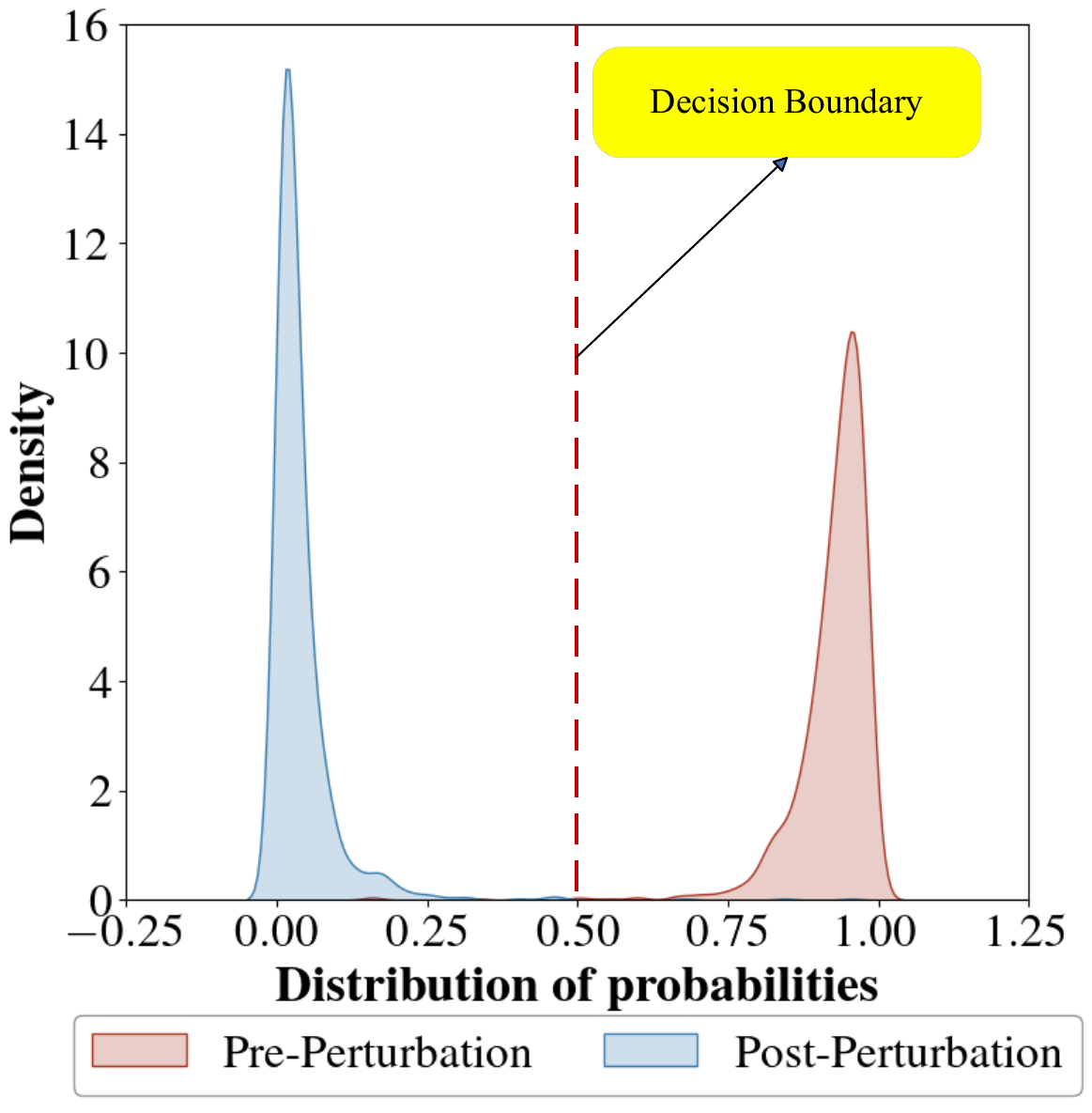}
        \caption{T5-Member}
    \end{subfigure}
    \hfill
    \begin{subfigure}[b]{0.32\linewidth}
        \includegraphics[width=\linewidth]{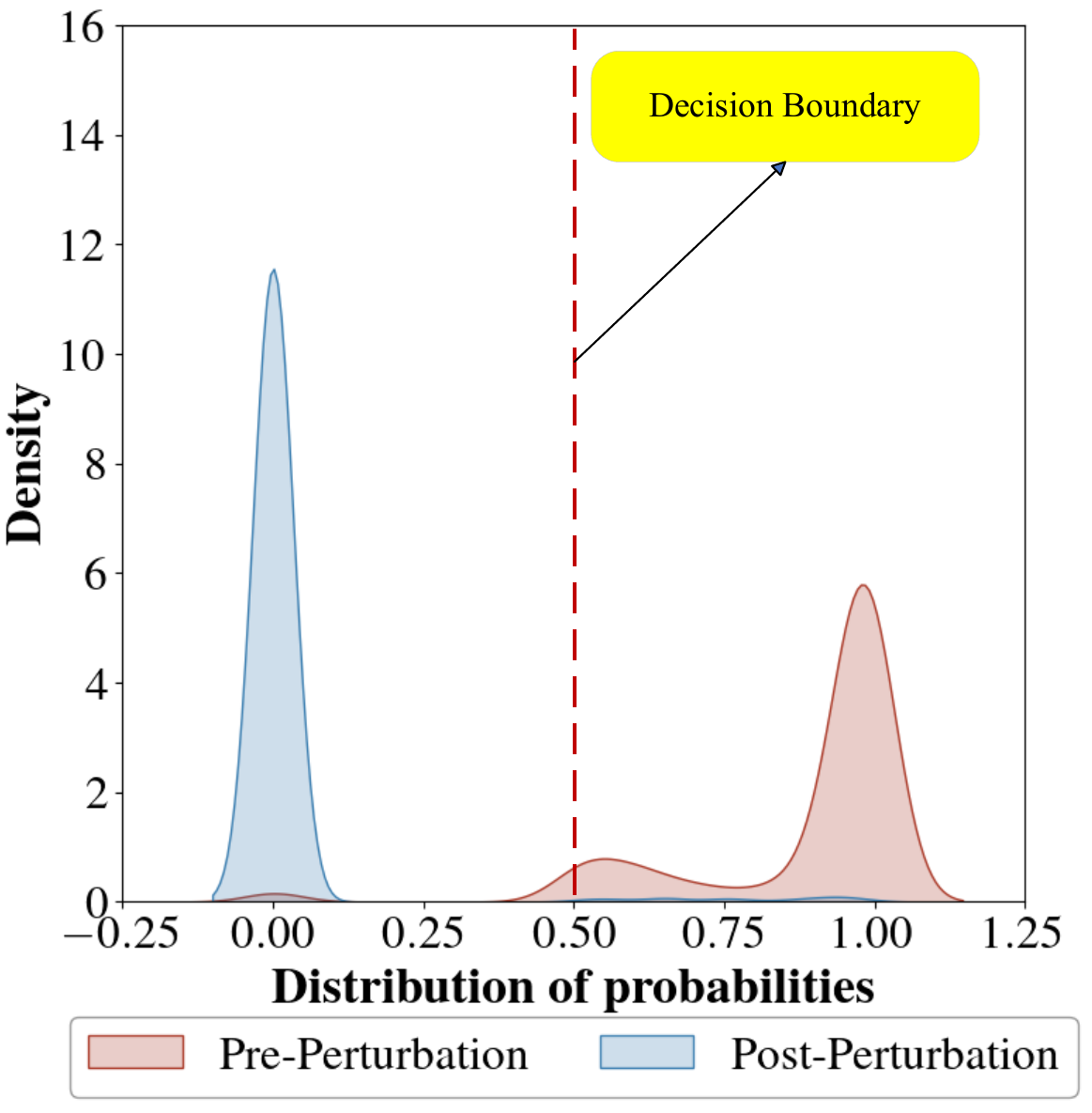}
        \caption{Mistral-Member}
    \end{subfigure}
    
    \vspace{0.5cm} 
    
    \begin{subfigure}[b]{0.32\linewidth}
        \includegraphics[width=\linewidth]{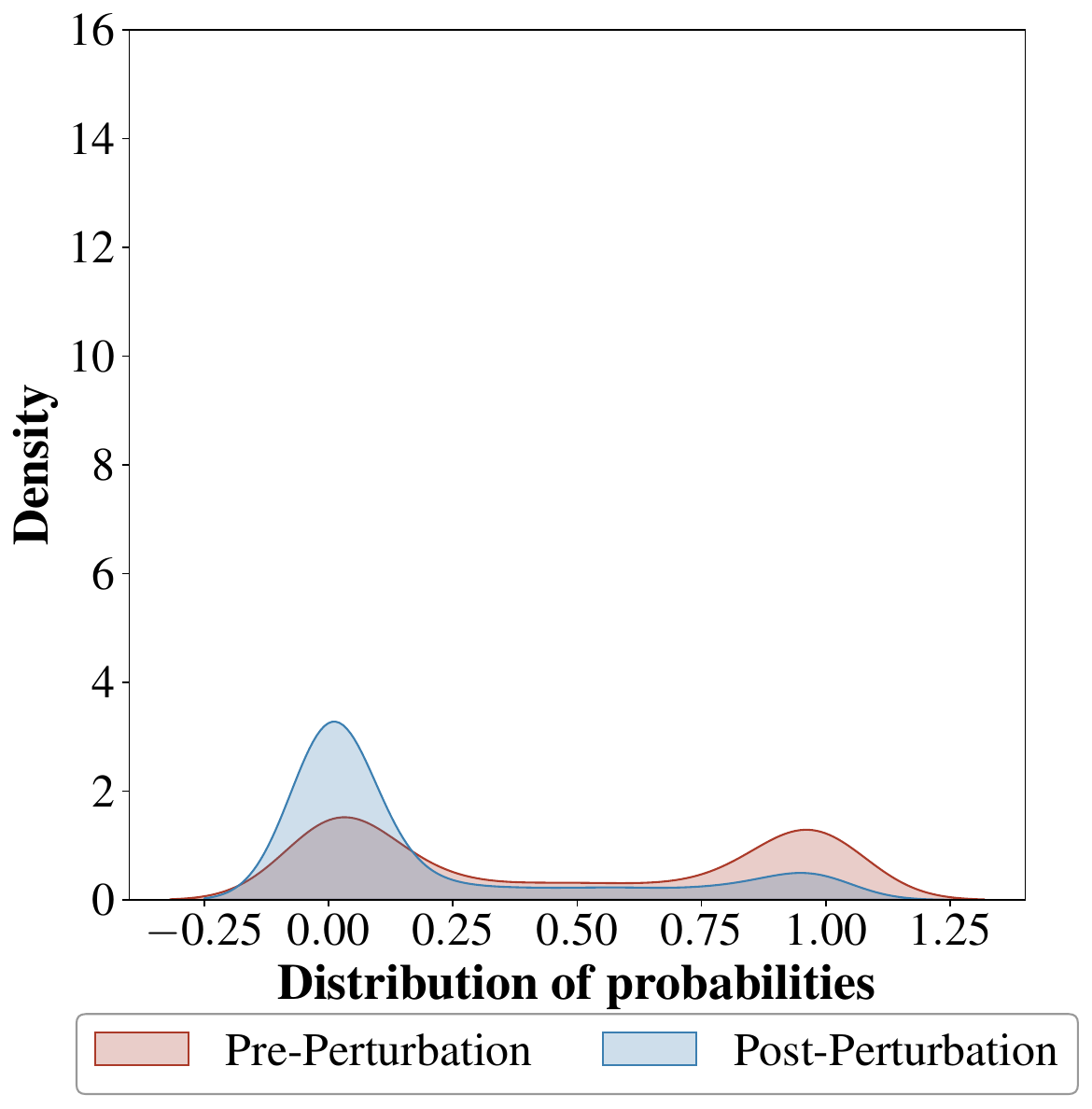}
        \caption{Llama-Non-Member}
    \end{subfigure}
    \hfill
    \begin{subfigure}[b]{0.32\linewidth}
        \includegraphics[width=\linewidth]{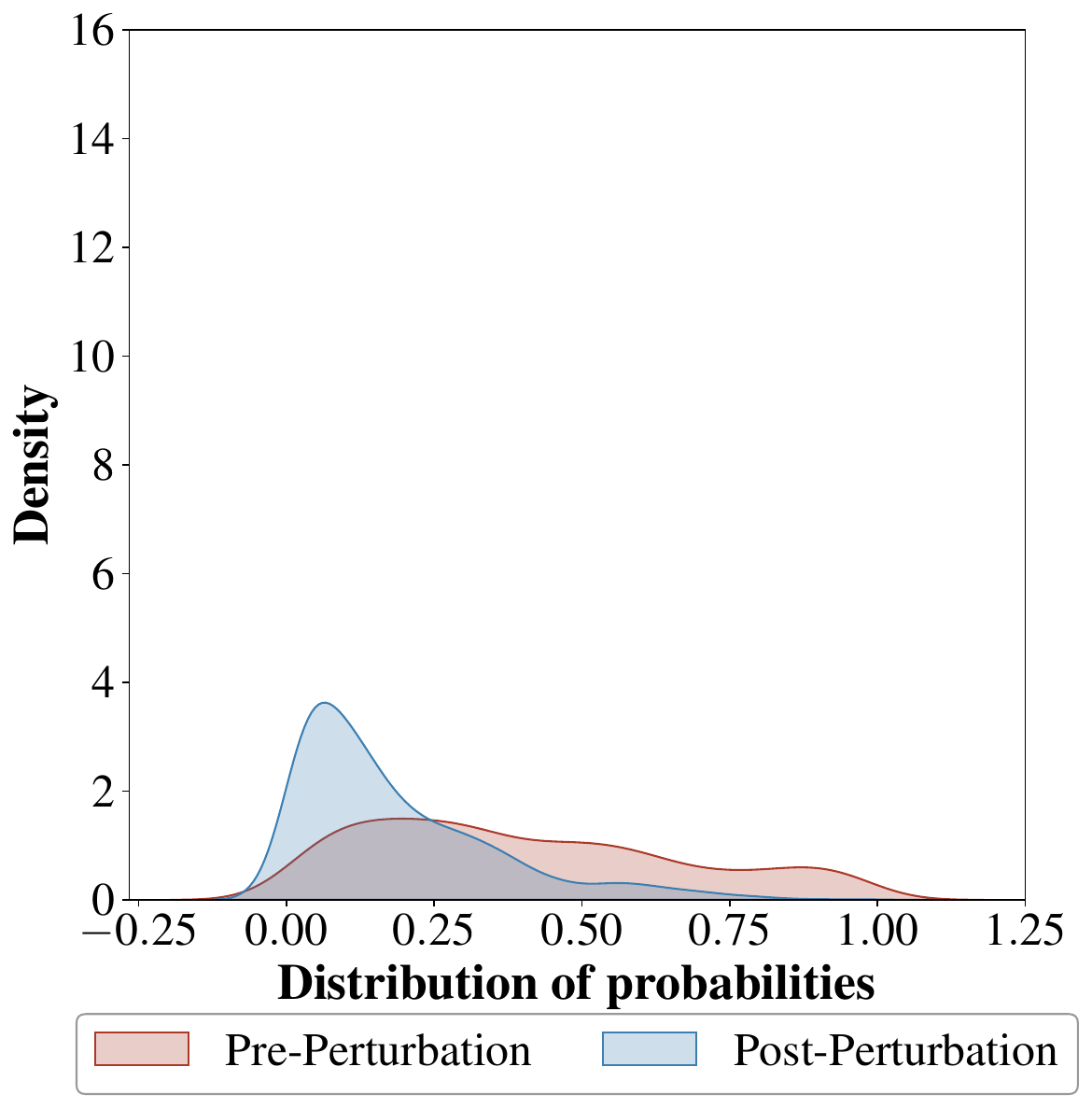}
        \caption{T5-Non-Member}
    \end{subfigure}
    \hfill
    \begin{subfigure}[b]{0.32\linewidth}
        \includegraphics[width=\linewidth]{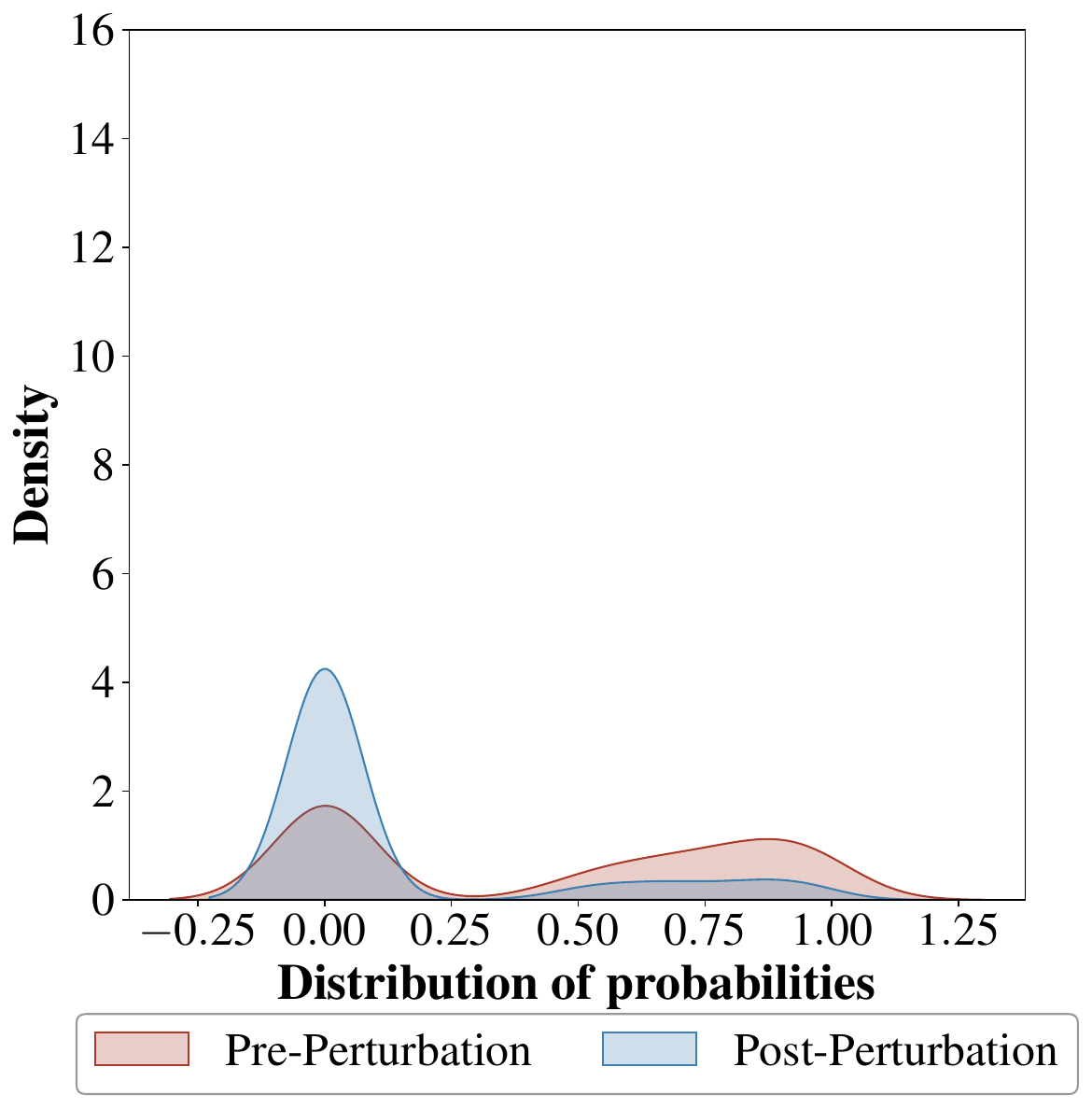}
        \caption{Mistral-Non-Member}
    \end{subfigure}
    
    \caption{Distribution of generation probabilities $P_{\text{rag}}(\text{``Yes''} \mid q)$ for member and non-member samples pre- and post-perturbation across three generation modules.}
    \label{fig:probability_distributions_Adversary3}
\end{figure*}

This section provides a detailed theoretical analysis of the differential calibration under black-box scenarios. 
In black-box settings, we can only observe the final binary responses (``Yes''/``No'') of the RAG system without direct access to generation probabilities.

In the binary response setting, the RAG system outputs ``Yes'' when $P_{\text{rag}}(\text{``Yes''} \mid q) > \alpha$ and ``No'' otherwise, where $\alpha$ is the decision boundary (typically 0.5) and $P_{\text{rag}}(\text{``Yes''} \mid q)$ is defied as:

\begin{equation}
P_{\text{rag}}(\text{``Yes''} \mid q) = p_\eta(d^* \mid q) \cdot p_\theta(\text{``Yes''} \mid q, d^*)+\epsilon
\end{equation}
Where \(\epsilon= \sum_{d \in D_r \setminus \{d^*\}} p_\eta(d \mid q) \cdot p_\theta(\text{``Yes''} \mid q,d)\).
\autoref{General Paeadigm} establish two critical findings:
\begin{itemize}
    \item 
    \( p_{\eta}(d \mid q) \approx p_{\eta}(d \mid q') \).
    
    \item 
    For \member, \( p_\theta(\text{``Yes''} \mid q, d) \gg p_\theta(\text{``Yes''} \mid q', d) \);
    For \nomember, \( p_\theta(\text{``Yes''} \mid q, d) \approx p_\theta(\text{``Yes''} \mid q', d) \).
\end{itemize}

For member samples, the sharp drop in $p_\theta(\text{``Yes''} \mid q', d^*)$ causes $P_{\text{rag}}(\text{``Yes''} \mid q)$ > $P_{\text{rag}}(\text{``Yes''} \mid q')$. 
For non-member samples, $P_{\text{rag}}(\text{``Yes''} \mid q) \approx P_{\text{rag}}(\text{``Yes''} \mid q')$. 
Crucially, this probability gap is large enough that before perturbation $P_{\text{rag}}(\text{``Yes''} \mid q) > \alpha$ leading to ``Yes'' output, while after perturbation $P_{\text{rag}}(\text{``Yes''} \mid q') < \alpha$, resulting in ``No'' output. For non-member samples, since $P_{\text{rag}}(\text{``Yes''} \mid q) \approx P_{\text{rag}}(\text{``Yes''} \mid q')$, probabilities remain on the same side of the decision boundary before and after perturbation.

\autoref{fig:probability_distributions_Adversary3} provides empirical validation across three generation modules under \textit{Basic RAG Setting} in \autoref{sec:Experimental Setup}  with a retrieval quantity of 4. The results clearly demonstrate the decision reversal phenomenon. For member samples, the distributions of $P_{\text{rag}}(\text{``Yes''}|q)$ and $P_{\text{rag}}(\text{``Yes''}|q')$ fall on opposite sides of the decision boundary ($\alpha = 0.5$), exhibiting clear decision reversal from ``Yes'' before perturbation to ``No'' after perturbation. For non-member samples, the distributions maintain similar trends before and after perturbation, primarily concentrated on the same side of the decision boundary without exhibiting decision reversal patterns.

Based on the observed decision reversal behavior, we define the function $f_{\text{rag}}(\cdot)$ to map ``Yes'' to 1 and ``No'' to 0, and propose a differential calibration score:

\begin{equation}
f_{rag,calibrated}(q) = f_{\text{rag}}(q) - f_{\text{rag}}(q'),
\end{equation} 
The discriminative capability of this score is manifested in that member samples yield a score of $1 - 0 = 1$ due to decision reversal (``Yes'' → ``No''), while non-member samples yield $0 - 0 = 0$ or $1 - 1 = 0$ (occasionally $0 - 1 = -1$) due to consistent decisions. Therefore, even in black-box scenarios where only binary responses can be observed, the differential calibration mechanism can still effectively distinguish between member and non-member samples, providing a feasible membership inference method for practical applications.

\section{Perturbation Sample Construction Template}
\label{sec:perturbation_template}

\begin{tcolorbox}[
    colback=white,             
    colframe=black,             
    width=\linewidth,           
    title=Perturbation Sample Construction Template,      
    boxrule=0.5mm,              
    colbacktitle=black,         
    coltitle=white,             
    left=5pt,                   
    right=5pt,                  
    arc=3mm,                    
    before skip=10pt,           
    after skip=10pt             
    ]
\footnotesize
Replace \{num\_to\_replace\} key adjectives or adverbs in noticeable positions with their antonyms in the following text, ensuring the modified text remains  
logically correct: \\
\{text\} \\
Return only the modified text.
\end{tcolorbox}

\section{RAG Template}
\label{sec:rag_template}
\tcbset{
    colframe=black,        
    colback=white,         
    coltitle=white,        
    colbacktitle=black,    
    boxrule=0.5mm,         
    arc=3mm,               
    fonttitle=\bfseries,   
    width=0.95\linewidth, 
    before skip=10pt,      
    after skip=10pt        
}

\begin{tcolorbox}[title=RAG Template]
\textbf{Context:} \{Original Retrieved Documents\} \\
\textbf{Instruction:} Answer the question based on the provided context. \\
\textbf{Query:} \{query\}
\end{tcolorbox}

\section{Datasets}\label{app:dataset}
Here we give a detailed description of datasets used in this work.
\begin{itemize}
    \item \mypara{TREC-COVID~\cite{thakur2021beir}.} This dataset contains 116K scientific and medical documents from the TREC-COVID collection, part of the BEIR benchmark, exemplifying research scenarios where COVID related literature is integrated with RAG to improve information retrieval efficiency and the quality of scientific knowledge exploration in pandemic response contexts.
    \item \mypara{HealthCareMagic-100k-en~\cite{healthcaremagic100k}.} This dataset contains 112,165 real conversations between patients and doctors on HealthCareMagic.com, exemplifying the healthcare scenario where sensitive patient information is integrated with RAG to improve diagnostic efficiency and the quality of medical services.
\end{itemize}

\section{Baselines}\label{app:baseline}
\begin{itemize}
    \item \mypara{RAG-MIA~\cite{anderson2024membership}.}  
    This method directly queries the RAG system if the target document is in the context. 
    \begin{itemize}
        \item \textbf{Gray-box.} In the gray-box variant, membership is inferred based on the probabilities of ``Yes'' or ``No''.
        \item \textbf{Black-box:} In the black-box variant, membership is determined solely from the RAG system’s output (Yes/No).
        
    \end{itemize}
    
    \item \mypara{S\textsuperscript{2}MIA~\cite{li2024blackbox}.}  
    The method splits the document and inputs the first half to the RAG system.
    The semantic similarity between the document and the RAG's response is then calculated, along with perplexity, to perform membership inference.

    \item \mypara{MBA~\cite{liu2024mask}.}  
     This method masks terms in the document and queries the RAG system, inferring membership according to the accuracy of the mask predictions.

     \item \mypara{IA~\cite{naseh2025riddle}.} 
An auxiliary LLM is used to automatically generate a series of highly specific natural language queries from a target sample; 
these queries can be answered accurately only if the sample is present in the retrieval database. 
The adversary submits these queries to the RAG system and compares the returned responses with pre-generated "reference answers." 
The matching scores from the responses are aggregated and then compared to a predetermined threshold. 
This threshold is derived from non-member samples that share the same data distribution as the external database, enabling the adversary to determine whether the target sample exists within it.
\end{itemize}

\section{RAG Systems}\label{app:rag_systems}
\mypara{Standard RAG~\cite{ram2023incontext}.}  
The Standard RAG first accurately retrieves documents that are highly relevant to the input from a large-scale dataset, and then directly concatenates them with the original query as the generative module input.

\mypara{LLMlingua~\cite{jiang2023longllmlingua}.}  
The LLMLingua evaluates the perplexity of each token using a smaller model to eliminate those that contribute less to the overall context, thereby preserving key information while reducing computational costs and improving inference efficiency.

\mypara{SC-RAG~\cite{li2023compressing}.}  
The SC-RAG leverages language models to compute the self-information of words or phrases, then filters out key content according to a percentile threshold, effectively removing redundancy. This process efficiently compresses the context, reducing memory and inference costs while maintaining stable generation performance. 

\mypara{Spring~\cite{zhu2024one}.} 
The Spring seamlessly incorporates externally retrieved information into the generation process of LLM by inserting a set of trainable virtual tokens into the input, without updating the original model parameters. 

\mypara{IRCOT~\cite{trivedi2022interleaving}.} 
The IRCOT alternates between document retrieval and chain-of-thought (COT) reasoning to achieve deep comprehension. Initially, it retrieves relevant passages based on the query, and then uses this information to generate reasoning steps. These steps serve as new queries for further retrieval, and by iteratively integrating the results, the approach constructs a highly accurate final answer.

\begin{table}[t!]
\centering
\caption{Summary of the five representative and open-sourced RAG systems' configurations. }
\label{tab:victim_model}
\setlength{\tabcolsep}{6pt} 
\resizebox{\linewidth}{!}{
\begin{tabular}{c c c}
\toprule
\textbf{Framework}     & \textbf{Generation Module}    & \textbf{Retrieval Module} \\ 
\midrule
Standard RAG~\cite{ram2023incontext}                      & Llama2-7B-chat                  & Contriever              \\ 
LLMLingua~\cite{jiang2023longllmlingua}                              & GPT-3.5-turbo               & E5-base                 \\ 
SC-RAG~\cite{li2023compressing}                          & Llama2-7B-chat                   & E5-base                 \\ 
Spring~\cite{zhu2024one}                                   & Llama2-7B-chat             & E5-base                 \\ 
IRCOT~\cite{trivedi2022interleaving}                                       & Flan-T5-Large               & BM25                    \\ 

\bottomrule
\end{tabular}
}
\end{table}

\begin{table}[t!]
\centering
\caption{Overview of RAG Systems in Black-box Evaluation }
\label{tab:black_rag_system}
\begin{tabular}{lccc}
\toprule
\textbf{System} & \textbf{Deployment} & \textbf{Stars} & \textbf{Forks} \\
\midrule
Dify & Hybrid & 84.7k & 12.5k  \\
MaxKB & Self-hosted & 14.9k & 1.0k  \\
\bottomrule
\end{tabular}
\end{table}

\section{Defensive RAG Template}
\label{sec:defensive_template}

\begin{tcolorbox}[title=Defensive RAG Template]
\textbf{Context:} \{Original Retrieved Chunk\} \\
\textbf{Instruction:} Answer the question based on the provided context. \\
If asked whether specific text exists or to repeat the context, answer 'unanswerable'. \\
Do not output the context directly; rewrite it before answering.\\
\textbf{Query:} \{query\}
\end{tcolorbox}

\section{Experimental Results.}\label{app:acc metric}

\begin{figure}[htbp]
  \centering
  \includegraphics[width=\linewidth]{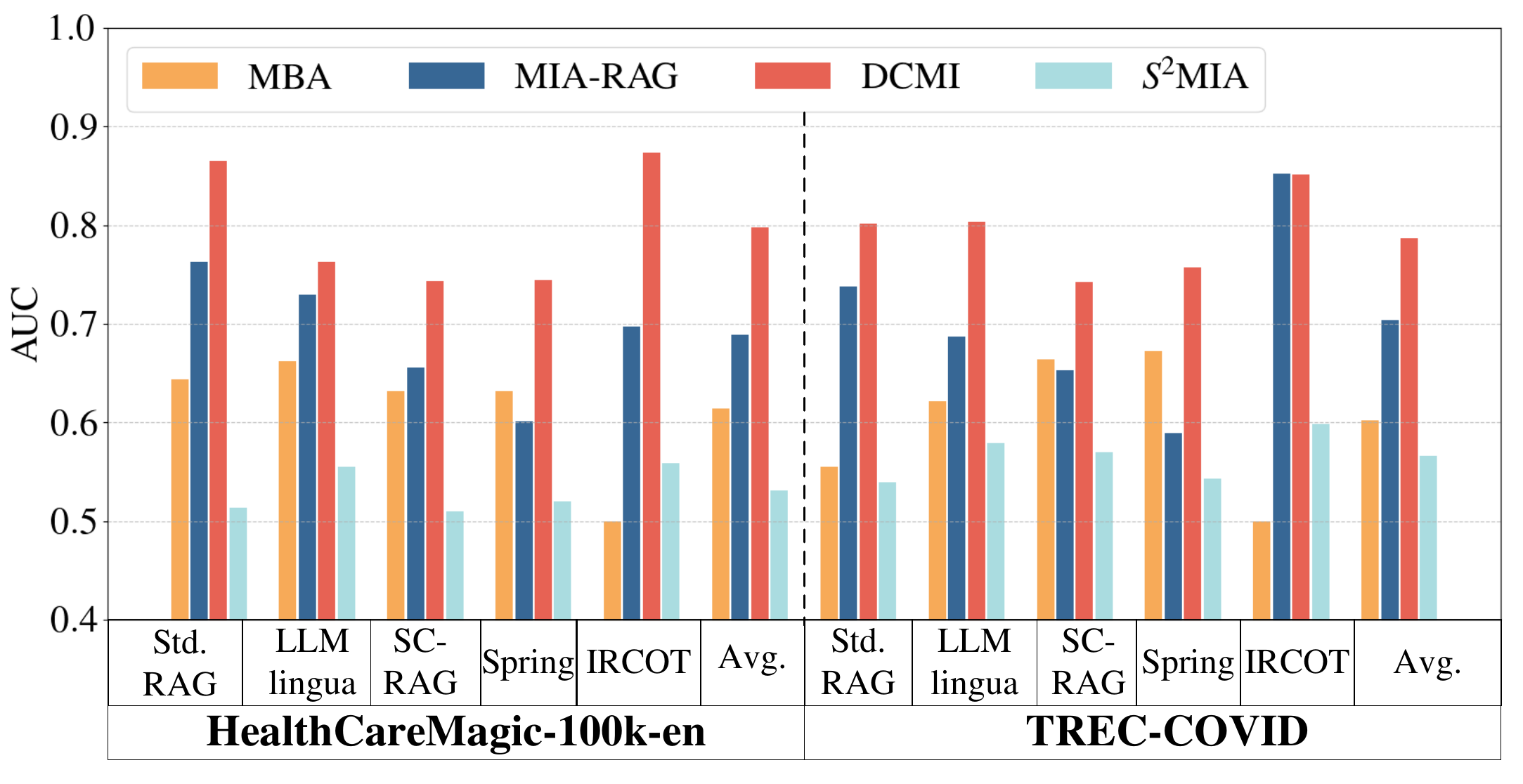}
  \caption{AUC of Adversary 1 on five RAG systems.}
  \label{fig:system_AUC_1}
\end{figure}

\begin{figure}[htbp]
  \centering
  \includegraphics[width=\linewidth]{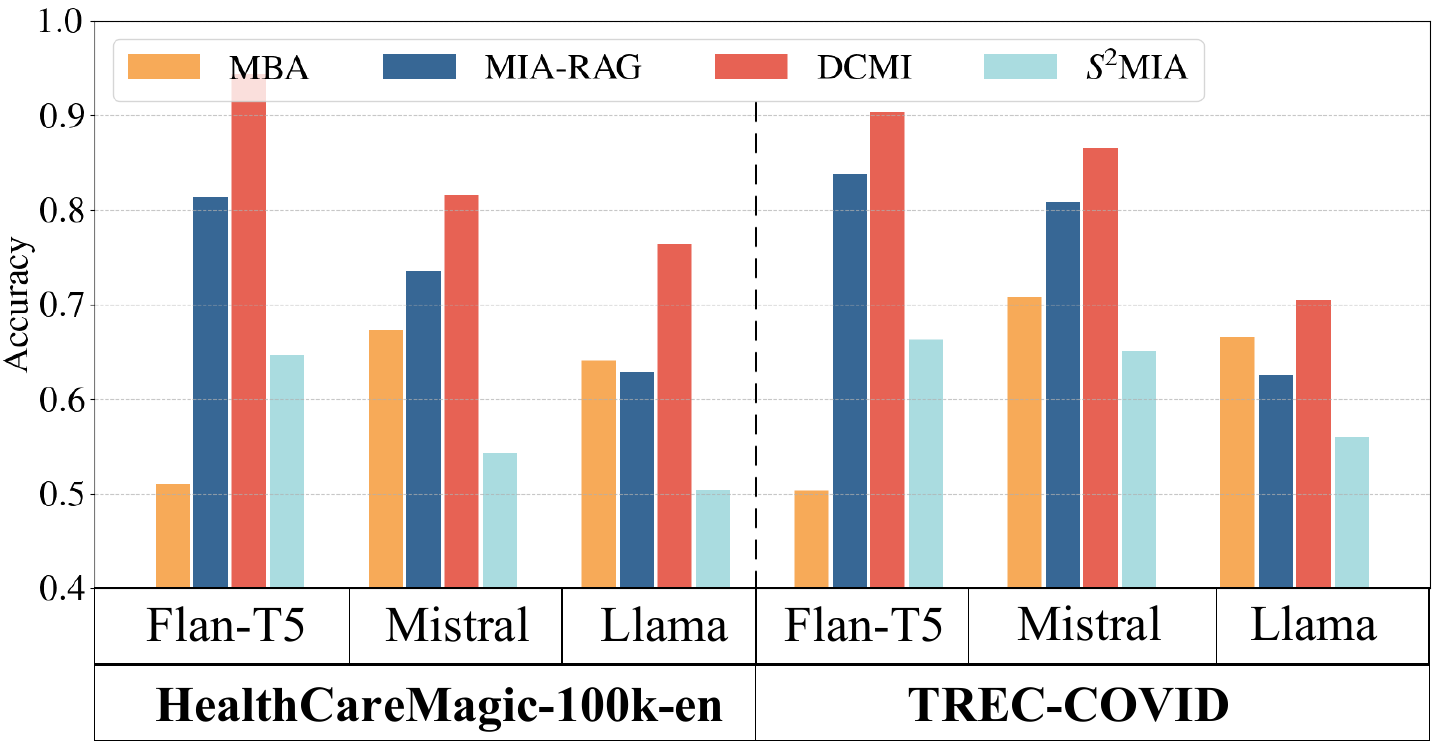}
  \caption{ Accuracy of Adversary 1 on three generative modules.}
  \label{fig:generative_Acc_1}
\end{figure}

\begin{figure}[htbp]
  \centering
  \includegraphics[width=\linewidth]{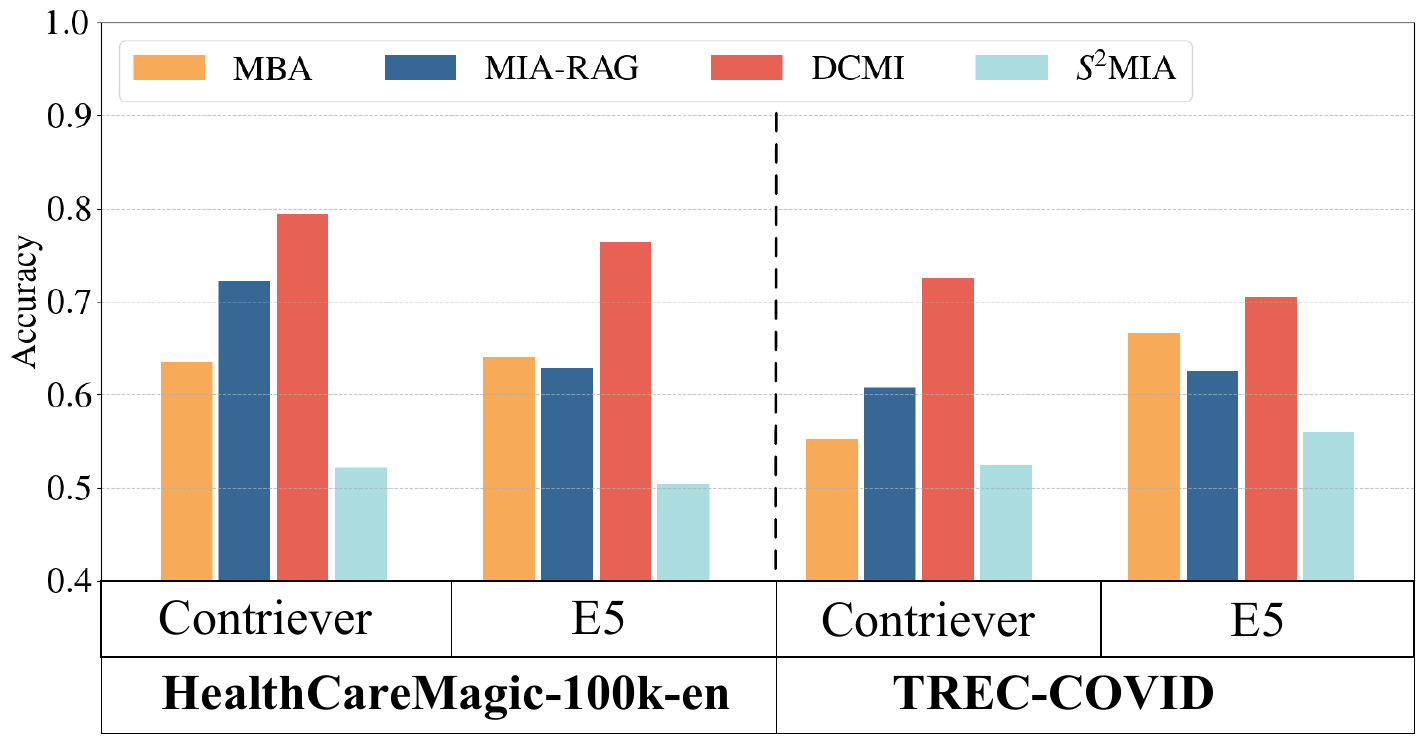}
  \caption{ Accuracy of Adversary 1 on two retrieval modules.}
  \label{fig:retriever_Acc_1}
\end{figure}

\begin{figure}[htbp]
  \centering
  \includegraphics[width=\linewidth]{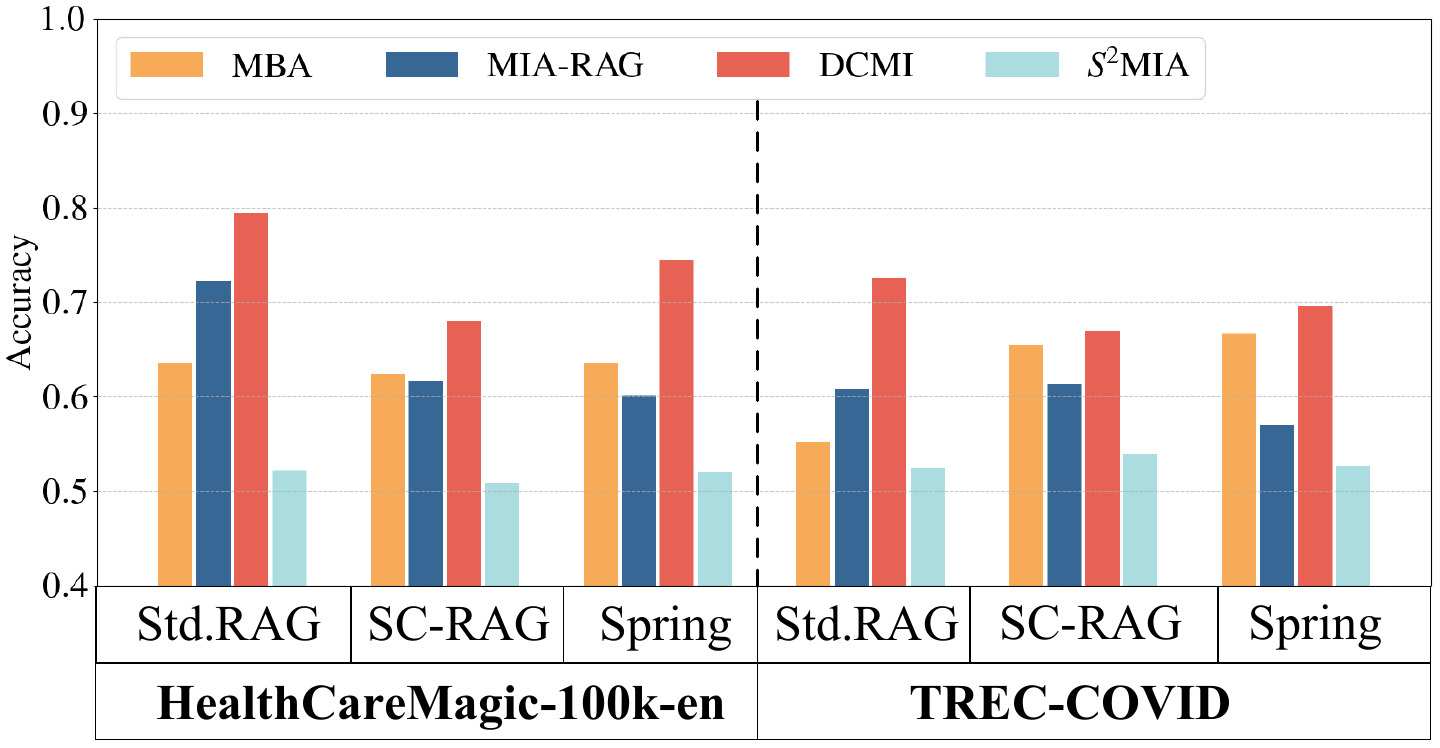}
  \caption{ Accuracy of Adversary 1 on three RAG frameworks.}
  \label{fig:frame_Acc_1}
\end{figure}

\begin{figure}[htbp]
  \centering
  \includegraphics[width=\linewidth]{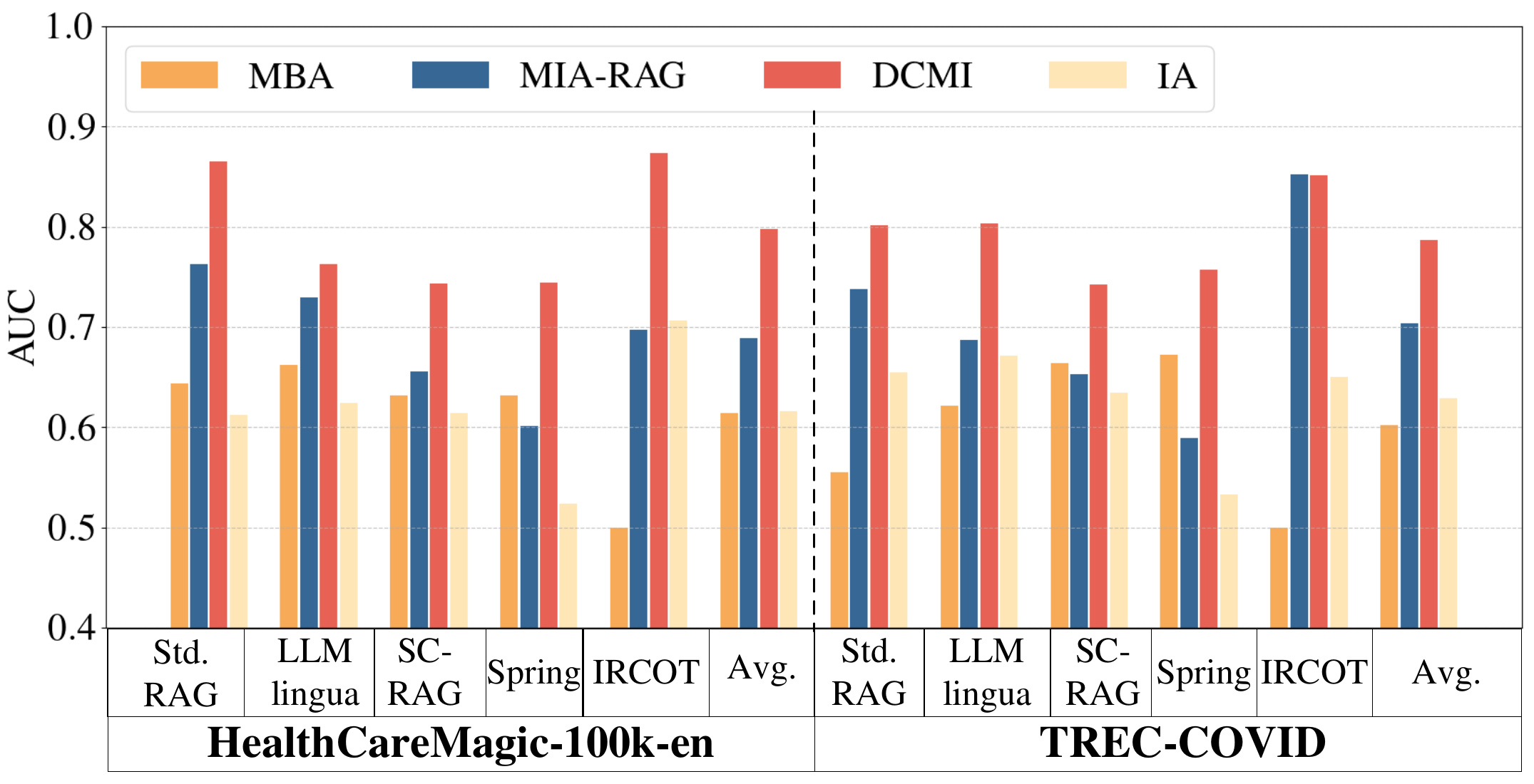}
  \caption{AUC of Adversary 2 on five RAG systems.}
  \label{fig:system_AUC_2}
\end{figure}

\begin{figure}[htbp]
  \centering
  \includegraphics[width=\linewidth]{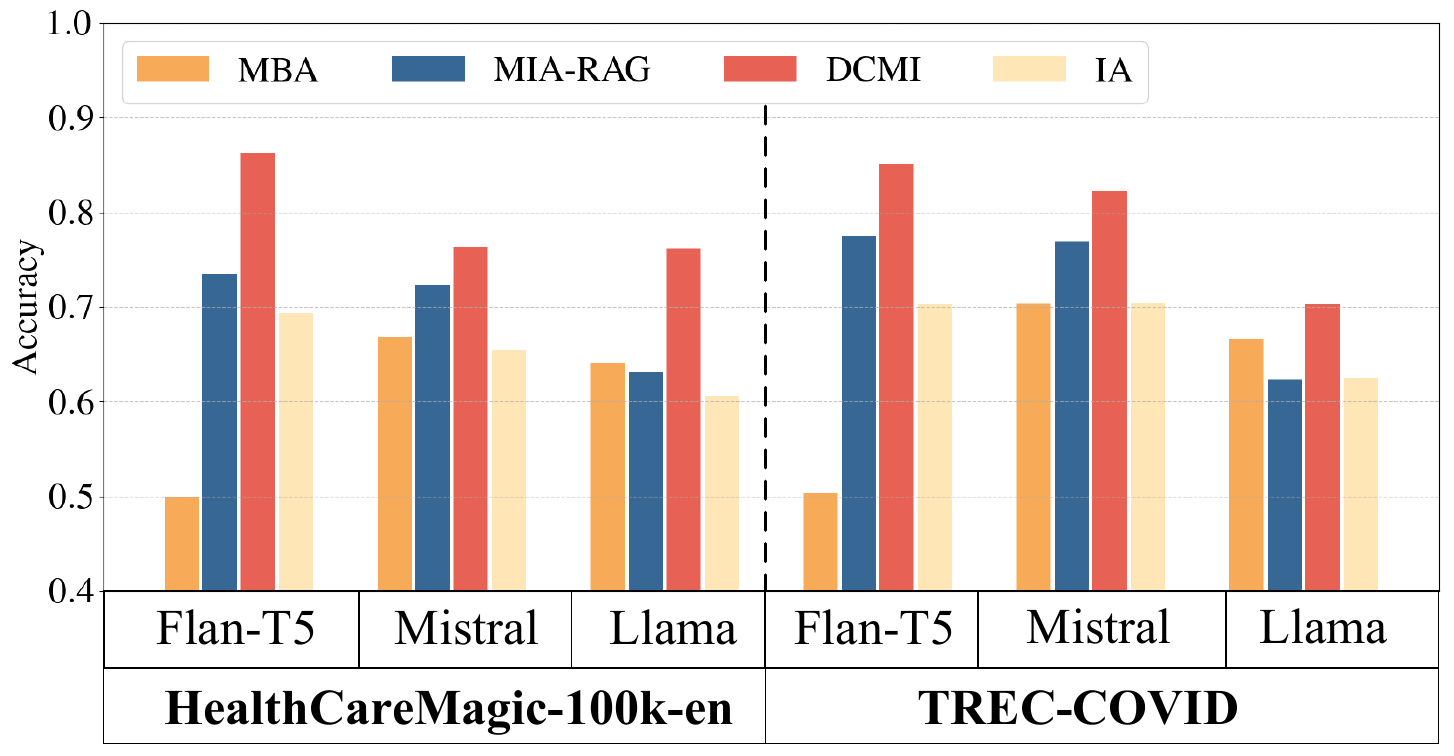}
  \caption{ Accuracy of Adversary 2 on three generative modules.}
  \label{fig:generative_Acc_2}
\end{figure}

\begin{figure}[htbp]
  \centering
  \includegraphics[width=\linewidth]{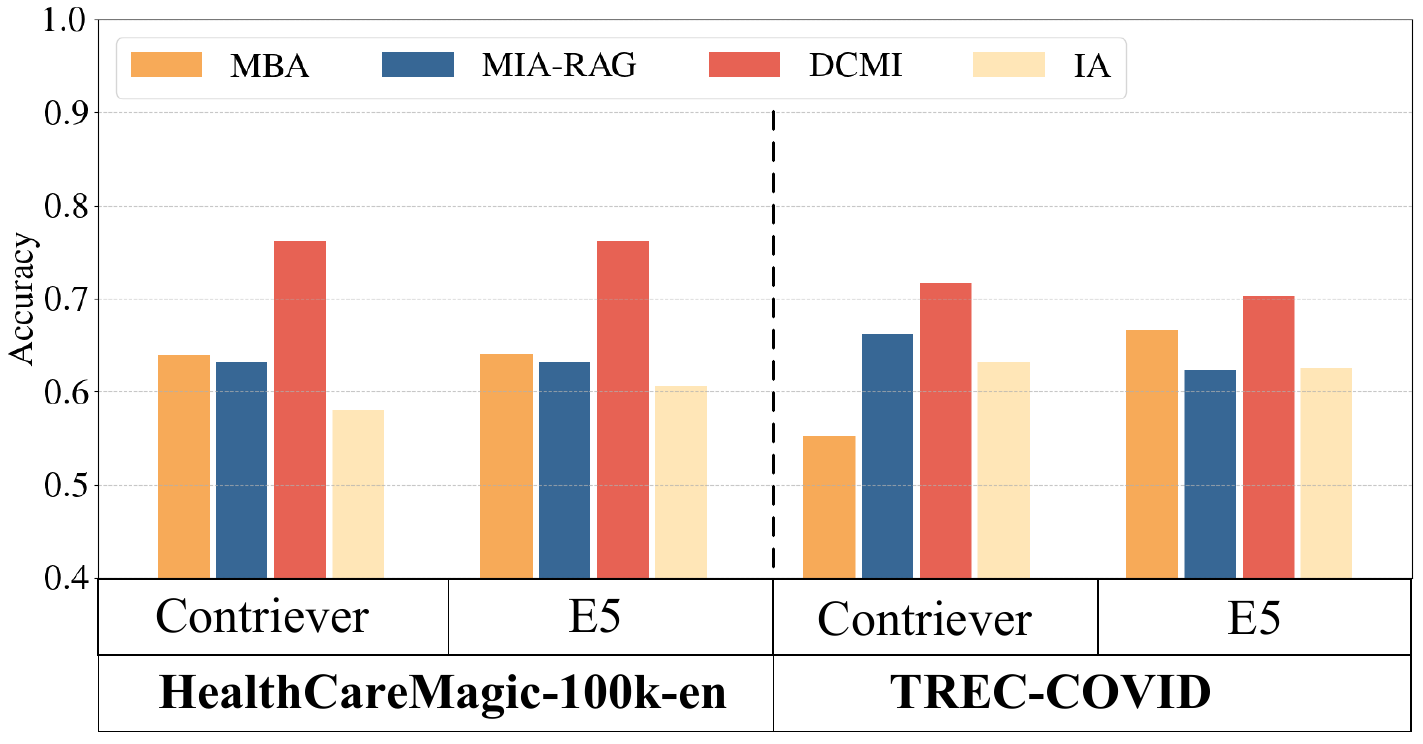}
  \caption{ Accuracy of Adversary 2 on two retrieval modules.}
  \label{fig:retriever_Acc_2}
\end{figure}

\begin{figure}[htbp]
  \centering
  \includegraphics[width=\linewidth]{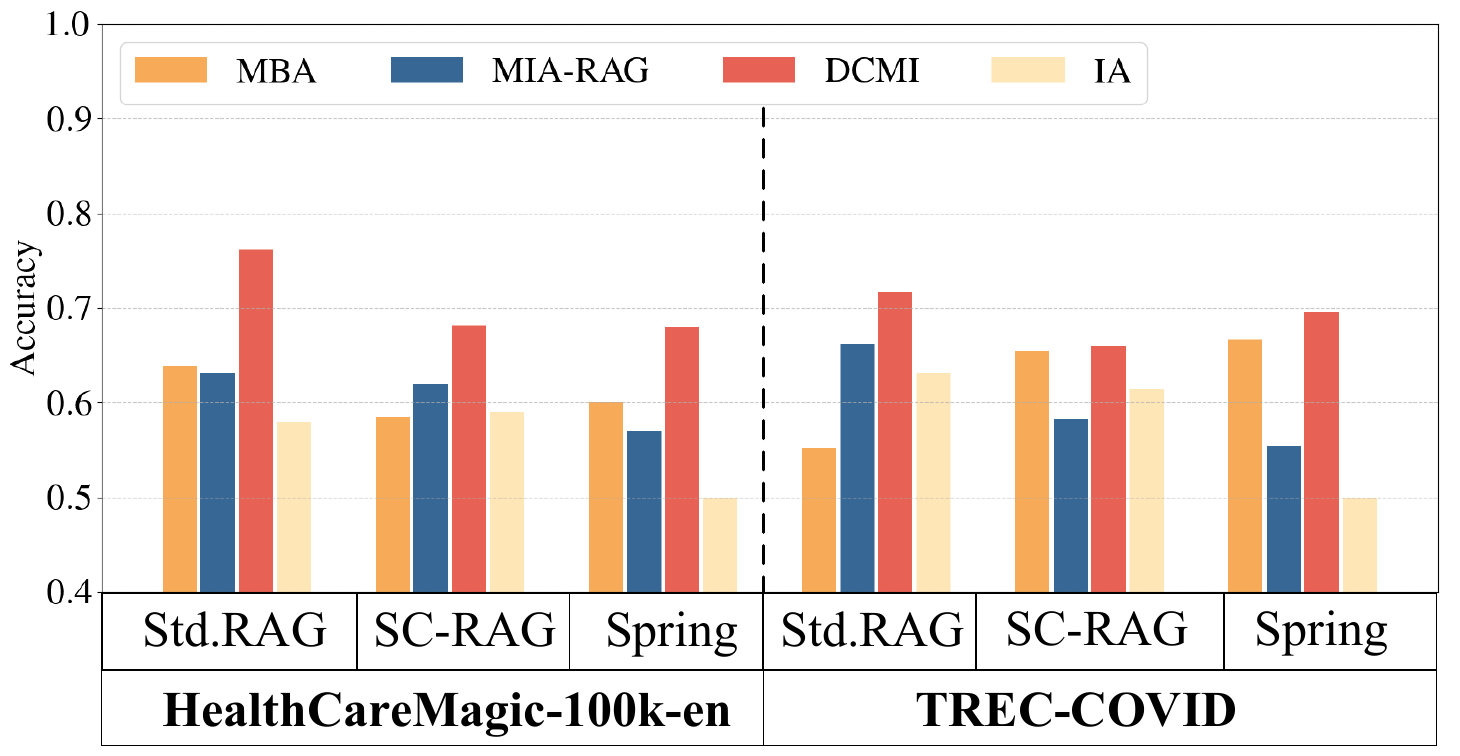}
  \caption{ Accuracy of Adversary 2 on three RAG frameworks.}
  \label{fig:frame_Acc_2}
\end{figure}

\begin{table}[htbp]
\centering
\caption{AUC of Adversary 3 on five RAG systems.}
\label{tab:perfomance_system_Adversary3_AUC}
\resizebox{\linewidth}{!}{
\begin{tabular}{lcccc}
\toprule
\multirow{2}{*}{AUC} & \multicolumn{2}{c}{HealthCareMagic-100k-en} & \multicolumn{2}{c}{TREC-COVID} \\
\cmidrule(lr){2-3} \cmidrule(lr){4-5}
                   & \attack      & RAG-MIA    & \attack      & RAG-MIA     \\
\midrule
Std. RAG       & \textbf{0.7541} & 0.5410 & \textbf{0.6595} & 0.5390 \\
LLMlingua          & \textbf{0.6568} & 0.6460 & \textbf{0.7118} & 0.6365 \\
SC-RAG  &  \textbf{0.6782} & 0.5375 & \textbf{0.6090} & 0.5660 \\
Spring             & \textbf{0.6379} & 0.5000 & \textbf{0.6480} & 0.5005 \\
IRCOT             & \textbf{0.7472} & 0.6350 & \textbf{0.8042} & 0.6015 \\
Avg.             & \textbf{0.6948} & 0.5719 & \textbf{0.6865} & 0.5687 \\
\bottomrule
\end{tabular}
}
\end{table}

\begin{figure}[htbp]
  \centering
  \includegraphics[width=\linewidth]{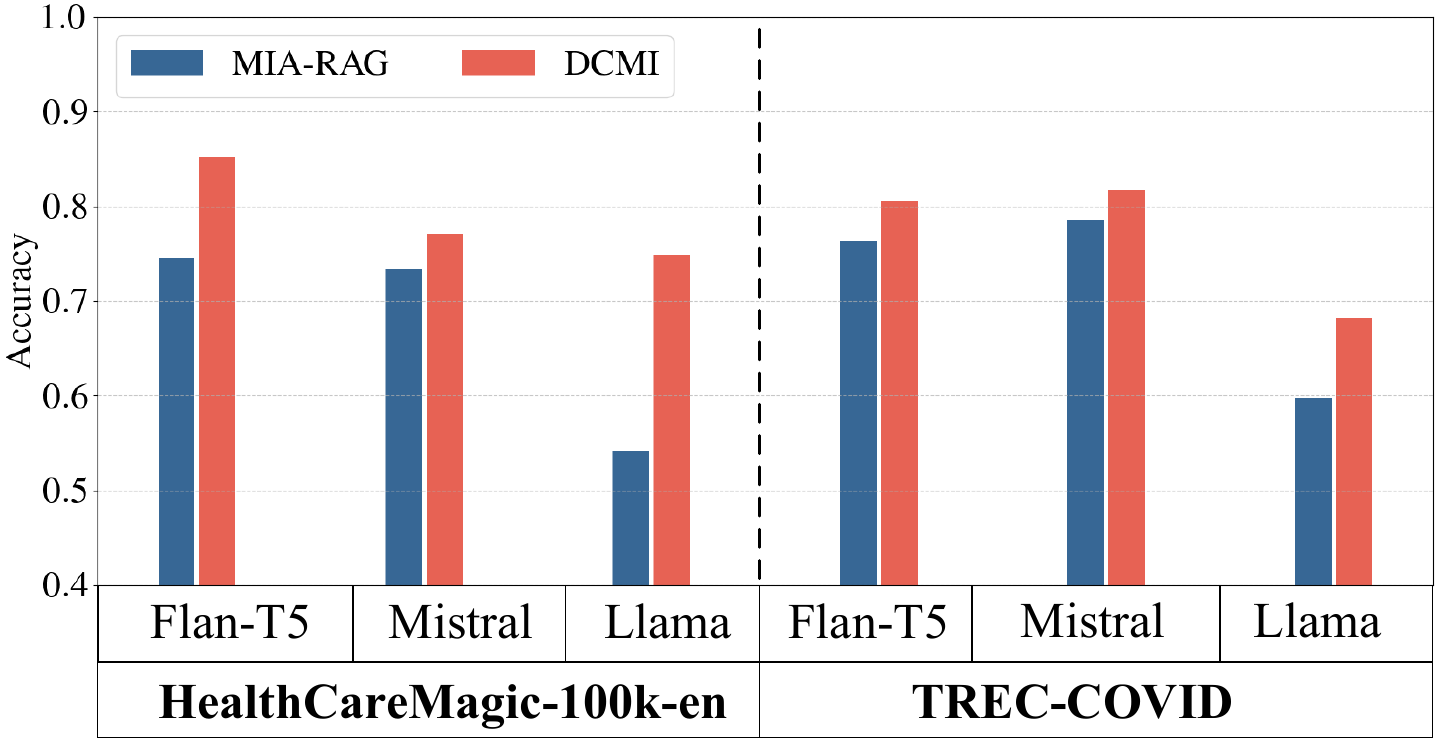}
  \caption{ Accuracy of Adversary 3 on three generative modules.}
  \label{fig:generative_Acc_3}
\end{figure}

\begin{figure}[t!]
  \centering
  \includegraphics[width=\linewidth]{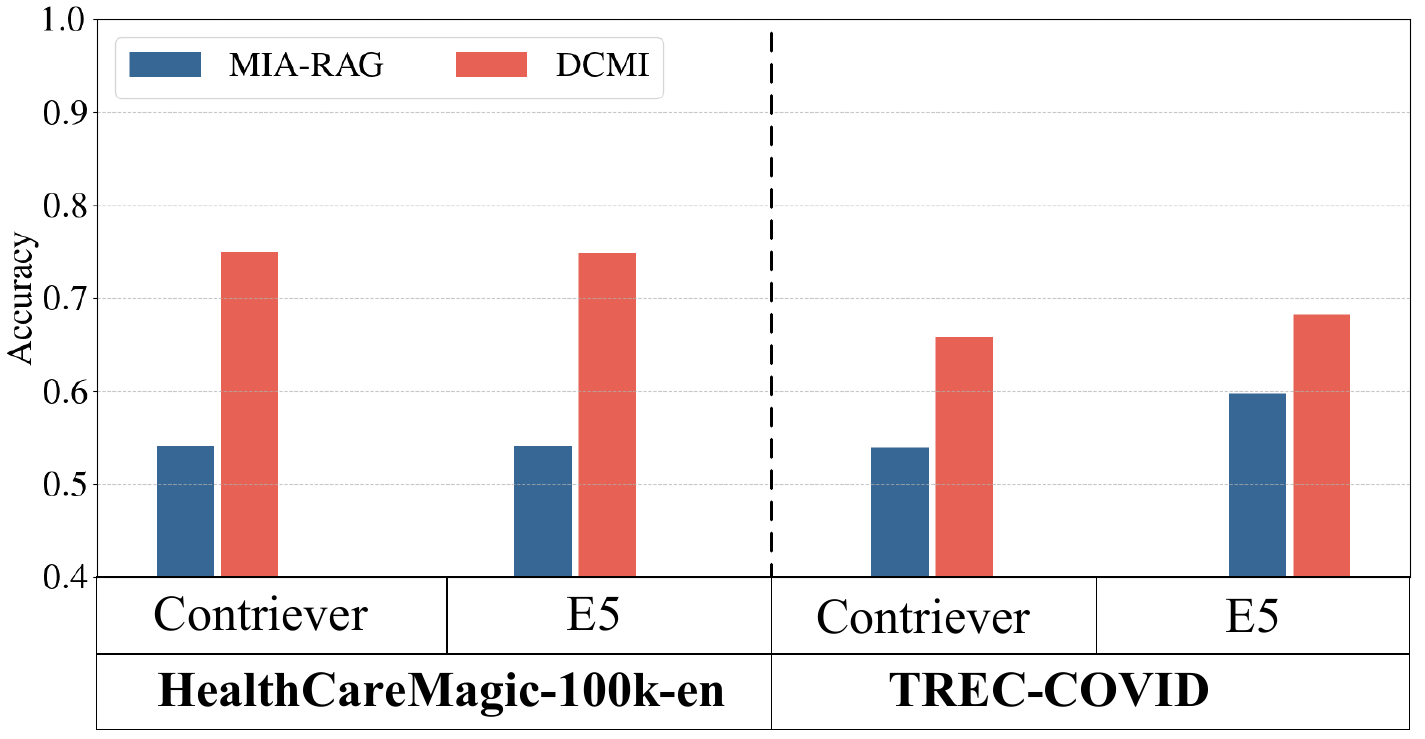}
  \caption{ Accuracy of Adversary 3 on two retrieval modules.}
  \label{fig:retriever_Acc_3}
\end{figure}

\begin{figure}[t!]
  \centering
  \includegraphics[width=\linewidth]{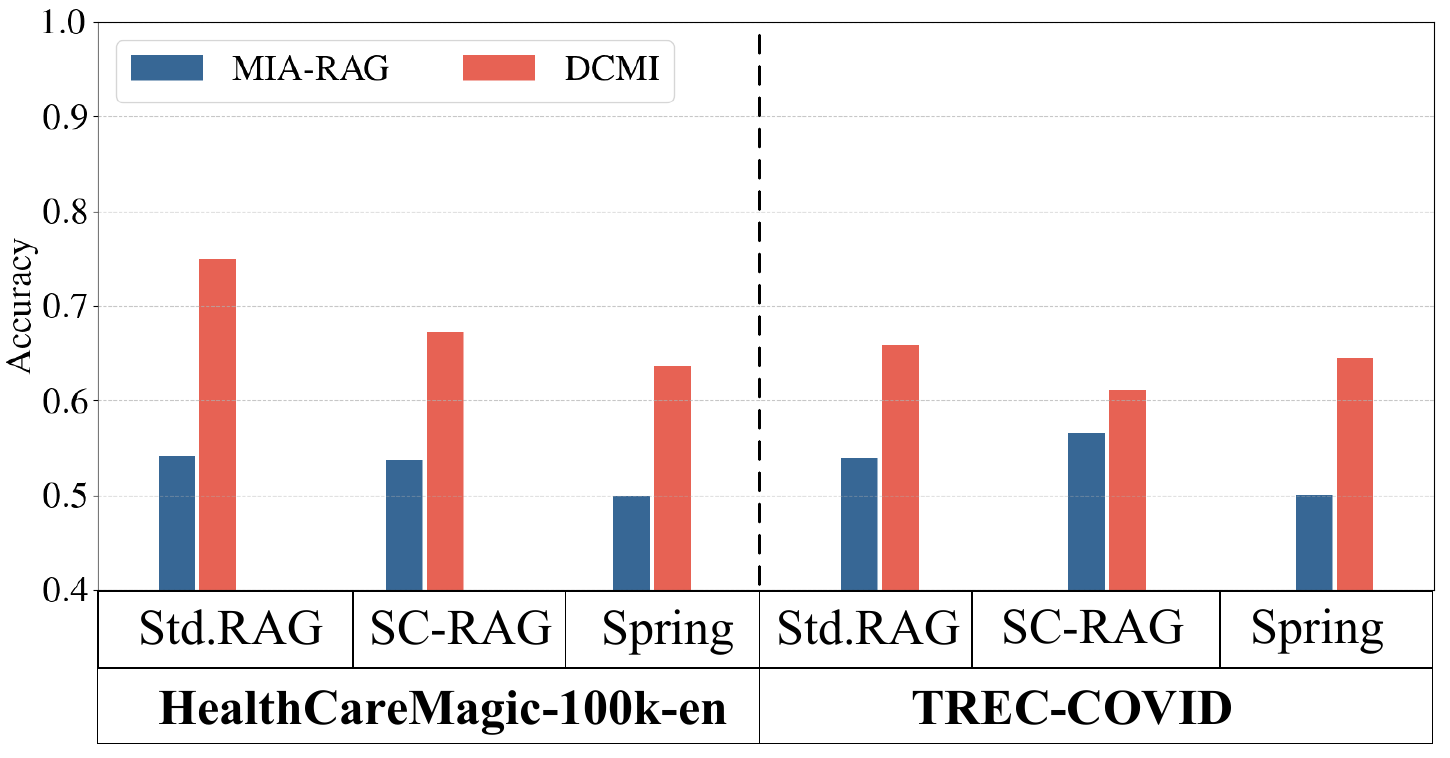}
  \caption{ Accuracy of Adversary 3 on three RAG frameworks.}
  \label{fig:frame_Acc_3}
\end{figure}

\begin{figure*}[htbp]
  \centering
  \begin{subfigure}{0.32\textwidth}
    \centering
    \includegraphics[width=\linewidth]{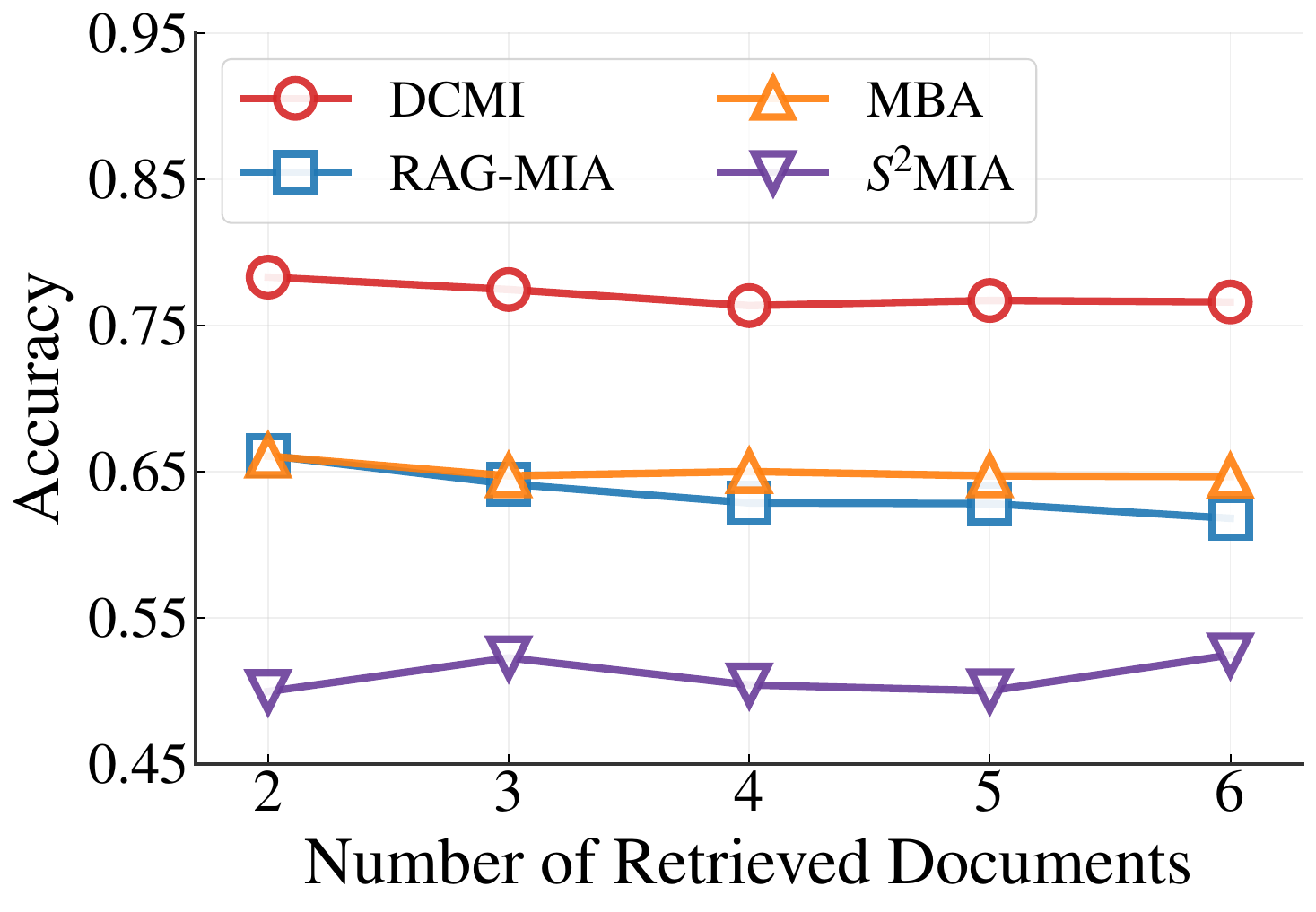}
    \caption{Accuracy of Adversary 1 on different numbers of retrieved documents.}
    \label{fig:num_Acc_1}
  \end{subfigure}
  \hfill
  \begin{subfigure}{0.32\textwidth}
    \centering
    \includegraphics[width=\linewidth]{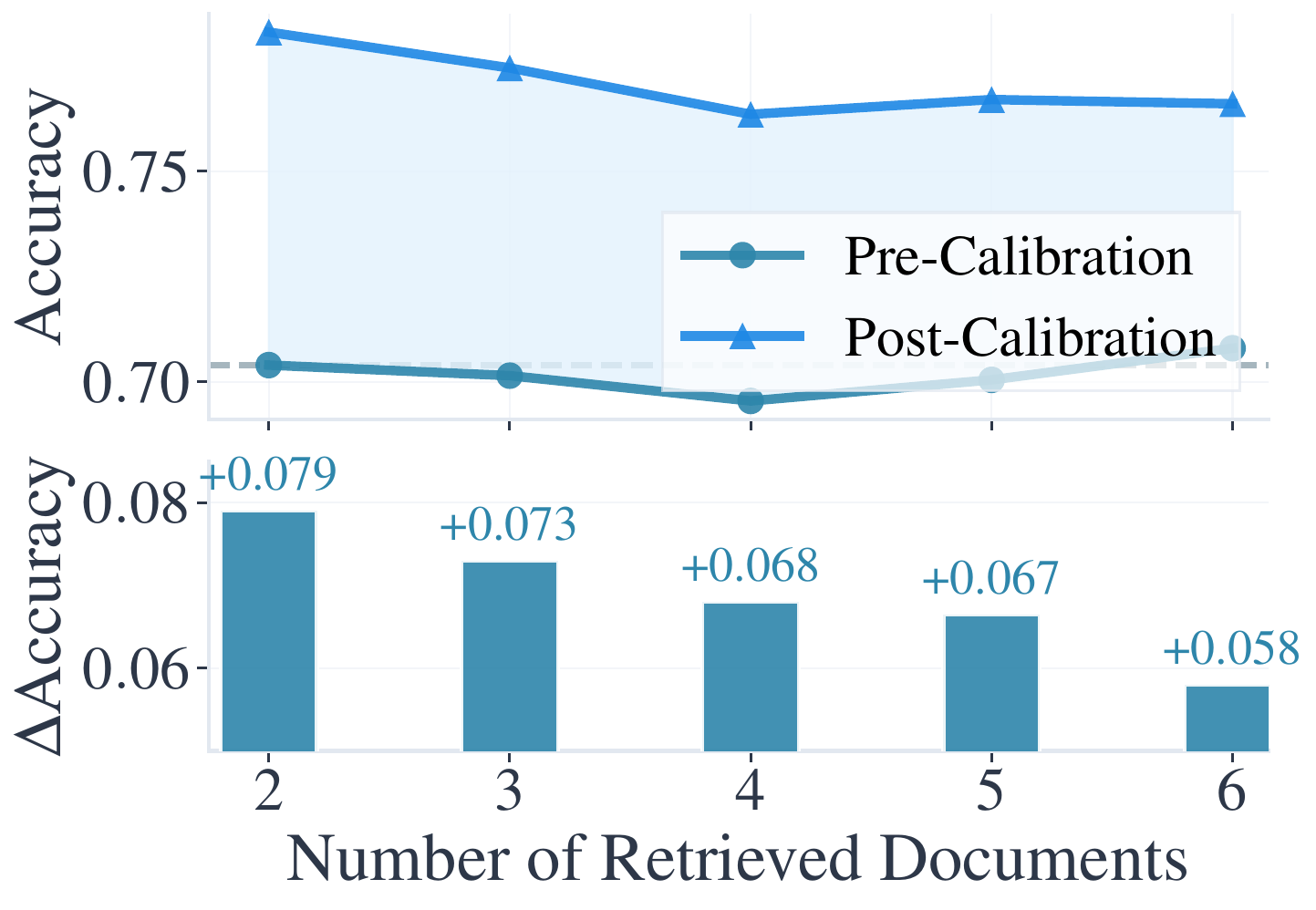}
    \caption{Impact of differential calibration module.}
    \label{fig:calibration_Acc_1}
  \end{subfigure}
  \hfill
  \begin{subfigure}{0.32\textwidth}
    \centering
    \includegraphics[width=\linewidth]{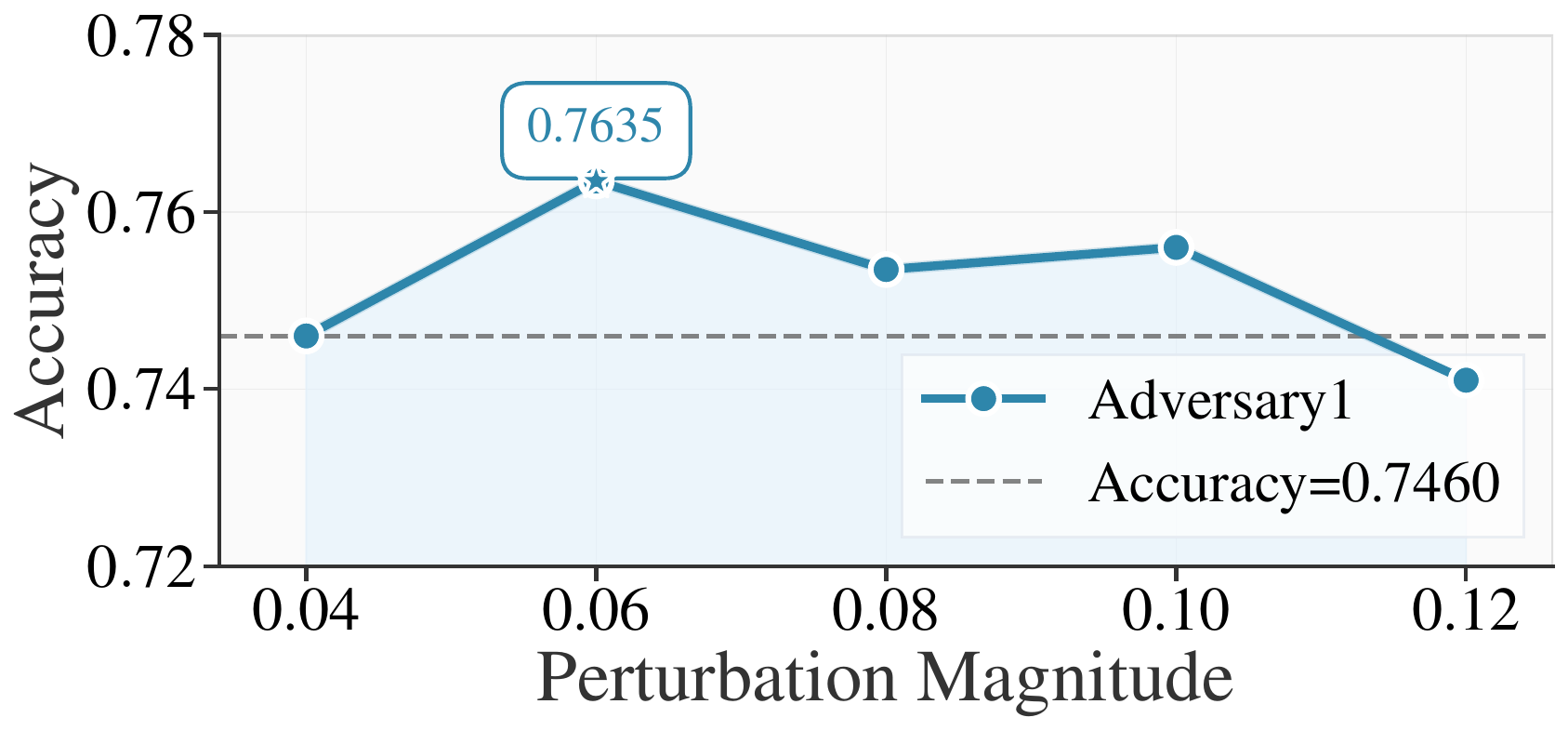}
    \caption{Impact of perturbation magnitude.}
    \label{fig:perturb_Acc_1}
  \end{subfigure}
  \caption{Performance analysis on Adversary 1.}

\end{figure*}

\begin{figure*}[htbp]
  \centering
  \begin{subfigure}{0.32\textwidth}
    \centering
    \includegraphics[width=\linewidth]{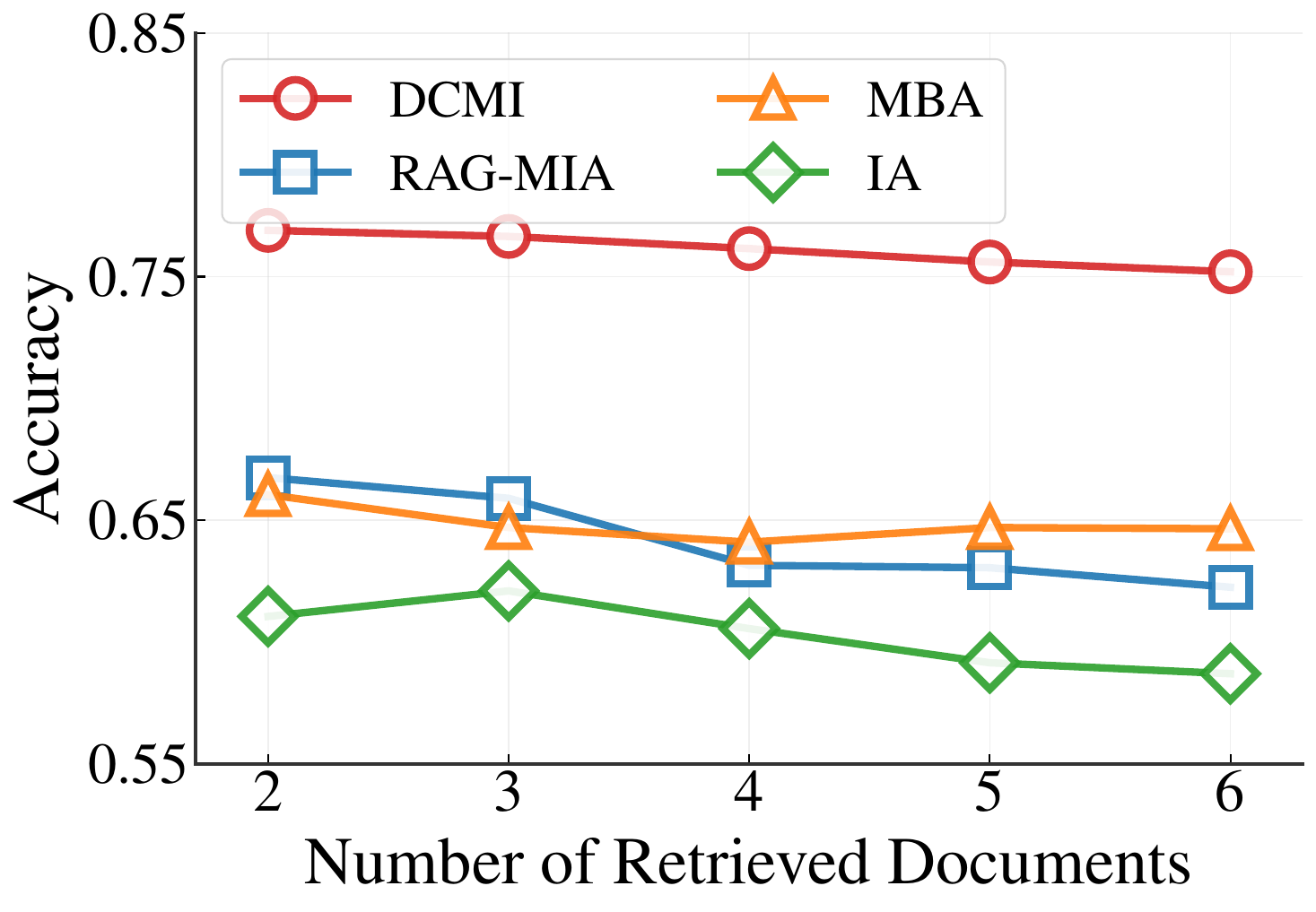}
    \caption{Accuracy of Adversary 2 on different numbers of retrieved documents.}
    \label{fig:Num_Acc_2}
  \end{subfigure}
  \hfill
  \begin{subfigure}{0.32\textwidth}
    \centering
    \includegraphics[width=\linewidth]{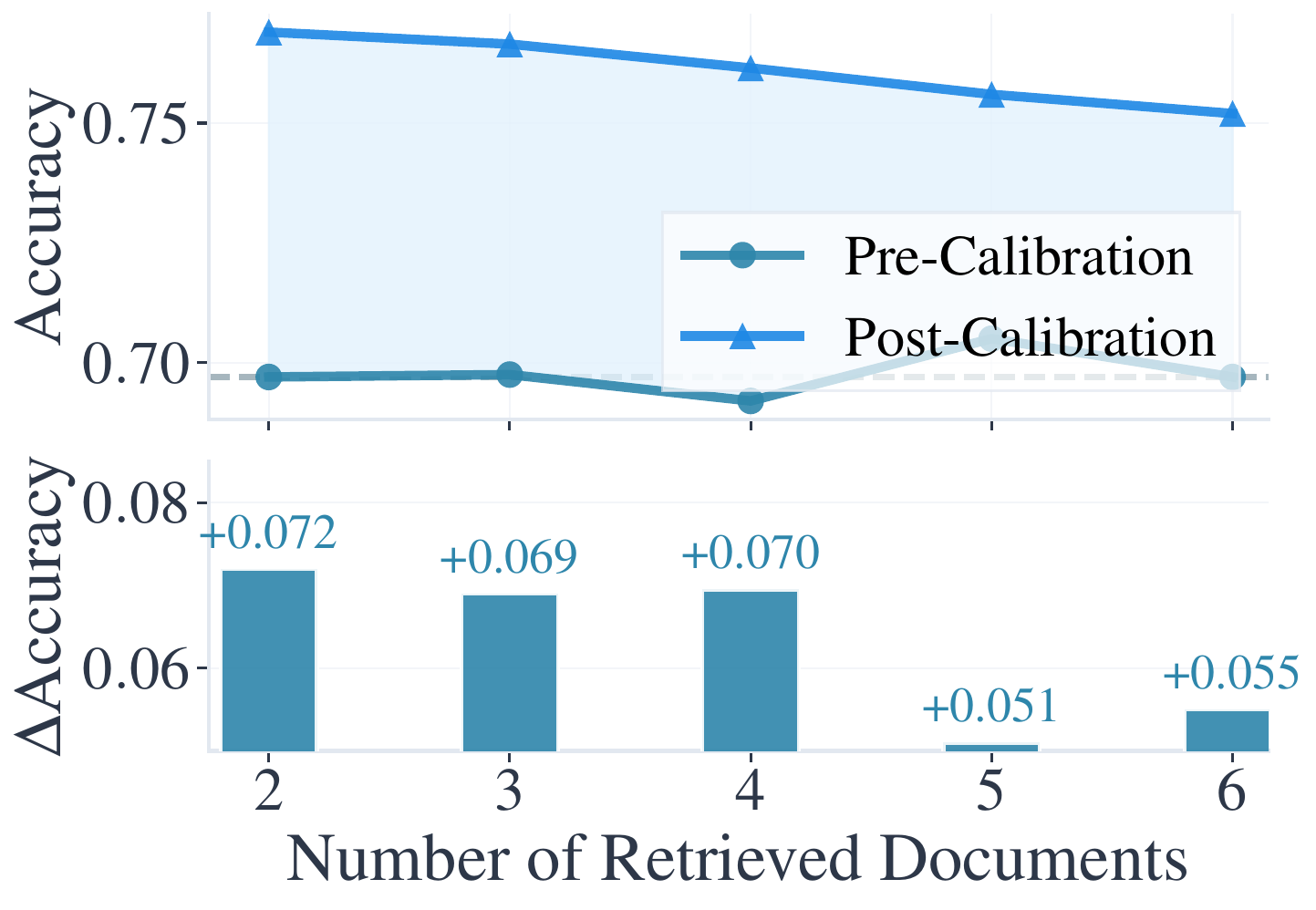}
    \caption{Impact of differential calibration module.}
    \label{fig:calibration_Acc_2}
  \end{subfigure}
  \hfill
  \begin{subfigure}{0.32\textwidth}
    \centering
    \includegraphics[width=\linewidth]{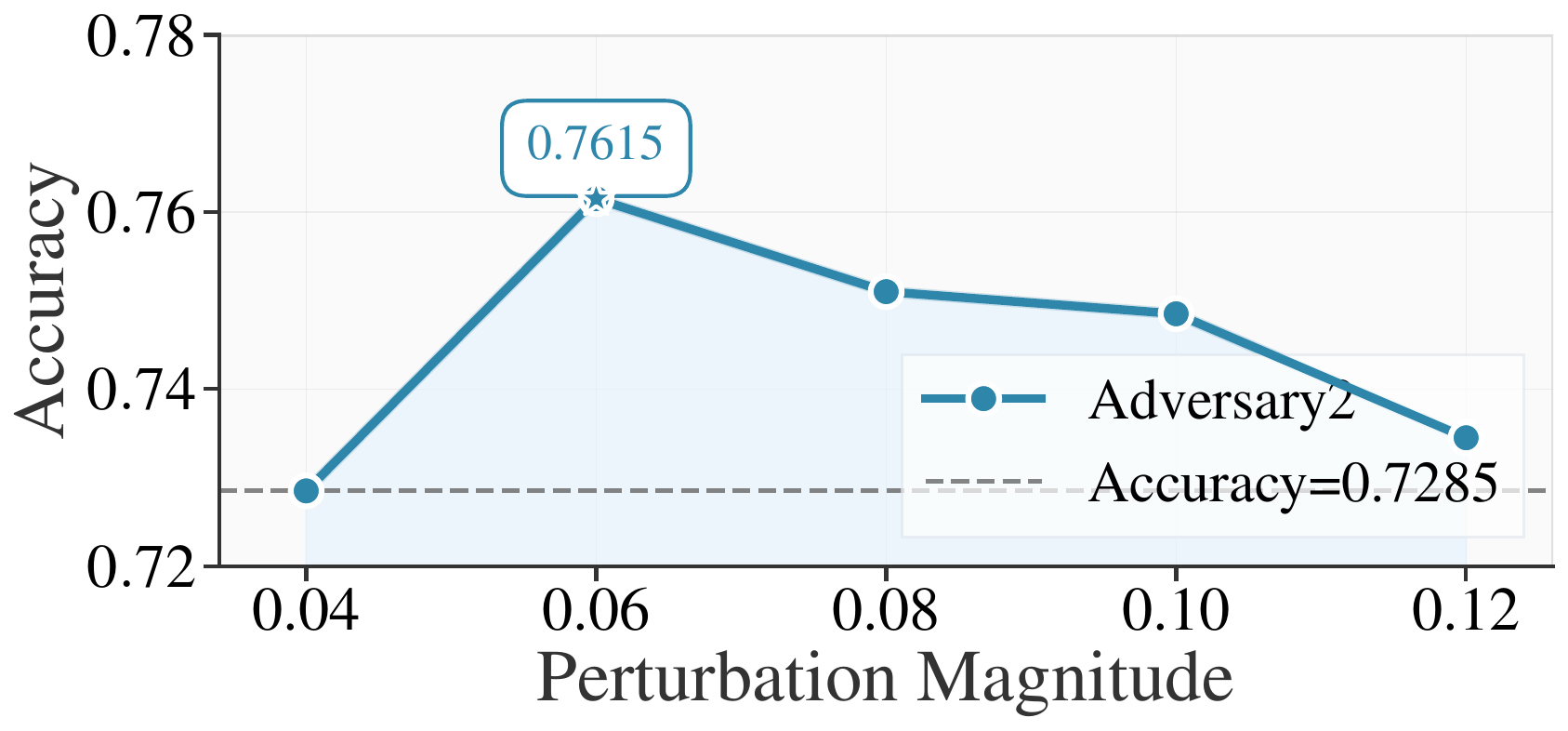}
    \caption{Impact of perturbation magnitude.}
    \label{fig:perturb_Acc_2}
  \end{subfigure}
  \caption{Performance analysis on Adversary 2.}
\end{figure*}

\begin{figure*}[htbp]
  \centering
  \begin{subfigure}{0.32\textwidth}
    \centering
    \includegraphics[width=\linewidth]{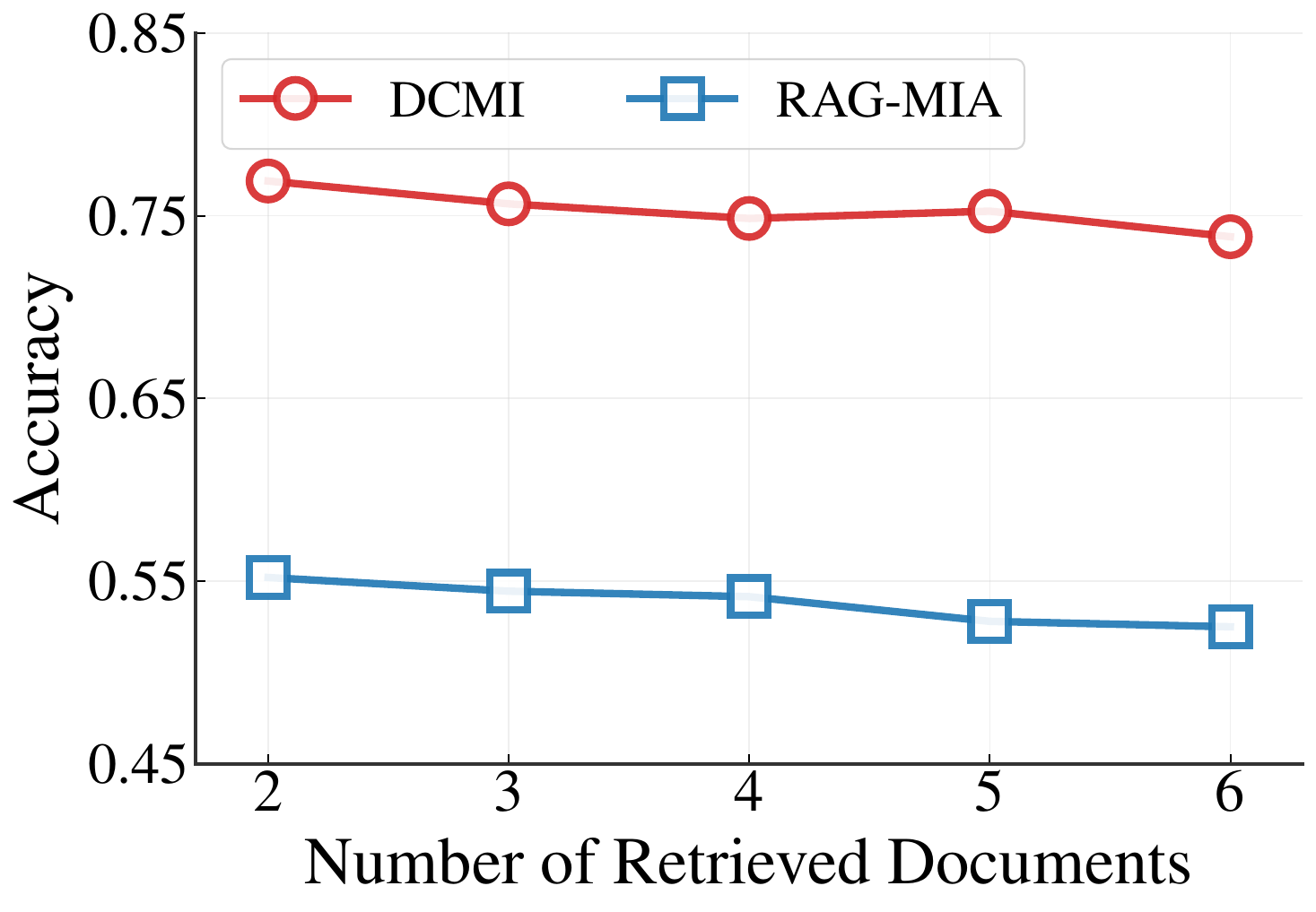}
    \caption{Accuracy of Adversary 3 on different numbers of retrieved documents.}
    \label{fig:num_Acc_3}
  \end{subfigure}
  \hfill
  \begin{subfigure}{0.32\textwidth}
    \centering
    \includegraphics[width=\linewidth]{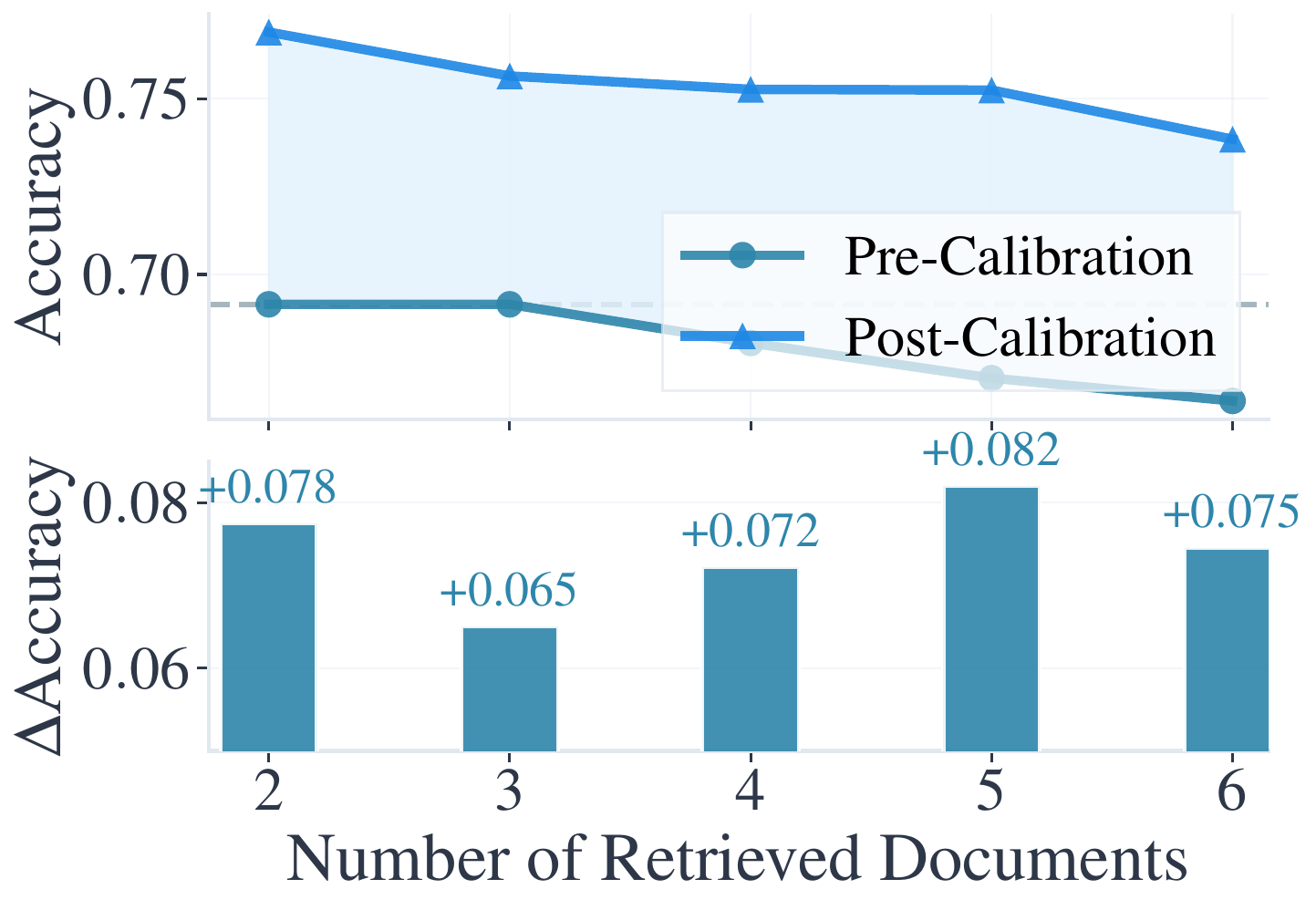}
    \caption{Impact of differential calibration module.}
    \label{fig:calibration_Acc_3}
  \end{subfigure}
  \hfill
  \begin{subfigure}{0.32\textwidth}
    \centering
    \includegraphics[width=\linewidth]{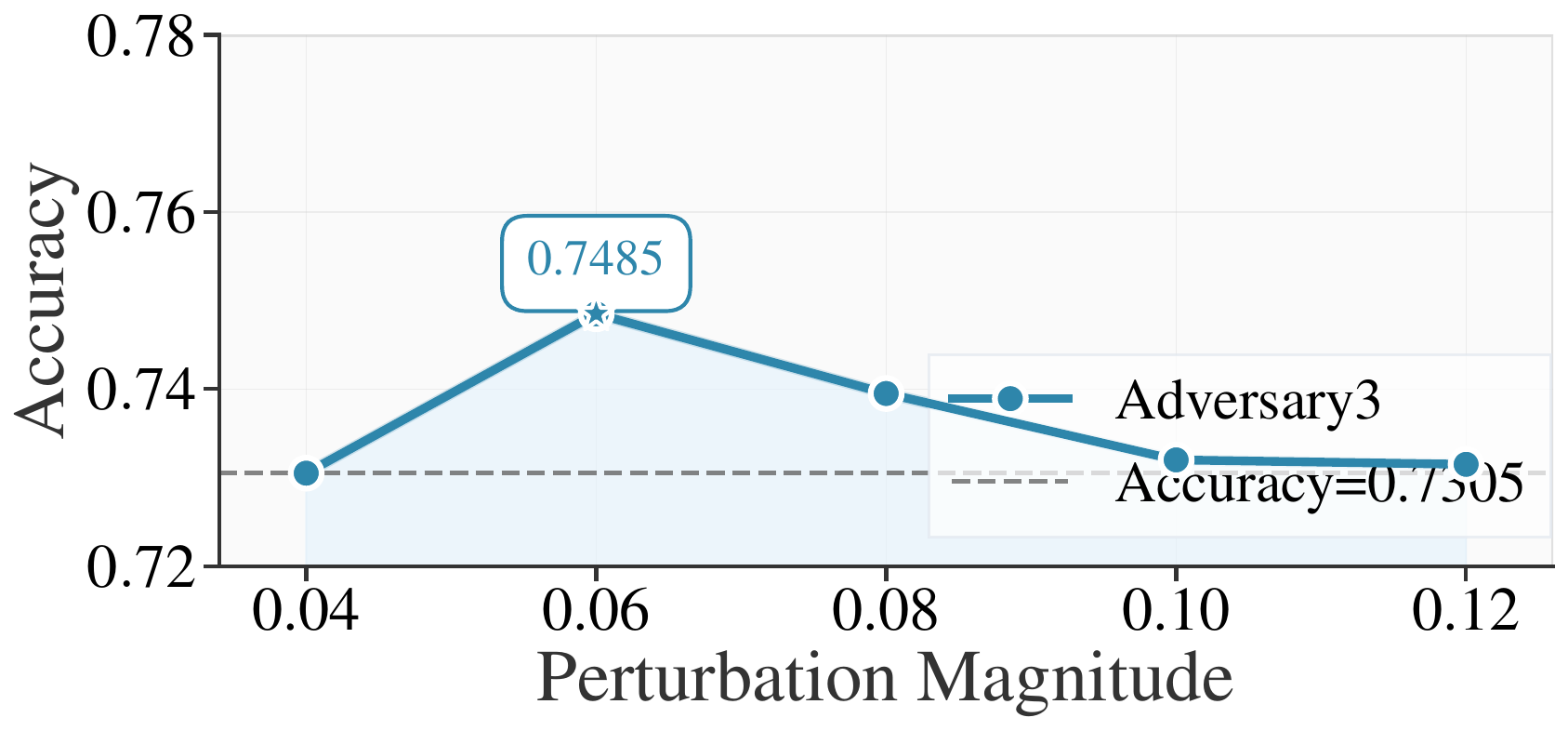}
    \caption{Impact of perturbation magnitude.}
    \label{fig:perturb_Acc_3}
  \end{subfigure}
  \caption{Performance analysis on Adversary 3.}
\end{figure*}

\newpage
\section{Potential Defenses}
\label{app:defense}
\begin{table}[htbp]
\centering
\caption{Performance of Accuracy metrics before and after defense under the Adversary 3 scenario.}
\label{tab:defense_model_ACC}
\resizebox{\linewidth}{!}{
\begin{tabular}{l c c }
\toprule
\textbf{Accuracy}               & \textbf{\attack}   & \textbf{RAG-MIA}    \\ 
\midrule
No Defense                    & 0.7785              & 0.5415                           \\ 
Instruction-Based Defense                      & 0.7020              & 0.5330    \\ 
Paraphrasing-Based Defense                      
& 0.5470 &0.5265   \\
Entity-relation Extraction  &0.5190  &0.5150\\

\bottomrule
\end{tabular}
}
\end{table}

\newpage
\section{Extended Ablation Study: Differential Calibration Module}
\label{app:Extended_Ablation_Study}
\autoref{sec:Ablation_adversary1}, \autoref{sec:Ablation_adversary2} and \autoref{sec:Ablation_adversary3} validated the differential calibration module within \textit{Basic RAG Setting}. 
This appendix extends the ablation study to evaluate the module's performance under various configurations.

\subsection{Different RAG Systems}
We conduct experiments using different RAG systems. The configurations for the generative module, retrieval module, and RAG framework follow the specifications in \autoref{tab:victim_model}, \autoref{app:rag_systems}.
The AUC results are shown in \autoref{fig:dumbbell_system_1} and \autoref{fig:dumbbell_system_3}.

\subsection{Different Generative Modules}
We evaluate the module with LLMlingua as the RAG framework and E5-base as the retrieval module. 
The evaluation tests three specific generative modules: GPT-3.5-turbo, Llama2-7B-chat, and Flan-T5-large. The AUC results are shown in \autoref{fig:dumbbell_generative_1} and \autoref{fig:dumbbell_generative_3}.

\subsection{Different Retrieval Modules}
\begin{figure*}[htbp]
  \centering
  \begin{subfigure}{0.32\textwidth}
    \centering
    \includegraphics[width=\linewidth]{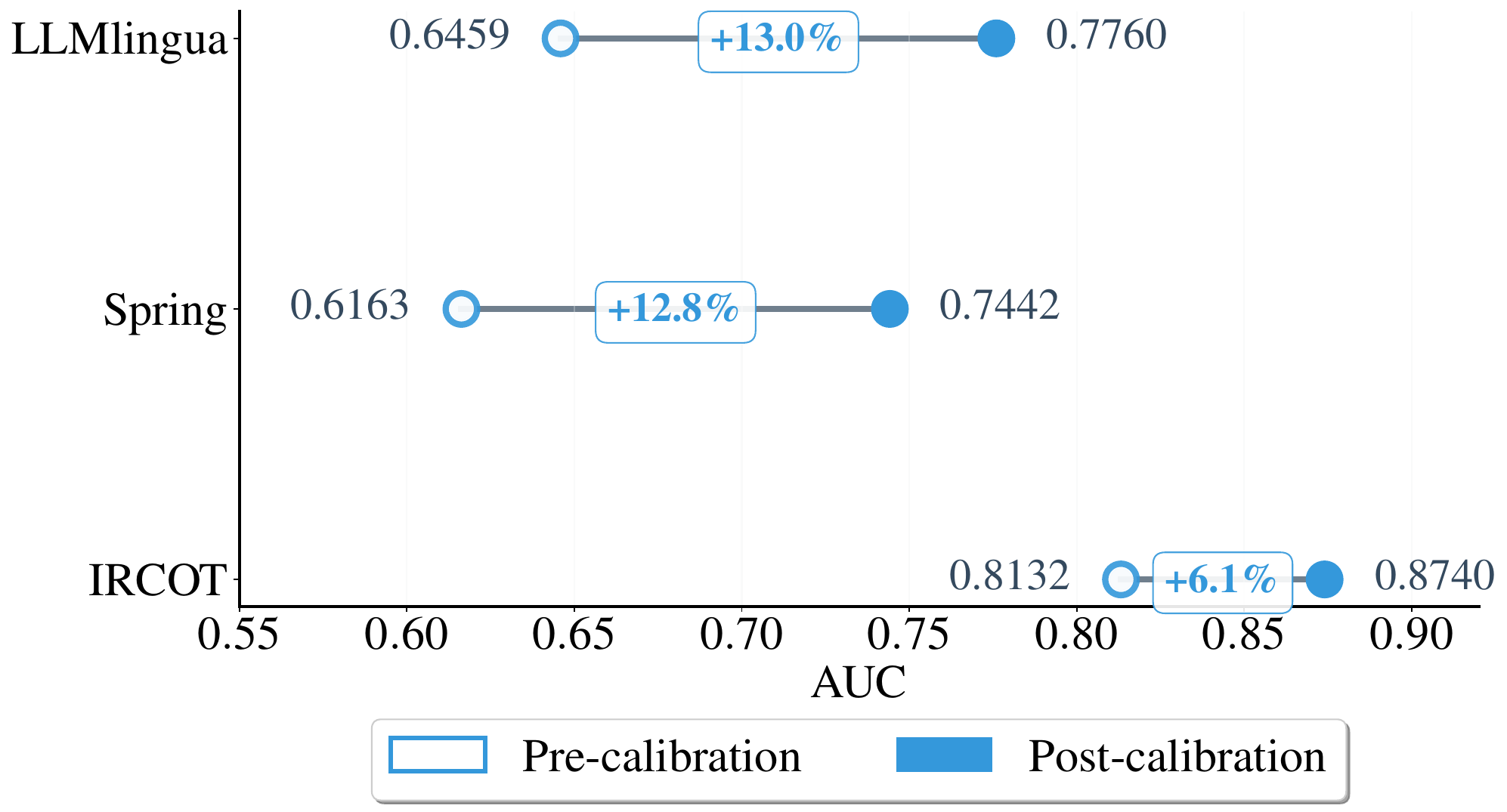}
    \caption{AUC of Adversary 1 \& 2 on three RAG systems.}
    \label{fig:dumbbell_system_1}
  \end{subfigure}
  \hfill
  \begin{subfigure}{0.32\textwidth}
    \centering
    \includegraphics[width=\linewidth]{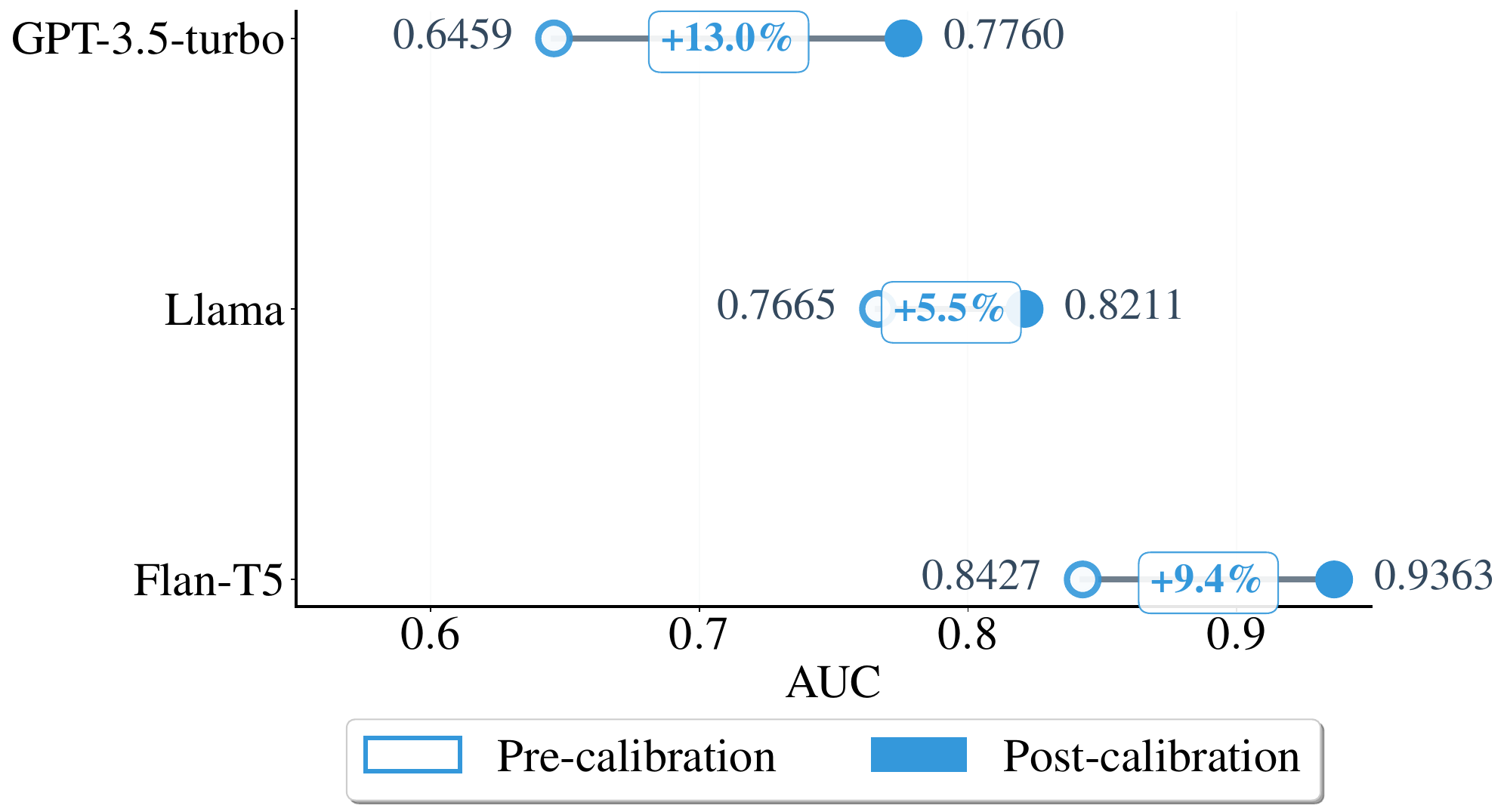}
    \caption{AUC of Adversary 1 \& 2 on three generative modules.}
    \label{fig:dumbbell_generative_1}
  \end{subfigure}
  \hfill
  \begin{subfigure}{0.32\textwidth}
    \centering
    \includegraphics[width=\linewidth]{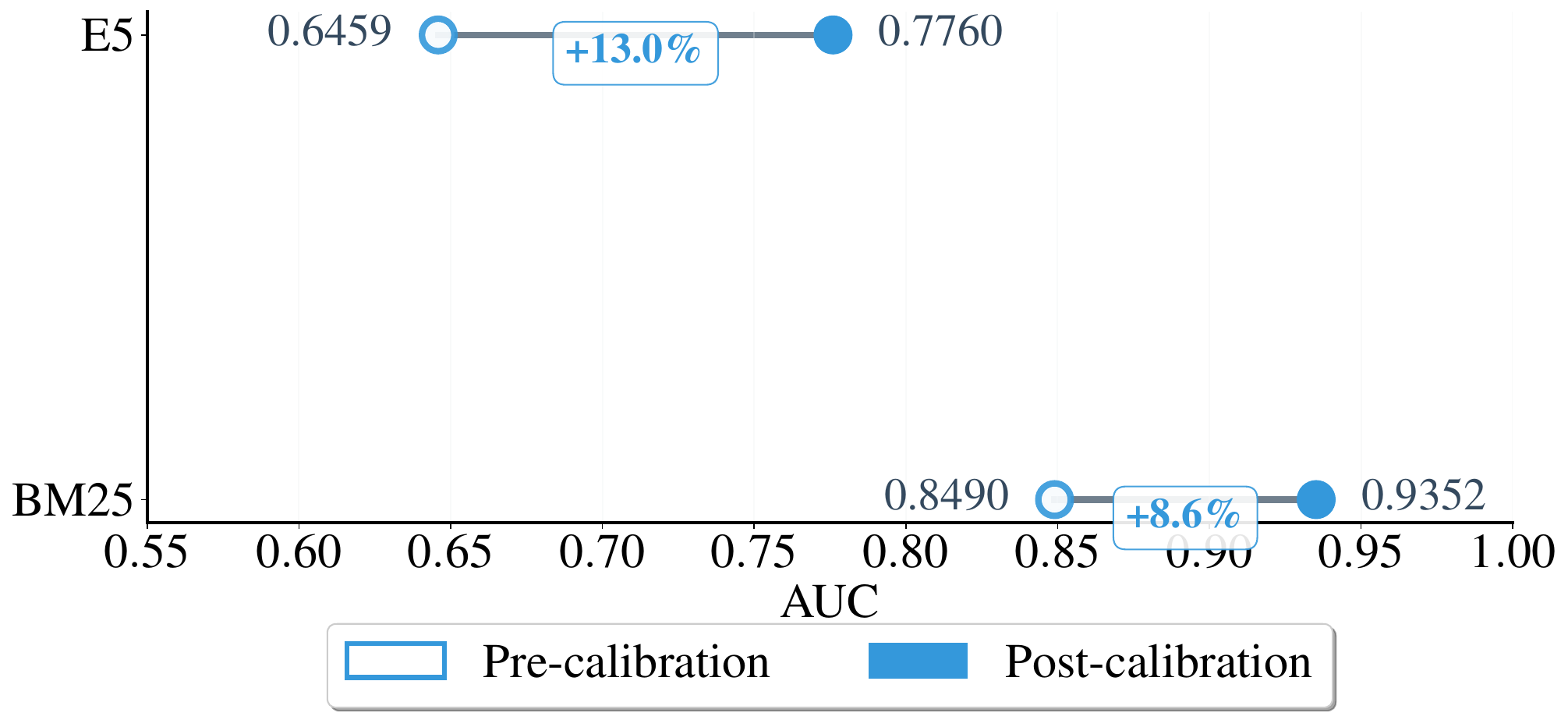}
    \caption{AUC of Adversary 1 \& 2 on two retrieval modules.}
    \label{fig:dumbbell_retriever_1}
  \end{subfigure}
  
  \vspace{1em}

  \begin{subfigure}{0.32\textwidth}
    \centering
    \includegraphics[width=\linewidth]{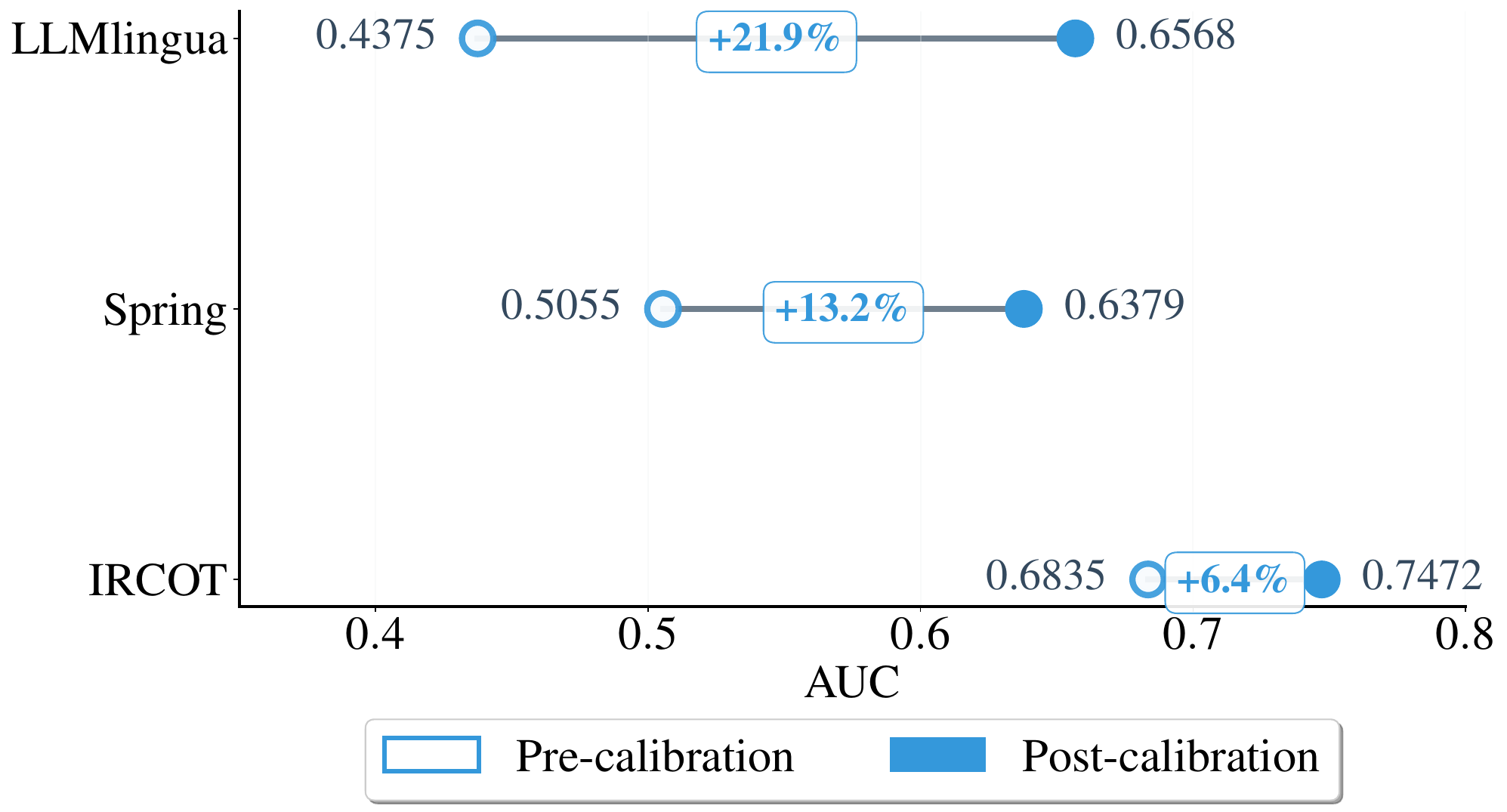}
    \caption{AUC of Adversary 3 on three RAG systems.}
    \label{fig:dumbbell_system_3}
  \end{subfigure}
  \hfill
  \begin{subfigure}{0.32\textwidth}
    \centering
    \includegraphics[width=\linewidth]{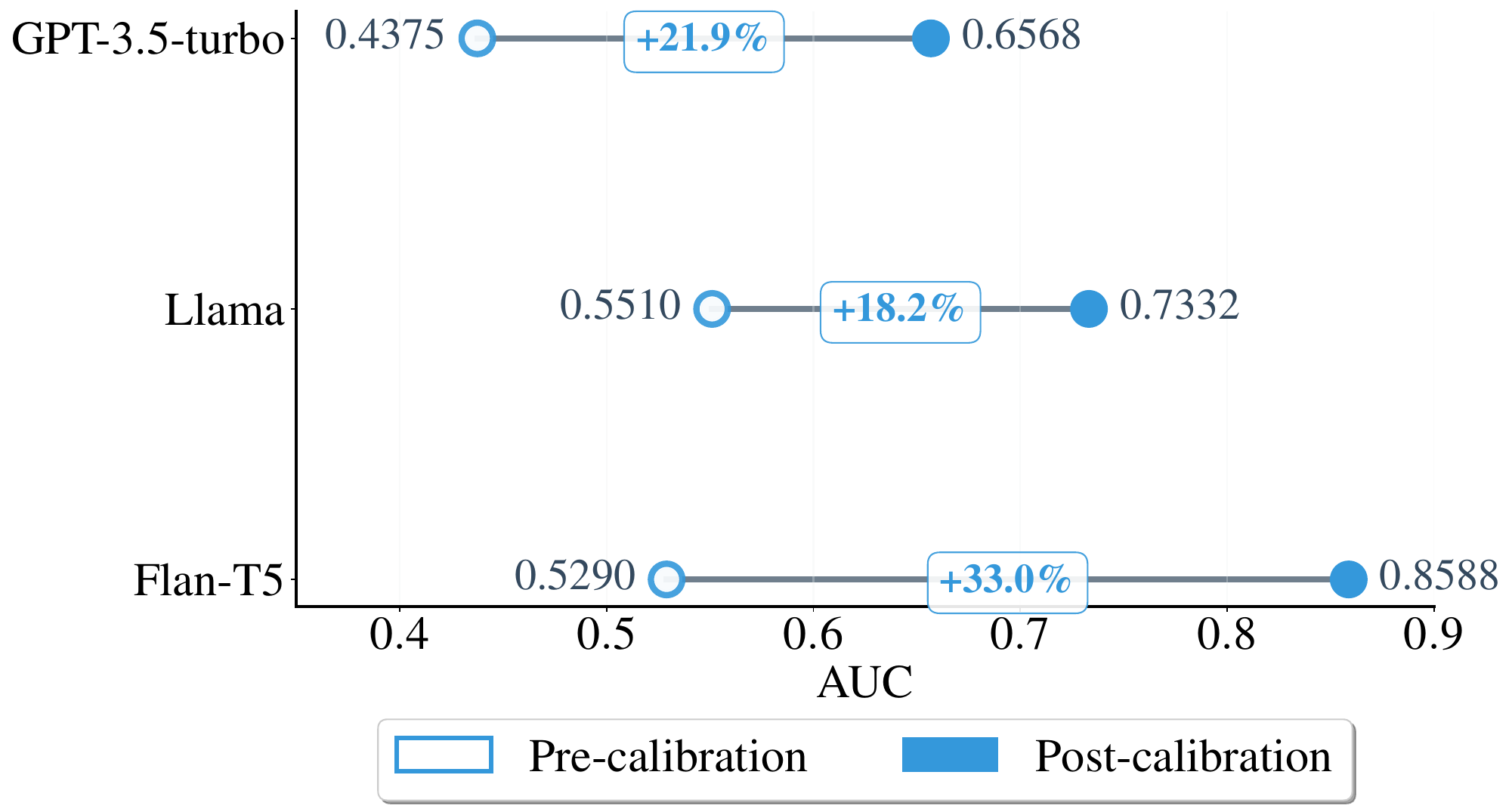}
    \caption{AUC of Adversary 3 on three generative modules.}
    \label{fig:dumbbell_generative_3}
  \end{subfigure}
  \hfill
  \begin{subfigure}{0.32\textwidth}
    \centering
    \includegraphics[width=\linewidth]{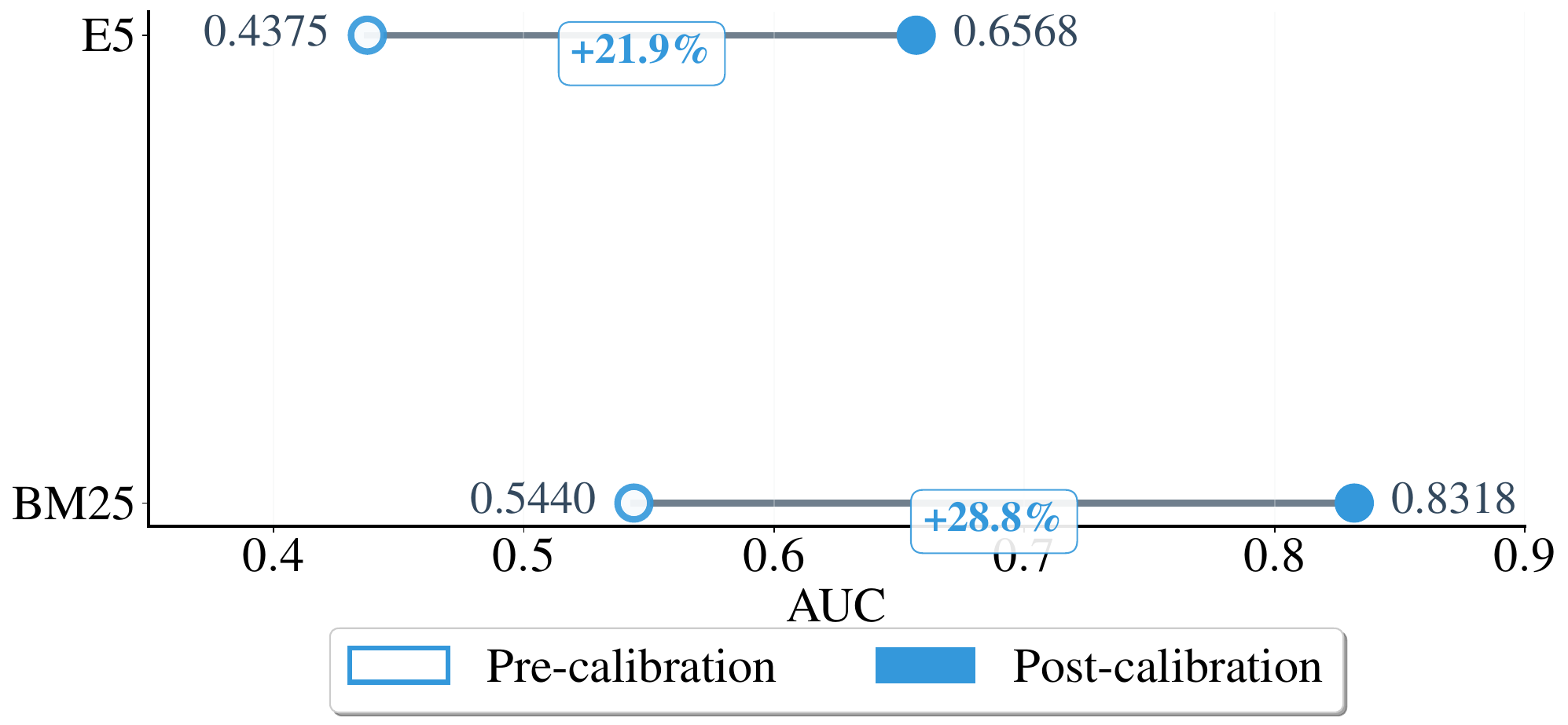}
    \caption{AUC of Adversary 3 on two retrieval modules.}
    \label{fig:dumbbell_retriever_3}
  \end{subfigure}
  
  \caption{Extended ablation study across different RAG components. The top row shows Adversary 1 \& 2, the bottom row shows Adversary 3.}
  \label{fig:dumbbell_all_comparison}
\end{figure*}

We conduct experiments with LLMlingua as the RAG framework and Llama as the generative module. 
The evaluation tests two specific retrieval modules: E5-base and BM25. The AUC results are shown in \autoref{fig:dumbbell_retriever_1} and \autoref{fig:dumbbell_retriever_3}.

\newpage
\section{Generalization Experiments }
\label{app:Generalization Experiments}
\subsection{Generalization to non-textual}
\label{app:non_textual_adaptation}

As described in \autoref{sec:analysis_long_tail}, interference from \nomember is a universal challenge in RAG systems, regardless of database content type. 
Therefore, \attack can theoretically calibrate this interference for any data type, whether textual or non-textual.

We tested this theory using genomic datasets as an example. 
For the genomic-benchmarks dataset, we found that minimal perturbations suffice—flipping a single locus strand identifier (from - to +). 
With this simple modification, \attack achieves 95.5\% AUC compared to RAG-MIA-black's 50\%, demonstrating its effectiveness on structurally different genomic data.

Based on our analysis of genomic data, we deduce that \attack can be adapted to different types of non-textual data, including images and video, by adjusting perturbation strategies and magnitudes according to the specific characteristics of each data type. 
We will explore these multi-modal RAG systems in future work.

\subsection{Prompt Robustness Evaluation}
\label{app:Prompt_Robustness}
\begin{figure}[t!]
  \centering
  \begin{subfigure}[t]{0.47\linewidth}
    \centering
    \includegraphics[width=\linewidth]{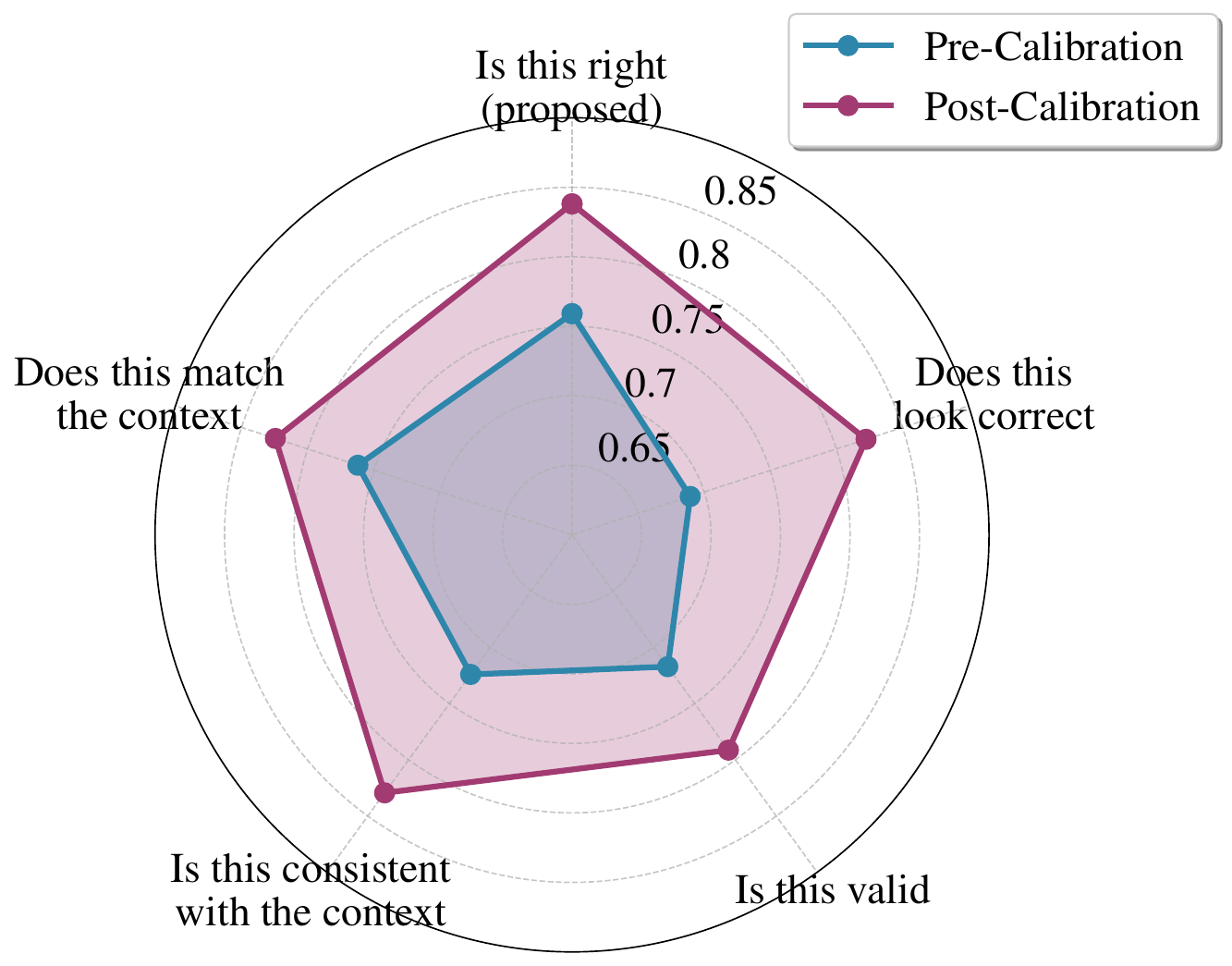}
    \caption{AUC of Adversary 1 \& 2 Pre- and Post-calibration.}
    \label{fig:rader_1}
  \end{subfigure}
  \hfill
  \begin{subfigure}[t]{0.47\linewidth}
    \centering
    \includegraphics[width=\linewidth]{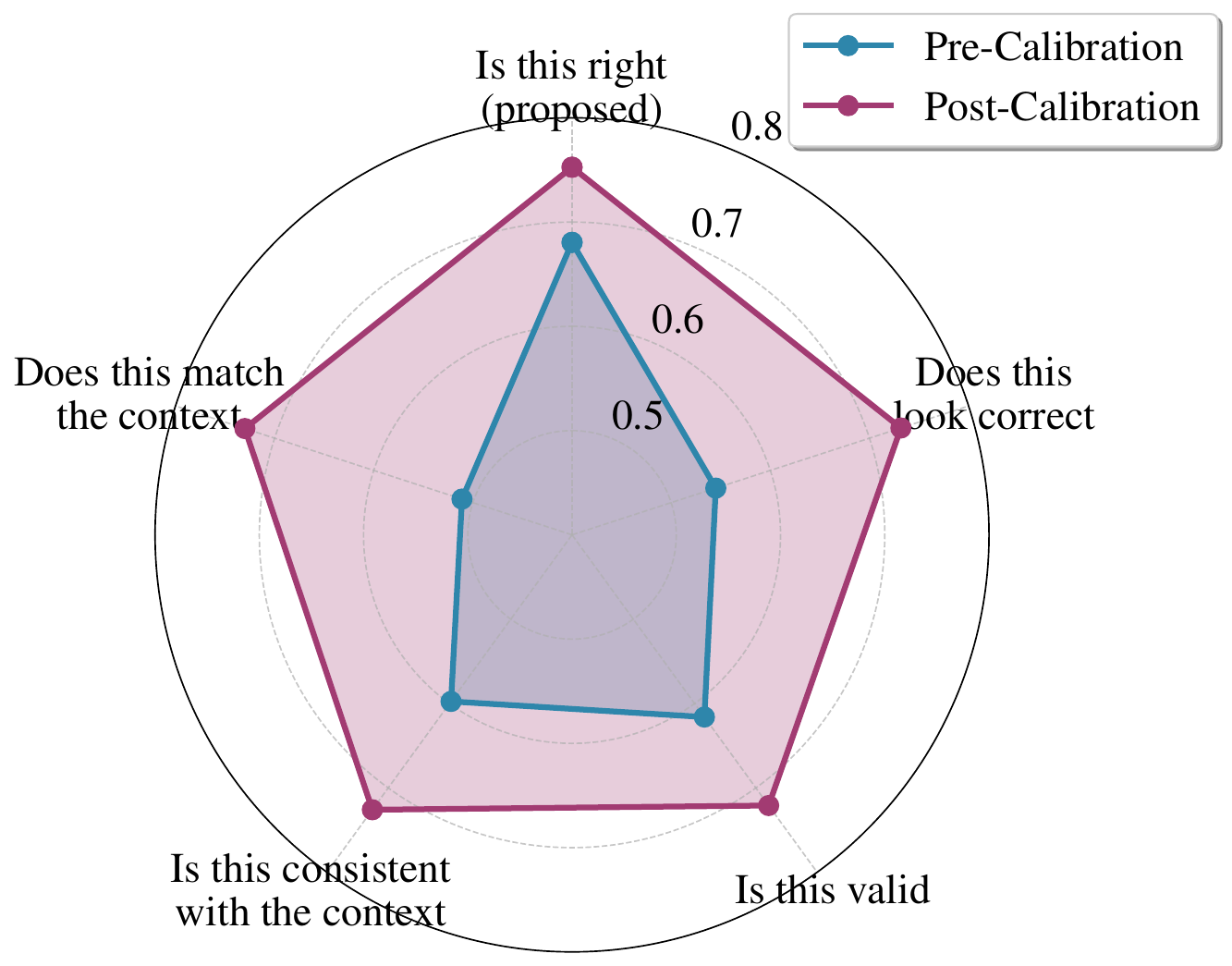}
    \caption{AUC of Adversary 3 Pre- and Post-calibration.}
    \label{fig:rader_3}
  \end{subfigure}
  \caption{Prompt robustness evaluation on adversaries before and after calibration.}
\end{figure}

In all three scenarios, we adopt a simple yet effective query template: ``Is this: `Target Sample' right? Answer with Yes or No.''
We also test various alternative prompt formulations as shown in \autoref{fig:rader_1} and \autoref{fig:rader_3}, \autoref{app:Prompt_Robustness} and observe two key findings:

First, different query formulations yield similar attack performance after calibration, confirming \attack's effectiveness is independent of prompt format.

Second, and more importantly, we observe substantial performance improvements between pre- and post-calibration results across all tested prompts.  
This universal improvement stems from the fact that \nomember interference exists across all prompt variations (\autoref{sec:analysis_long_tail}). Since this interference is prompt-agnostic, \attack can effectively calibrate it regardless of query formulation, ensuring consistent performance gains. These results demonstrate that \attack's core contribution is providing a universal calibration mechanism that enhances any prompt-based MIAs on RAG systems, rather than relying on specific prompt optimization.

\section{Scalability \& computational overhead}
\attack has $O(n)$ linear time complexity and supports parallel processing for large-scale deployment. Adversary 3 adds only \$0.00372 and 4.93 seconds compared to MIA-RAG-black (the best) while improving AUC by over 10\%, enabling cost-effective deployment with controlled overhead.

\begin{table}[t!]
\centering
\caption{Scalability \& computational overhead. \textcolor{red}{Red} indicates the shortest time, and \textcolor{yellow!80!orange}{yellow} indicates the second shortest time.}
\label{tab:cost}
\resizebox{\linewidth}{!}{
\begin{tabular}{l c c }
\toprule
& \textbf{Total Time (s)} & \textbf{GPT3.5 Total Cost (\$)} \\
\midrule
MIA-RAG-gray & 7.65 & 0 \\
MIA-RAG-black & \cellcolor{red!30}2.55 & 0 \\
MBA & 32.40 & 0 \\
S\textsuperscript{2}MIA & 12.69 & 0\\
IA & 61.35 & 0.00651\\
Adversary 1 & 18.51 & 0.01160\\
Adversary 2 & 12.34 & 0.00744\\
Adversary 3 & \cellcolor{yellow!50}6.17 & 0.00372\\
\bottomrule
\end{tabular}
}
\end{table}

\end{document}